\documentclass{elsart}
\usepackage{amsmath}
\usepackage{graphicx}

\begin{document}
\begin{frontmatter}
\title{Vortex generation in the RSP game on the triangular lattice}
\author{Hiroyuki Nishiuchi$^\mathrm{a}$, Naomichi Hatano$^\mathrm{b, 1}$, Kenn Kubo$^\mathrm{a, 2}$}
 
\thanks{e-mail: hatano@iis.u-tokyo.ac.jp}
\thanks{e-mail: kkubo@phys.aoyama.ac.jp}
\address{
$^{\rm a}$Department of Physics and Mathematics, Aoyama Gakuin University\\
5-10-1 Fuchinobe, Sagamihara, Kanagawa 229-8558, Japan}
\address{
$^{\rm b}$Institute of Industrial Science, University of Tokyo\\
4-6-1 Komaba, Meguro, Tokyo 153-8505, Japan}


\begin{abstract}
A new model of population dynamics on lattices is proposed.
The model consists of players on lattice points, each of which plays the RSP game with neighboring players.
Each player chooses the next hand from the hand of the neighboring player with the maximum point.
The model exhibits a steady pattern with pairs of vortices and sinks on the triangular lattice.
It is shown that the stationary vortex is due to the frustrations on the triangular lattice.
A frustration is the three-sided situation where each of the three players around a triangle chooses the rock, the scissors and the paper, respectively.
\end{abstract}
\end{frontmatter}

\section{Introduction}
\label{sec1}

The RSP game~\cite{Hopbauer98} is a game where players take their move simultaneously, each choosing a hand from the rock (R), the scissors (S) and the paper (P).
The cyclic strength relation of the three hands determines the win and the loss;
the rock crushes the scissors, the scissors cut the paper and the paper wraps up the rock.
The cyclic competition of the RSP game can mimic various relations in reality, particularly in the population dynamics;
\textit{e.g.}\ a colony of three competing mutations of \textit{E.\ coli}~\cite{Kerr02} and a three-morph mating system of a lizard~\cite{Sinervo96}.

The present study proposes a model of the RSP game on lattices.
Each player on a lattice point chooses the next hand from the hand of the neighboring player with the maximum point.
(We refer to such a player as a copy player.)
We found interesting spatial patterns, such as vortices and sinks, appearing particularly on the triangular lattice.
The spatial pattern with vortices and sinks appears as a coexisting steady state on the triangular lattice.

\begin{figure}
\begin{center}
\includegraphics[width=0.5\textwidth]{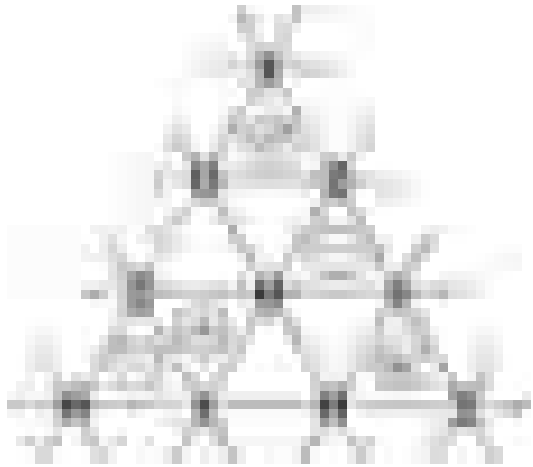}
\end{center}
\caption{RSP players on the triangular lattice. The hands~0, 1, and~2 can form frustrations.}
\label{fig1}
\end{figure}

As far as we know, the previous studies considered the RSP game either on the square lattice~\cite{Tainaka88,Tainaka89,Tainaka94,Fisch90,Durrett98,Ravasz04,Szolnoki05} or on various networks~\cite{Szabo04,Szolnoki04b,Lieberman05,Masuda05,Masuda06}.
It is, however, easy to imagine that triangles appear in the population dynamics in reality, \textit{e.g.}~clusters in complex networks.
It is also known that elementary properties can be very different in many-body systems on non-bipartite lattices and on bipartite lattices.
The vortex pattern that we observe in the present study is, in fact, due to the frustration of the triangular lattice (Fig.~\ref{fig1}), a three-sided situation where each of the three players around a triangle chooses the rock, the scissors and the paper, respectively.
(Hereafter, we refer to the three hands simply as the hands~2, 1 and~0, respectively.)

The existence of vortices was pointed out by some studies in the past~\cite{Tainaka89,Tainaka94,Fisch90}.
We here stress the importance of the frustration as the cause of the stationary vortex pattern.
We show that the stationary vortex pattern does not appear on the square lattice nor on the honeycomb lattice.

The paper is organized as follows.
In Sec.~\ref{sec2}, we introduce the new model and discuss its elementary properties.
We show that pairs of vortices and sinks can appear as spatial patterns.
We also argue that players close a vortex core scores a high point while players close to a sink scores a low point.
We report the results of our simulation on the triangular lattice in Sec.~\ref{sec3} and on the square and honeycomb lattices in Sec.~\ref{sec4}.
We confirm that the spatial pattern with vortices and sinks is stationary on the triangular lattice, while it is not on the square nor honeycomb lattices.
In Sec.~\ref{sec5}, we introduce a random player, who chooses its hand randomly.
We show that a random player can be a source in the spatial pattern.

\section{Lattice RSP model}
\label{sec2}
\subsection{Definition of the model}
\label{sec2-1}

We propose a model where players residing on lattice points repeatedly play the RSP game with the nearest neighbors.
We hereafter consider the triangular lattice, the square lattice and the honeycomb lattice, with an emphasis on the triangular lattice, whose frustration generates stationary vortices in the course of the RSP game.

All players on the lattice points make their moves all at once, which constitutes one time step.
A move is either 0, 1 or 2.
The hand~1 wins over the hand~0, the hand~2 wins over the hand~1, and the hand~0 wins over the hand~2.
A win, a draw or a loss are determined between each pair of the nearest neighbors of the lattice.
Each player scores one point for a win, zero point for a draw and minus one point for a loss.
Hence, a player can score $z$ points at most and minus $z$ points at least in each time step, where $z$ is the number of the nearest neighbors on the lattice ($z=6$ for the triangular lattice, $z=4$ for the square lattice and $z=3$ for the honeycomb lattice). 
This is a zero-sum game; that is, the sum of the scores of all the players is always zero.

Particularly on the triangular lattice, we define the frustration and its sign (Fig.~\ref{fig1}).
We refer as a positive frustration to the situation where the hands~2, 1, and~0 appear in this order when we circle around a triangle counterclockwise.
On the other hand, a negative frustration is the situation where the hands~0, 1, and~2 appear in this order when we circle around a triangle counterclockwise.
We will argue in the next subsection that a positive frustration generates a counterclockwise vortex, whereas a negative frustration generates a clockwise vortex.

All the players choose their hands at random in the initial time step with an equal probability.
Each player adopts the copy strategy or the random strategy afterwards.
The copy strategy is to choose a hand of the player who, of all the nearest neighbors and the player itself, marked the highest score in the last time step.
We refer to a player adopting the copy strategy as a copy player hereafter.
If there are more than a player of the highest score with different hands, a copy player chooses a hand from their hands randomly.
The random strategy is to choose a hand at random.
We refer to a player adopting the random strategy as a random player.
In the present study, we consider only the case where each player is either a copy player or a random player all through the game.

We mostly consider copy players hereafter.
We show that copy players on the triangular lattice exhibit vortex structure.
In Sec.~\ref{sec5}, we discuss an impact of random players on the structure as impurities.

\subsection{Vortices, sinks and sources}

Before showing the simulation results, let us argue that two spatial patterns typically appear.
One is a vortex and the other is a sink. They are logical consequences of the combination of the RSP game and the copy strategy.

Note first that copy players tend to form domains of the same hands.
A copy player well inside a domain of, say, the hand~0, will keep the hand~0 in the next time step because all its neighbors are of the hand~0 and their scores are all zero; hence the bulk of the domain is stable.
Copy players on the boundary of a domain, on the other hand, may change their hands in the next time step, and hence the boundary moves.

\begin{figure}
\begin{minipage}[b]{0.45\textwidth}
\vspace{0pt}
\includegraphics[width=\textwidth]{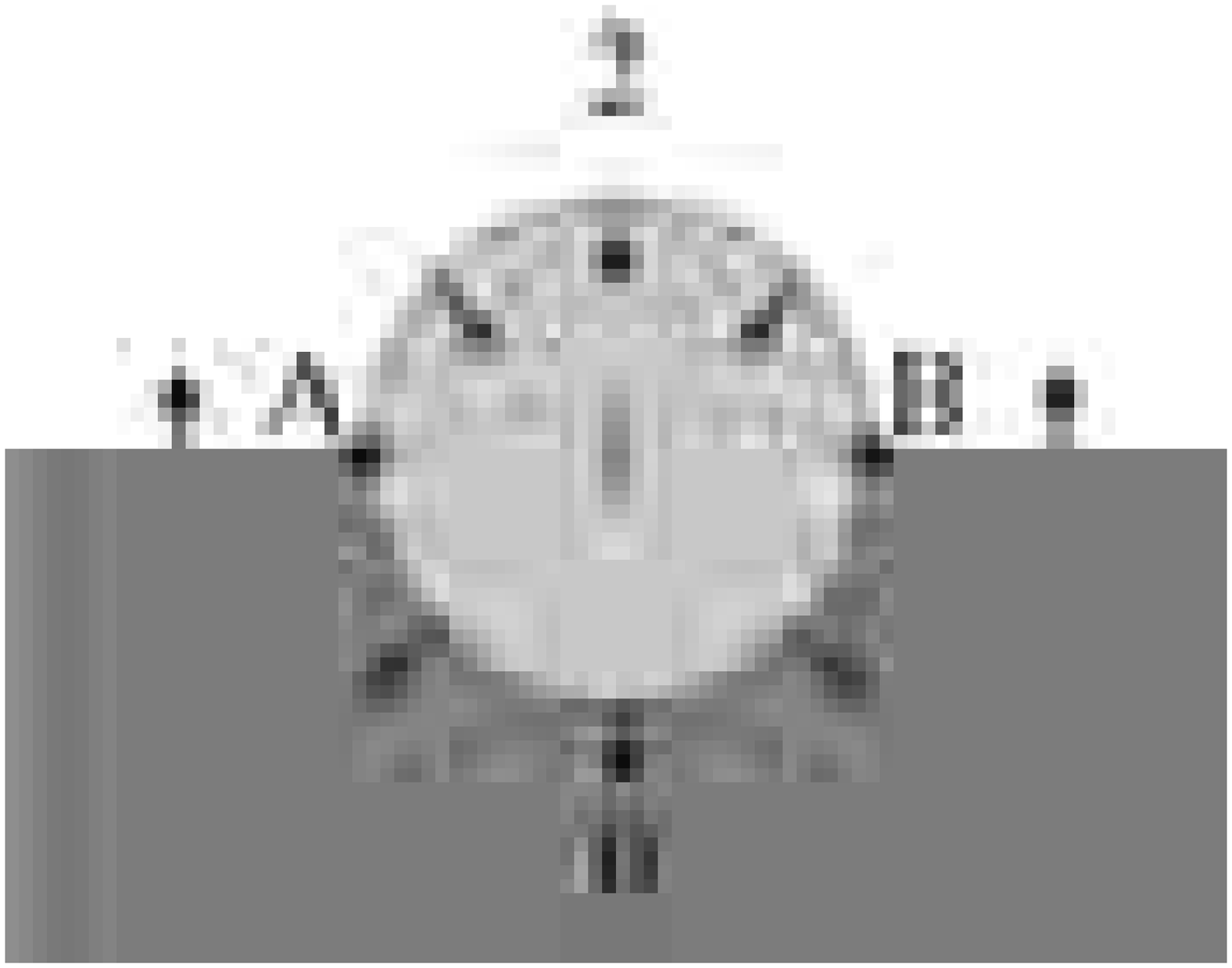}
\centering (a)
\end{minipage}
\hfill
\begin{minipage}[b]{0.4\textwidth}
\vspace{0pt}
\includegraphics[width=\textwidth]{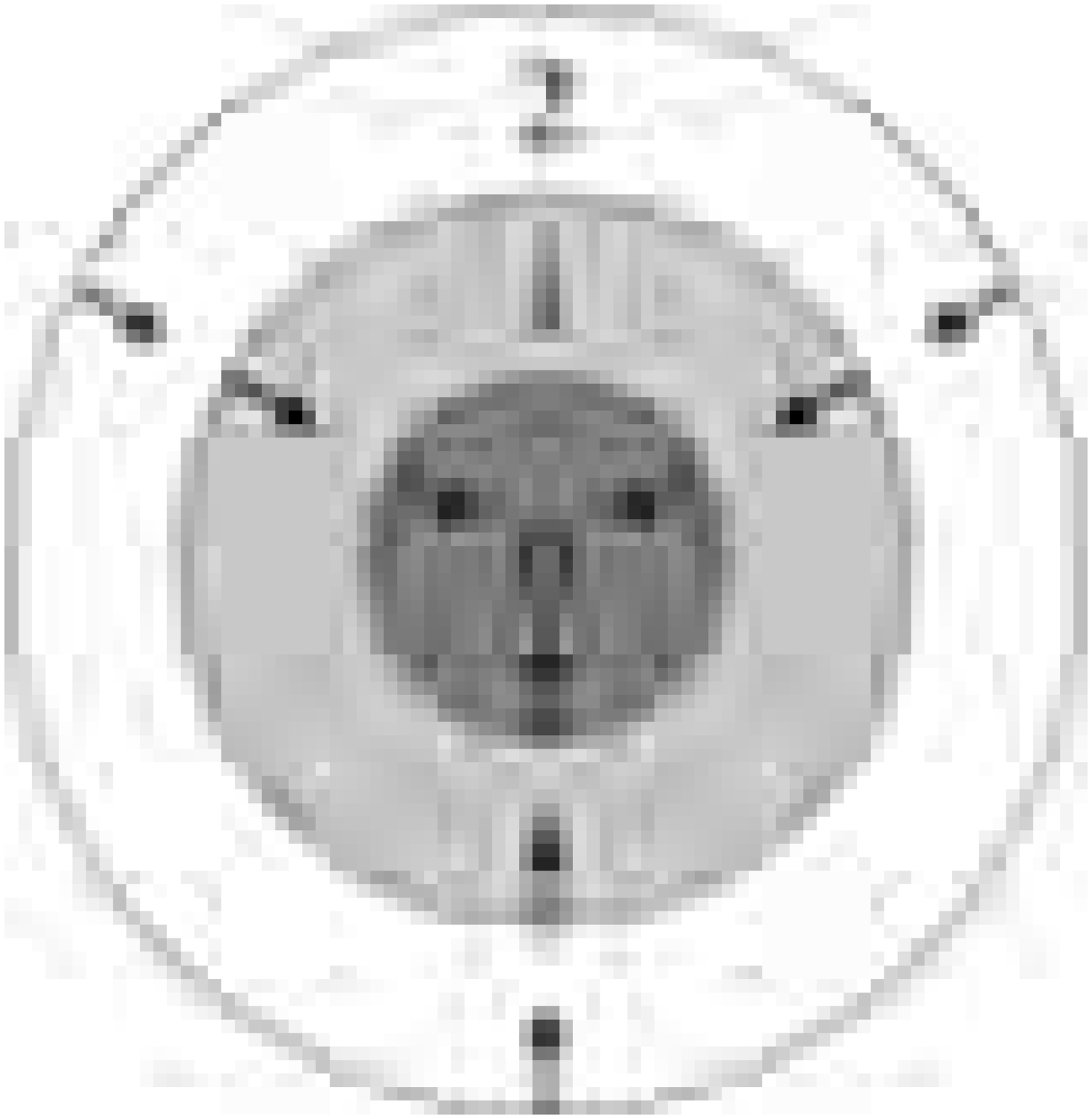}
\centering (b)
\end{minipage}
\\
\begin{center}
\includegraphics[width=0.7\textwidth]{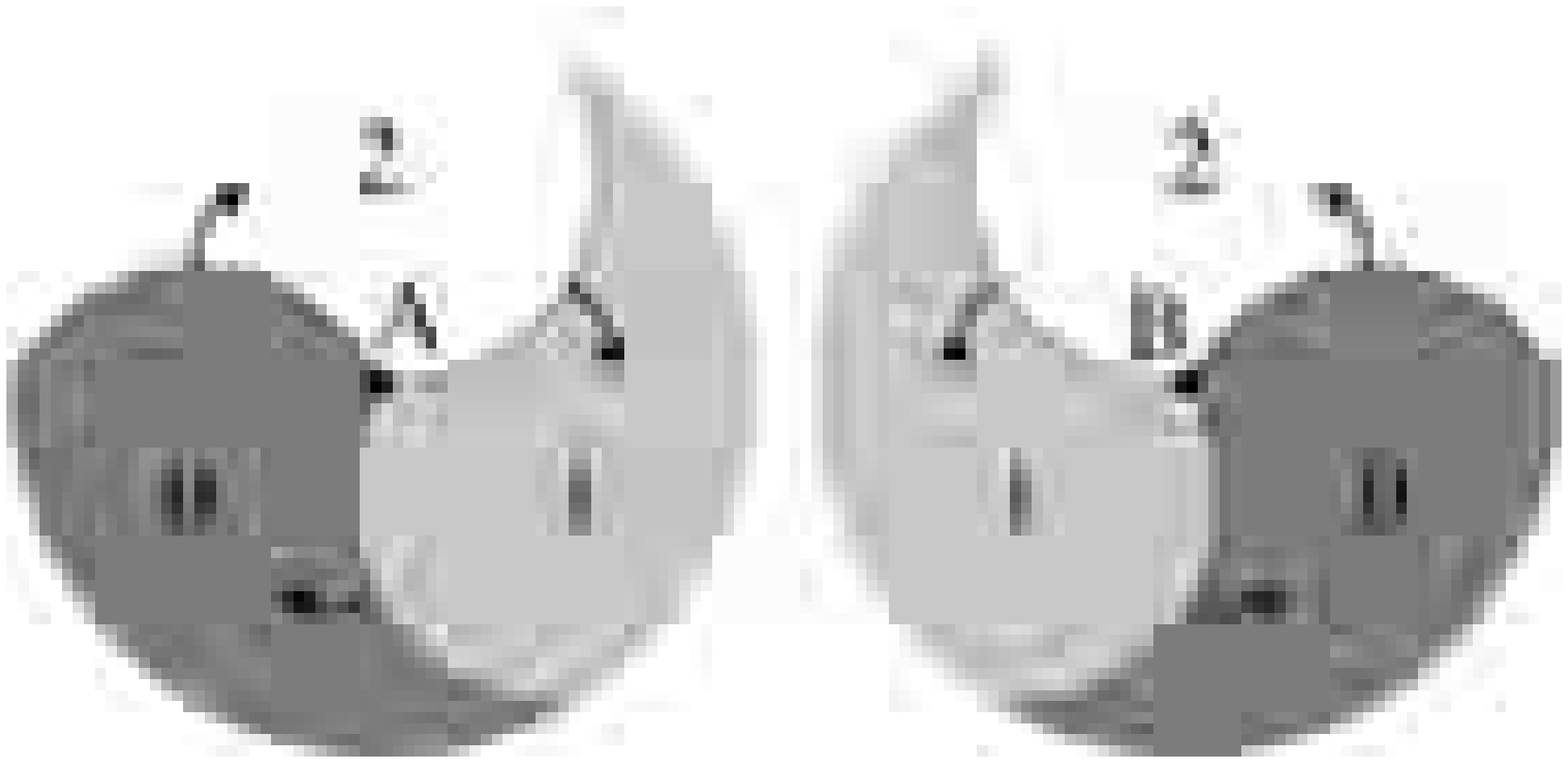}

(c)
\end{center}
\caption{(a) and (b) Two possible topologies of the three domains.
(c) The topology of (a) generates a pair of vortices.}
\label{fig2}
\end{figure}

Let us argue how the boundary moves in the following two cases.
The three domains of the hands 0, 1 and 2 can have either the topology of Fig.~\ref{fig2}~(a) or~(b).

In the topology of Fig.~\ref{fig2}~(a), there is a negative frustration around the point A and a positive frustration around the point B.
A copy player of, say, the hand~0, located just outside the domain of the hand~1, tends to choose the hand~1 in the next time step because the neighbors with the hand~1 get high scores.
Thus the boundary between the domains of the hand~0 and the hand~1 moves onto the the domain of the hand~0, so that the domain of the hand~1 expands.
Likewise, the boundary between the domains of the hand~1 and the hand~2 moves onto the domain of the hand~1 and the boundary between the domains of the hand~2 and the hand~0 moves on to the domain of the hand~2.
Hence the boundaries rotate clockwise around the negative frustration at the point A and counterclockwise around the positive frustration at the point B.
We will indeed show below in Section~\ref{sec3-2} that the boundaries take a configuration schematically illustrated in Fig.~\ref{fig2}~(c).
That is, the topology of Fig.~\ref{fig2}~(a) generates a pair of vortices of moving boundaries.
(A similar argument for a different model can be found in Refs.~\cite{Szabo99,Szolnoki04}.)

We refer to the counterclockwise vortex as a positive vortex and the clockwise vortex as a negative vortex.
In short, a positive frustration of a configuration generates a positive vortex of moving boundaries and a negative frustration generates a negative vortex.

In the topology of Fig.~\ref{fig2}~(b), on the other hand, the circular boundaries shrink toward the center;
the players with the hand~0 just inside the boundary mimic the players with the hand~1 just outside the boundary.
The central domain of the hand~0 collapses eventually.
Then the domain of the hand~1 becomes the central domain and will collapse after a while.
Thus the topology of Fig.~\ref{fig2}~(b) generates a sink.

Finally, a source does not appear when there are only copy players, because a new domain is never generated inside a domain.
It is spontaneously generated only when some players adopt strategies other than the copy strategy.
Specifically, a random player can be a source as is shown in Sec.~\ref{sec5}.

\subsection{Scores in vortices and sinks}
\label{sec2-3}

\begin{figure}
\begin{minipage}[b]{0.45\textwidth}
\vspace{0pt}
\includegraphics[width=\textwidth]{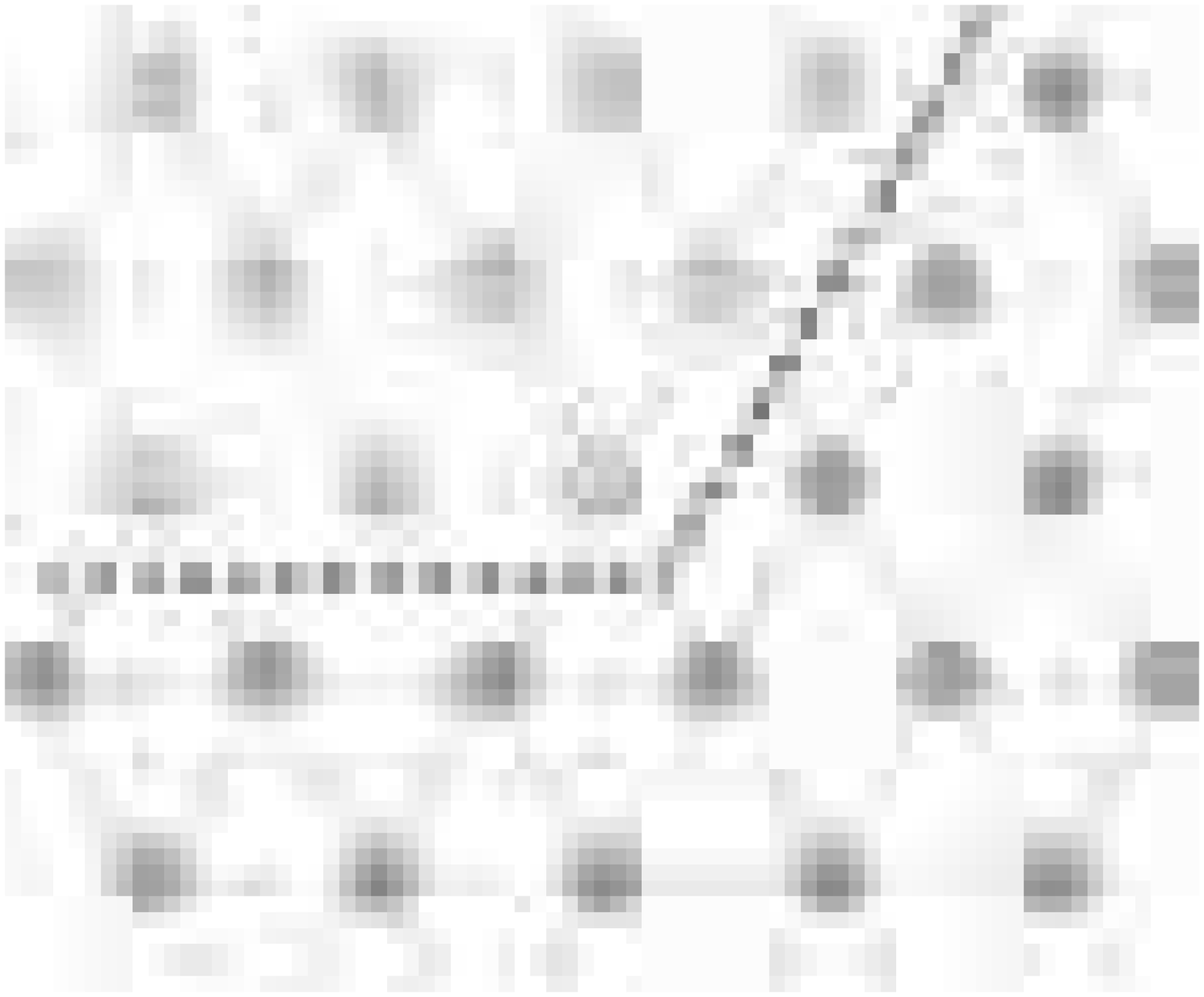}
\centering (a)
\end{minipage}
\hfill
\begin{minipage}[b]{0.45\textwidth}
\vspace{0pt}
\includegraphics[width=\textwidth]{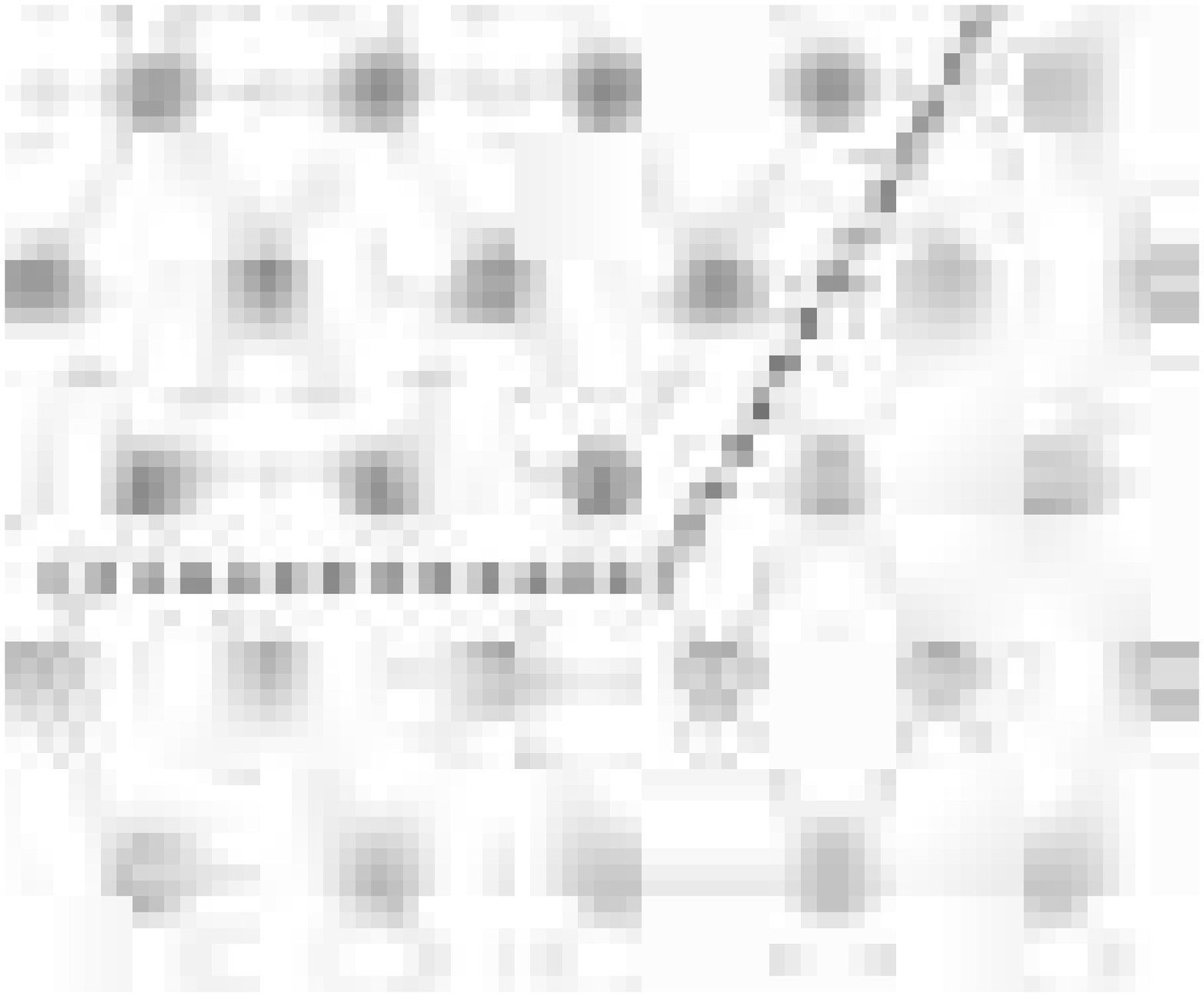}
\centering (b)
\end{minipage}
\\
\begin{center}
\includegraphics[width=\textwidth]{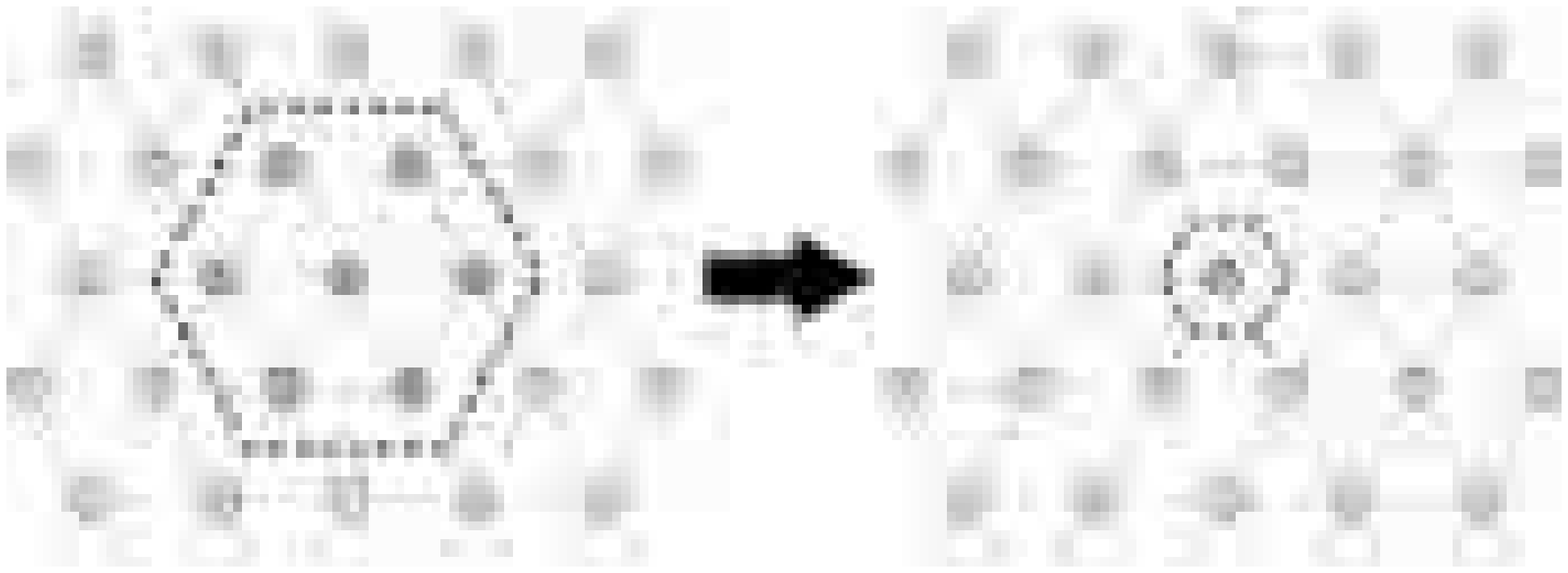}
\\
(c)
\end{center}
\caption{A part of the boundaries in Fig.~\ref{fig2}~(c) and~(b):
(a) A bend around the frustration A of Fig.~\ref{fig2}~(c);
(b) A bend around the sink of Fig.~\ref{fig2}~(b).
(c) A simple case of two subsequent steps.}
\label{fig3}
\end{figure}

We here argue that the scores of the players near a vortex core are high, while those near a sink are low.
Both in Fig.~\ref{fig2}~(b) and~(c), the boundaries are not straight.
Near the vortex cores in Fig.~\ref{fig2}~(c), the boundary is convex from the viewpoint of the winners (the hand~2 in Fig.~\ref{fig3}~(a)) and concave from the viewpoint of the losers (the hand~1 Fig.~\ref{fig3}~(a)).
Near the sink in Fig.~\ref{fig2}~(b), on the other hand, the boundary is convex from the viewpoint of the losers (the hand~1 in Fig.~\ref{fig3}~(b)) and concave from the viewpoint of the winners (the hand~2 in Fig.~\ref{fig3}~(b)).

Around the bend of the boundary in Fig.~\ref{fig3}~(a), the number of the winners  (the hand~2) is one less than the number of the losers (the hand~1).
Since this is a zero-sum game even locally, the total of the positive scores of the winners is equal to the total of the negative scores of the losers.
Therefore, the time-averaged positive score of a winner is greater than the time-averaged negative score of a loser.
For example, the player with the hand~2 at the corner of the boundary scores $+3$, whereas the player with the hand~1 at the corner of the boundary scores $-1$.
In other words, each player wins a high score when it is a winner and loses a low score when it is a loser.
Since each player spends about equal time as a winner and as a loser over a long time, the time-averaged score is positive.
In short, the time-averaged score of a player around a vortex is positive.
The closer a player is to the vortex core, the more often the bend appears in the boundary, and the higher the time-averaged score of the player is.

The situation is the opposite near a sink.
Around the bend of the boundary in Fig.~\ref{fig3}~(b), the number of the winners (the hand~2) is one greater than the number of the losers (the hand~1).
Each player, as a winner, share the total positive score with more players and, as a loser, share the total negative score with less players.
Hence the time-averaged score of a player around a sink is negative.
The closer a player is to the center of the sink, the lower the time-averaged score of the player is.

For example, let us calculate the time-averaged score over the two steps of Fig.~\ref{fig3}~(c).
The score of the central player is zero in the first step and $-6$ in the next step.
The time-averaged score over the two steps is $-3$ for the central player.
The score of the player next to the central player is $-3$ in the first step while $+1$ in the next step.
The time-averaged score over the two steps is $-1$ for the player next to the central player.

\section{Simulation of the society of copy players: triangular lattice}
\label{sec3}

In this section, we show results of our simulations on the triangular lattice.
We simulated the society of copy players on a triangular lattice  with $2^{14}$ players.
We imposed periodic boundary conditions.
We demonstrate that the stationary vortex structure appear.

\subsection{Convergence to a steady pattern}

\begin{figure}
\begin{minipage}[b]{0.45\textwidth}
\vspace{0pt}
\includegraphics[width=\textwidth]{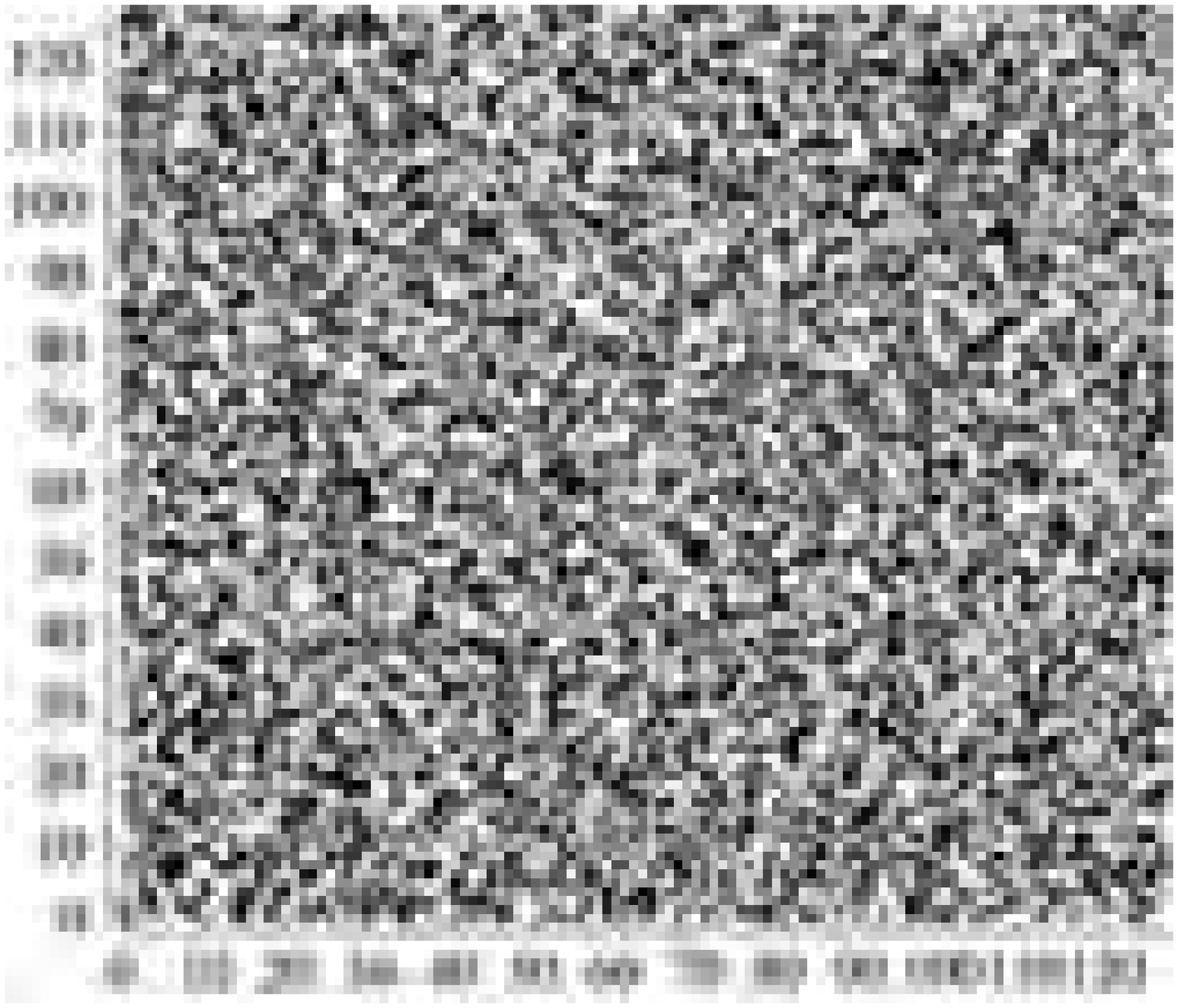}
\centering (a)
\end{minipage}
\hfill
\begin{minipage}[b]{0.45\textwidth}
\vspace{0pt}
\includegraphics[width=\textwidth]{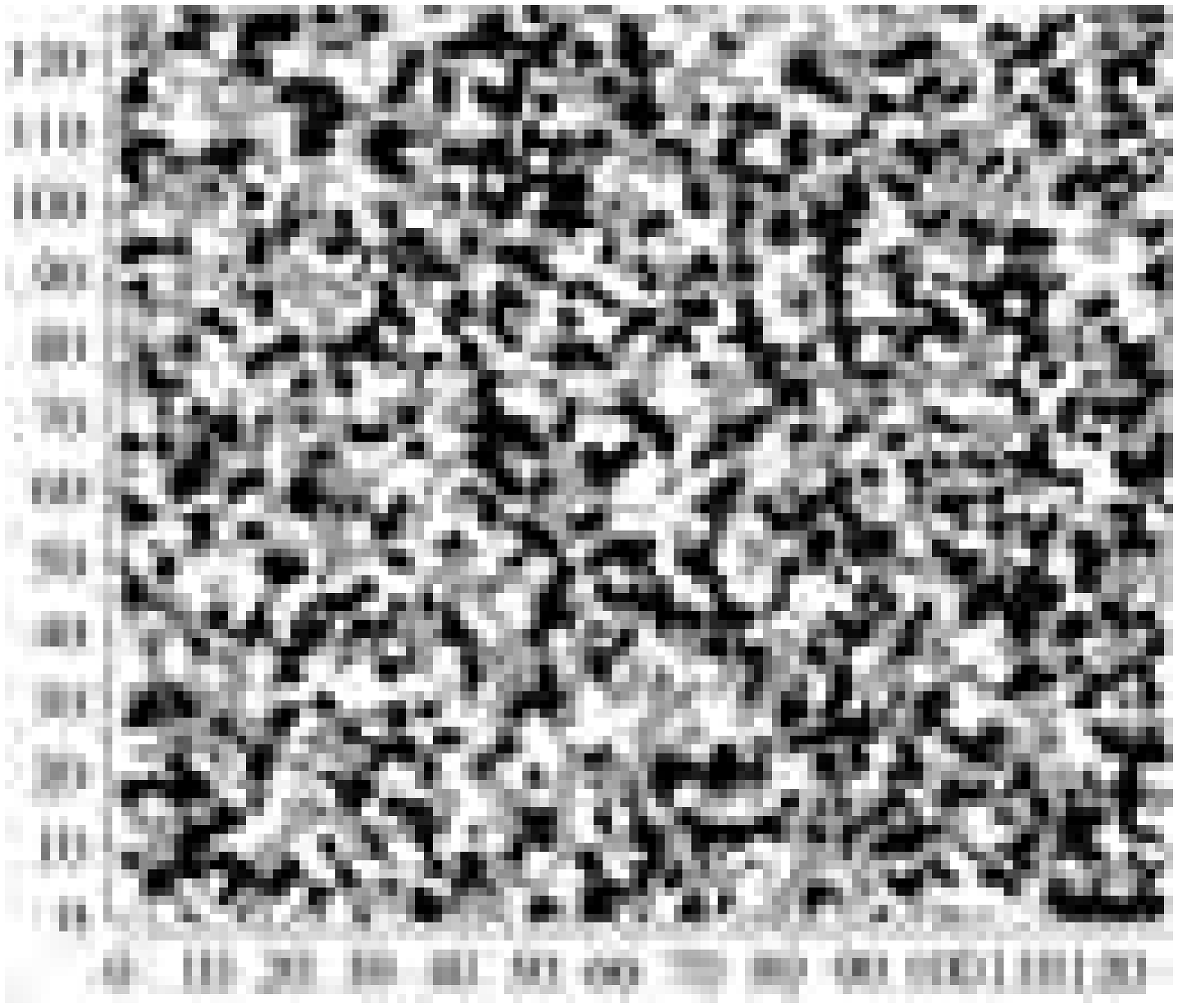}
\centering (b)
\end{minipage}
\\
\begin{minipage}[b]{0.45\textwidth}
\vspace{\baselineskip}
\includegraphics[width=\textwidth]{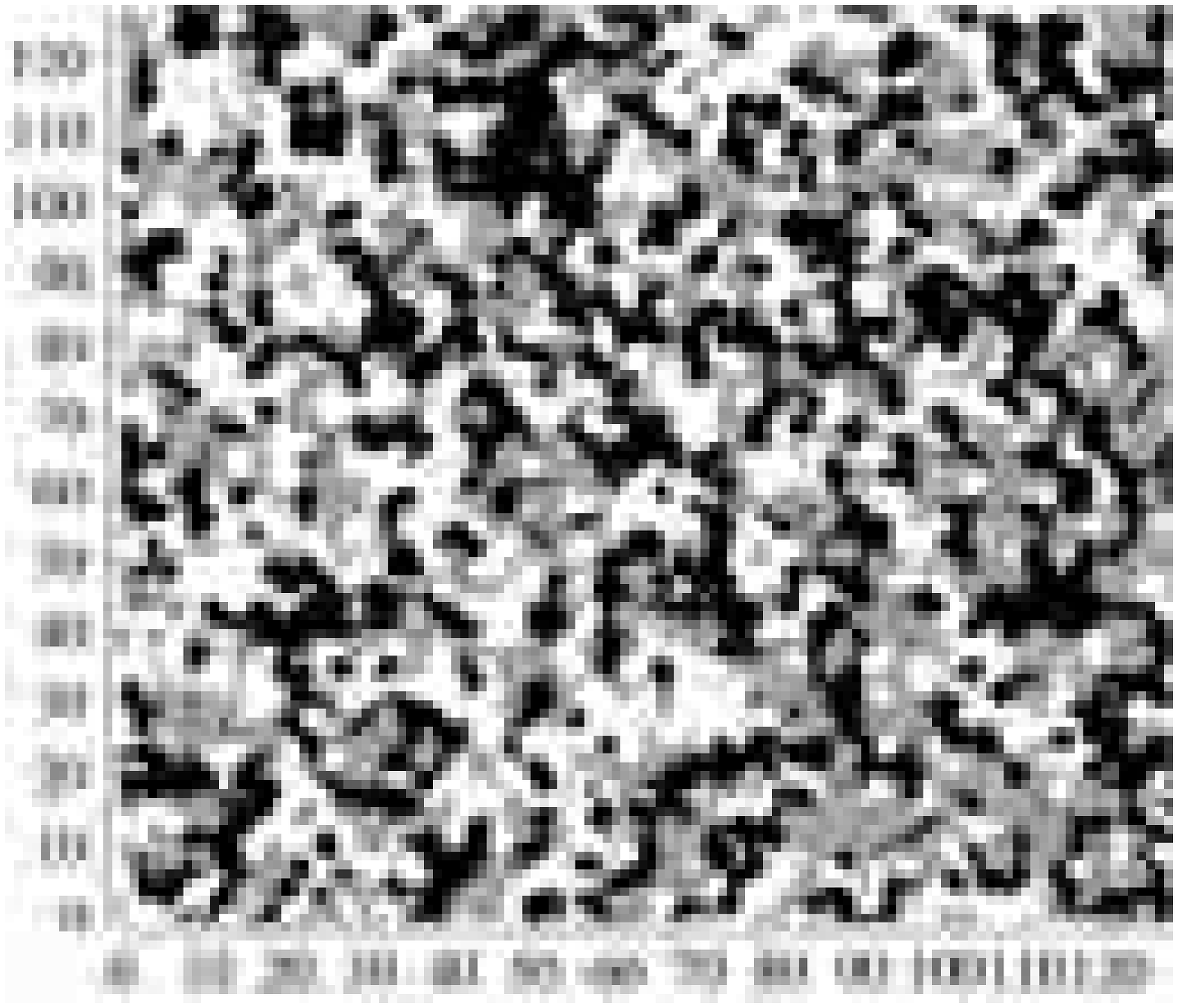}
\centering (c)
\end{minipage}
\hfill
\begin{minipage}[b]{0.45\textwidth}
\vspace{\baselineskip}
\includegraphics[width=\textwidth]{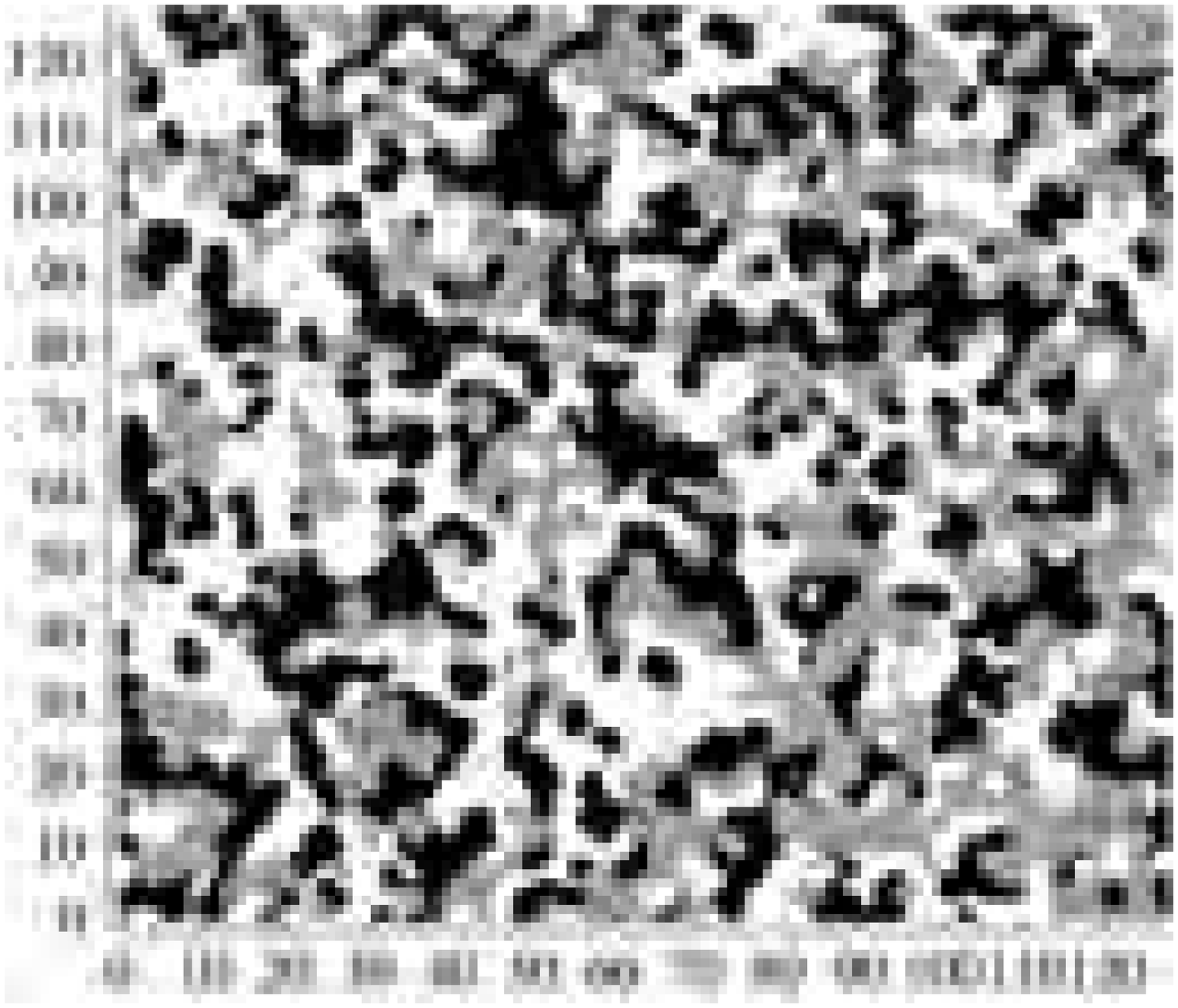}
\centering (d)
\end{minipage}
\caption{The initial configuration (a) and the configurations (b), (c) and (d) at the first three steps of a simulation on the triangular lattice with $2^{14}$ players.
The black hexagons denote the player with the hand~0, the gray hexagons the hand~1, and the white hexagons the hand~2.}
\label{fig4}
\end{figure}

\begin{figure}
\begin{minipage}[b]{0.45\textwidth}
\vspace{0pt}
\includegraphics[width=\textwidth]{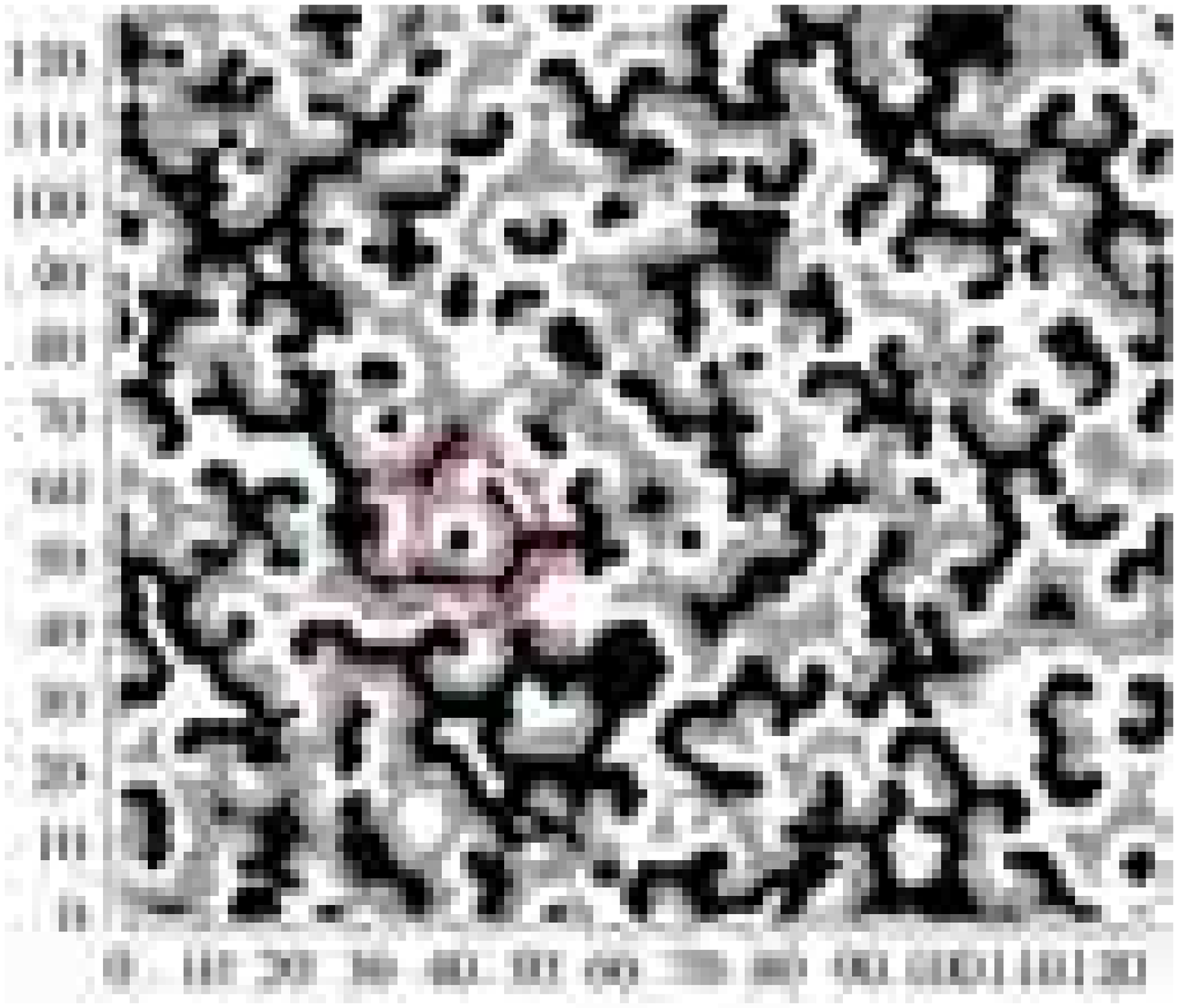}
\centering (a)
\end{minipage}
\hfill
\begin{minipage}[b]{0.45\textwidth}
\vspace{0pt}
\includegraphics[width=\textwidth]{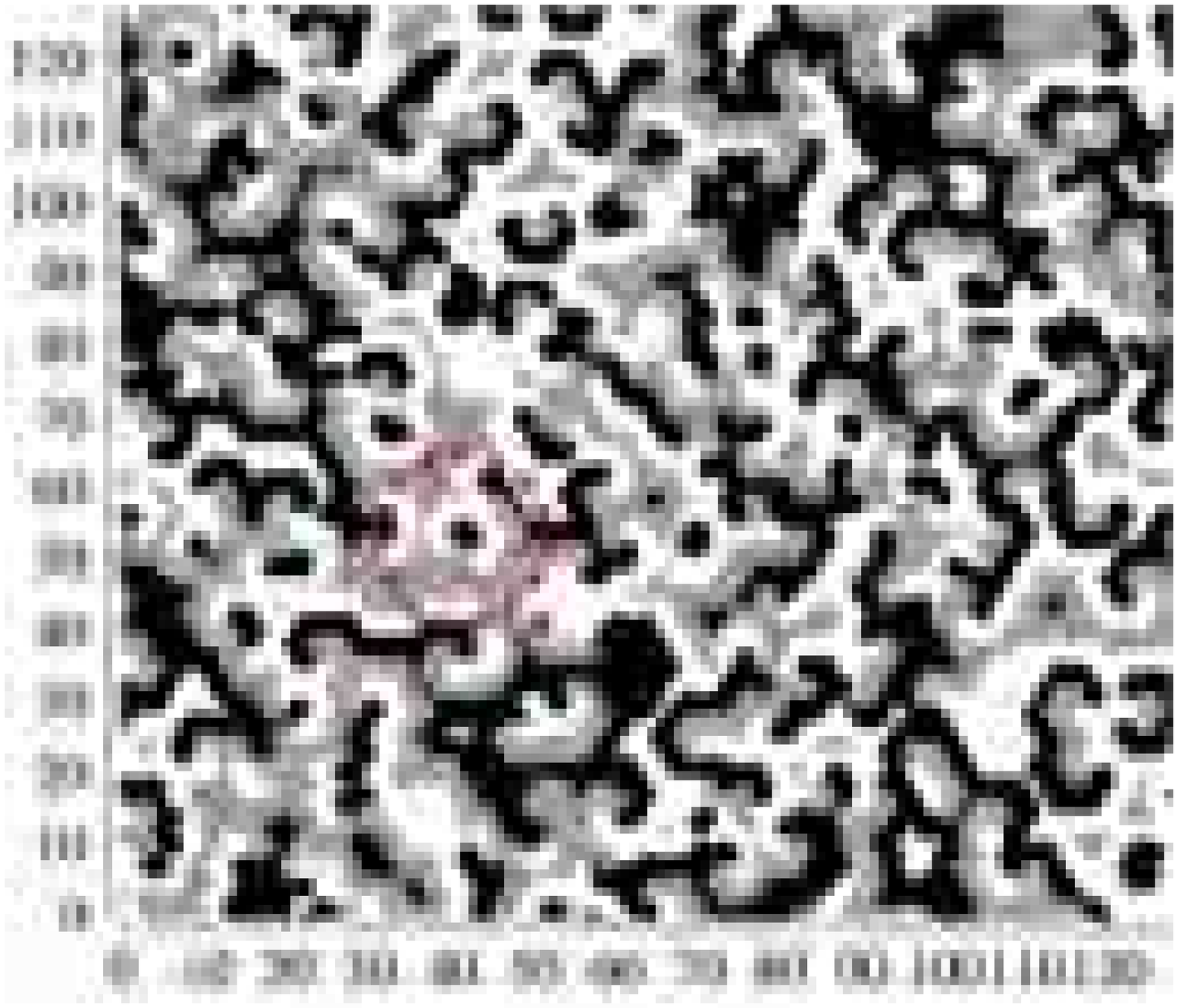}
\centering (b)
\end{minipage}
\\
\begin{minipage}[b]{0.45\textwidth}
\vspace{\baselineskip}
\includegraphics[width=\textwidth]{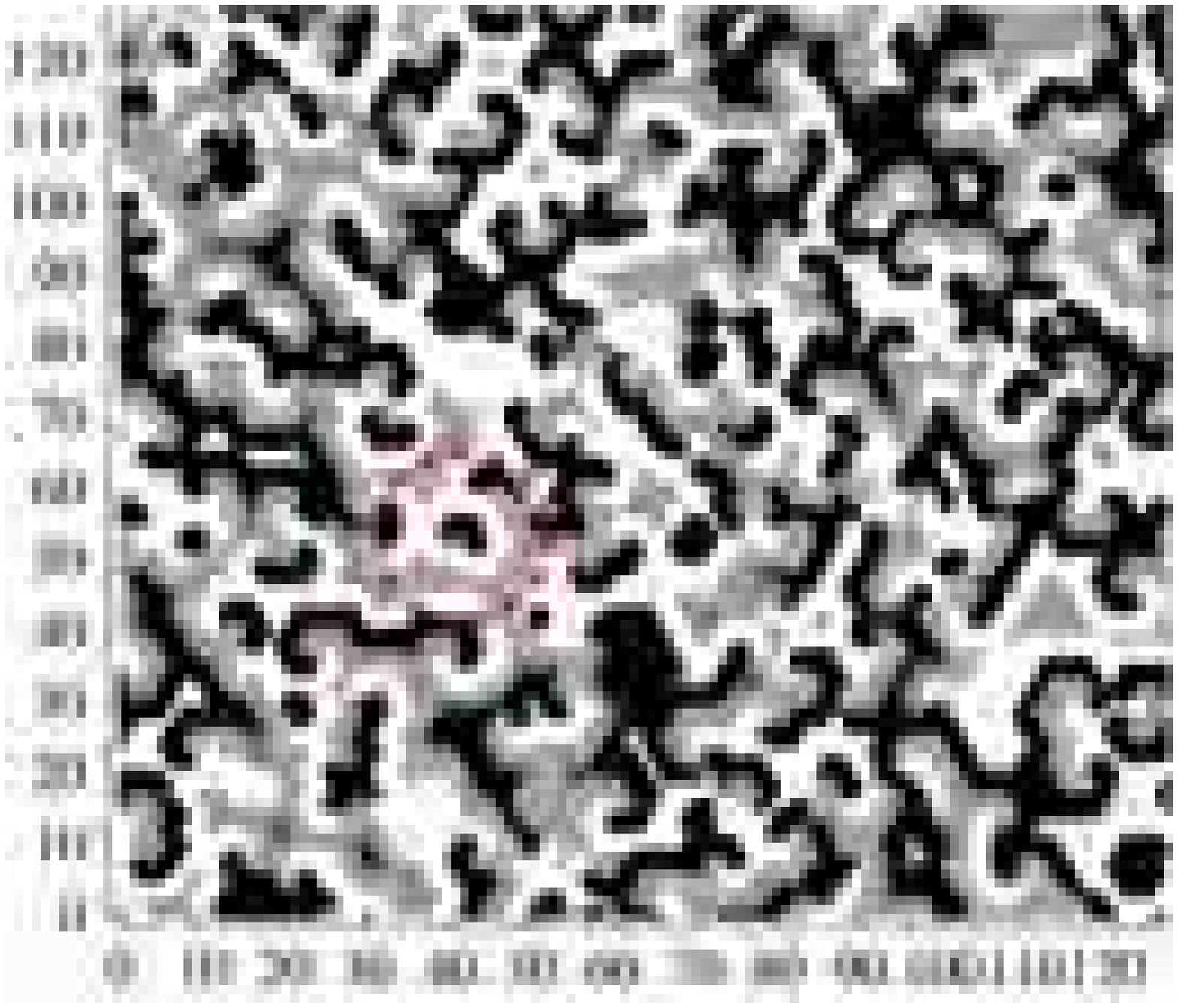}
\centering (c)
\end{minipage}
\hfill
\begin{minipage}[b]{0.45\textwidth}
\vspace{\baselineskip}
\includegraphics[width=\textwidth]{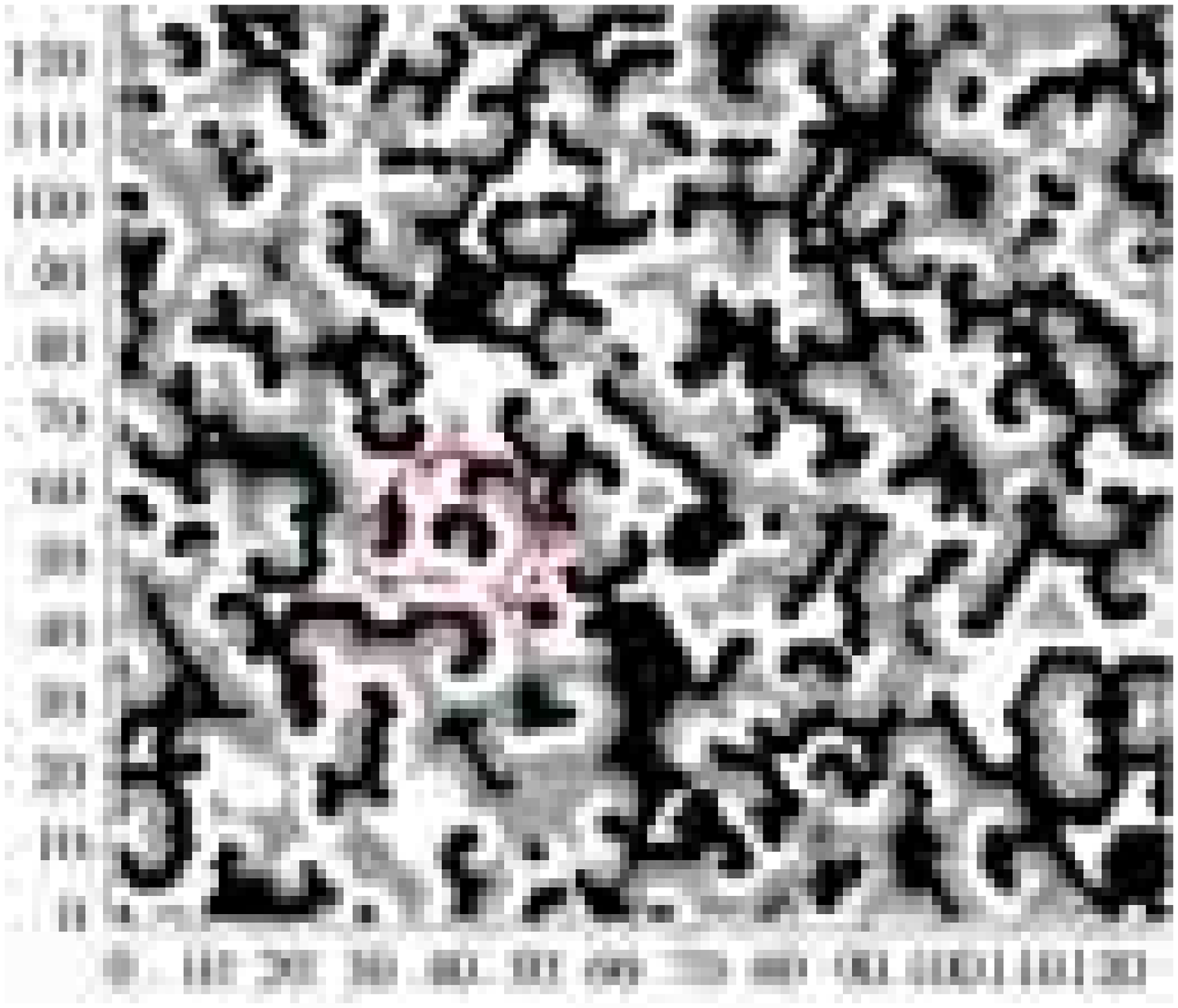}
\centering (d)
\end{minipage}
\\
\begin{minipage}[b]{0.45\textwidth}
\vspace{\baselineskip}
\includegraphics[width=\textwidth]{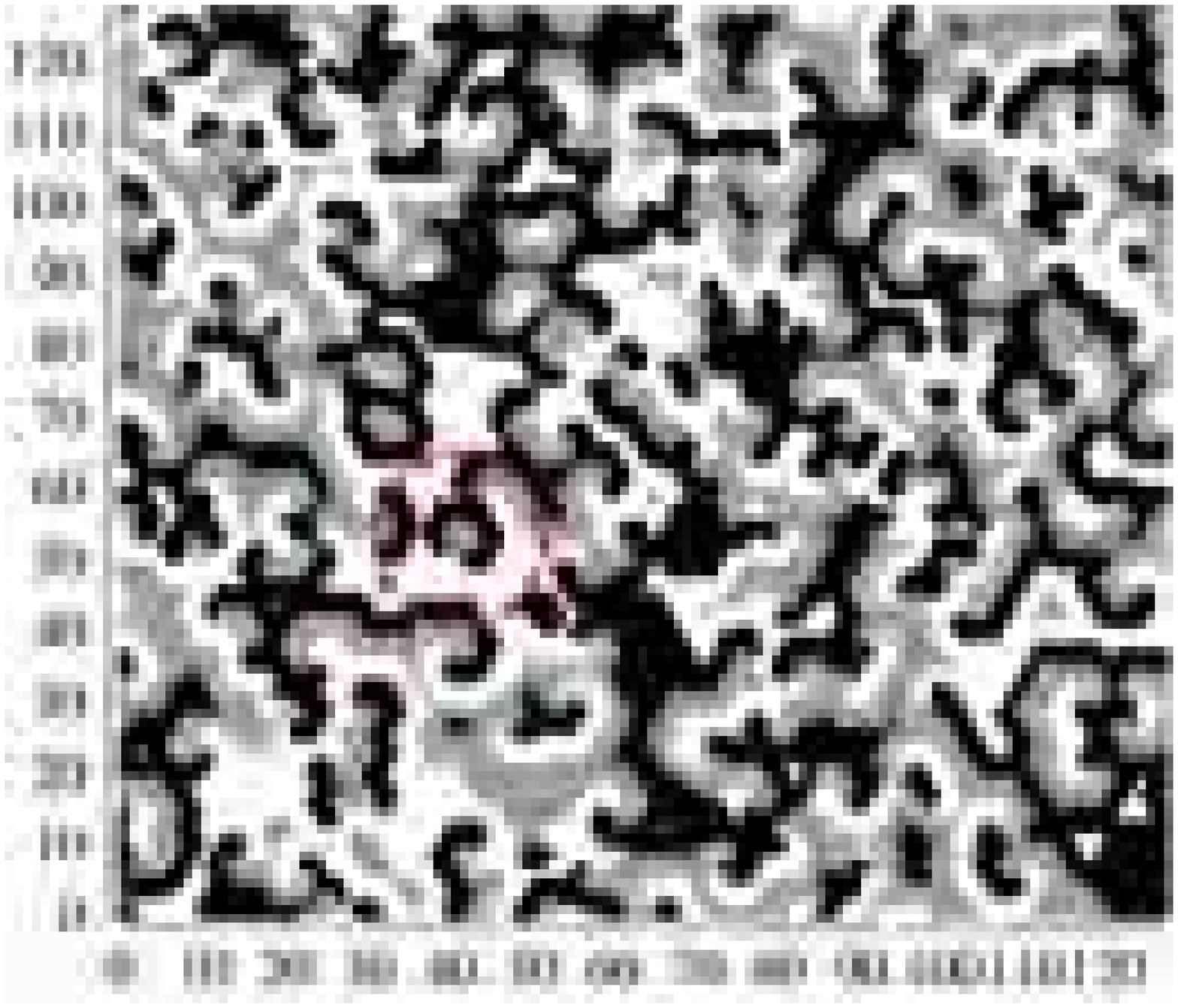}
\centering (e)
\end{minipage}
\hfill
\begin{minipage}[b]{0.45\textwidth}
\vspace{\baselineskip}
\includegraphics[width=\textwidth]{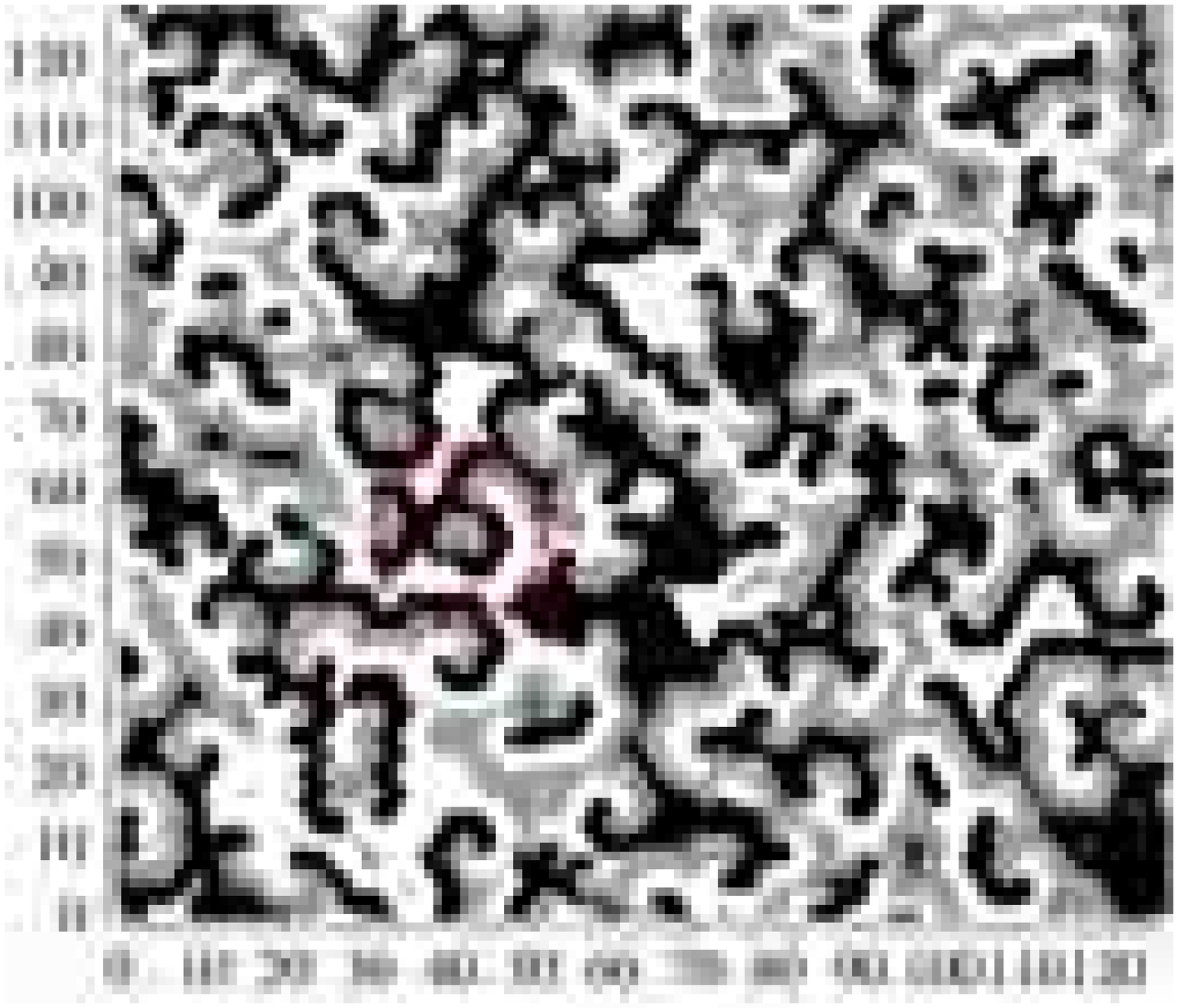}
\centering (f)
\end{minipage}
\caption{The configurations at the 17th step (a), the 18th step (b), the 19 step (c), the 20th step (d), the 21st step (e) and the 22nd step (f) of a simulation on the triangular lattice with $2^{14}$ players.
The black hexagons denote the player with the hand~0, the gray hexagons the hand~1, and the white hexagons the hand~2.}
\label{fig5}
\end{figure}

\begin{figure}
\begin{minipage}[b]{0.45\textwidth}
\vspace{0pt}
\includegraphics[width=\textwidth]{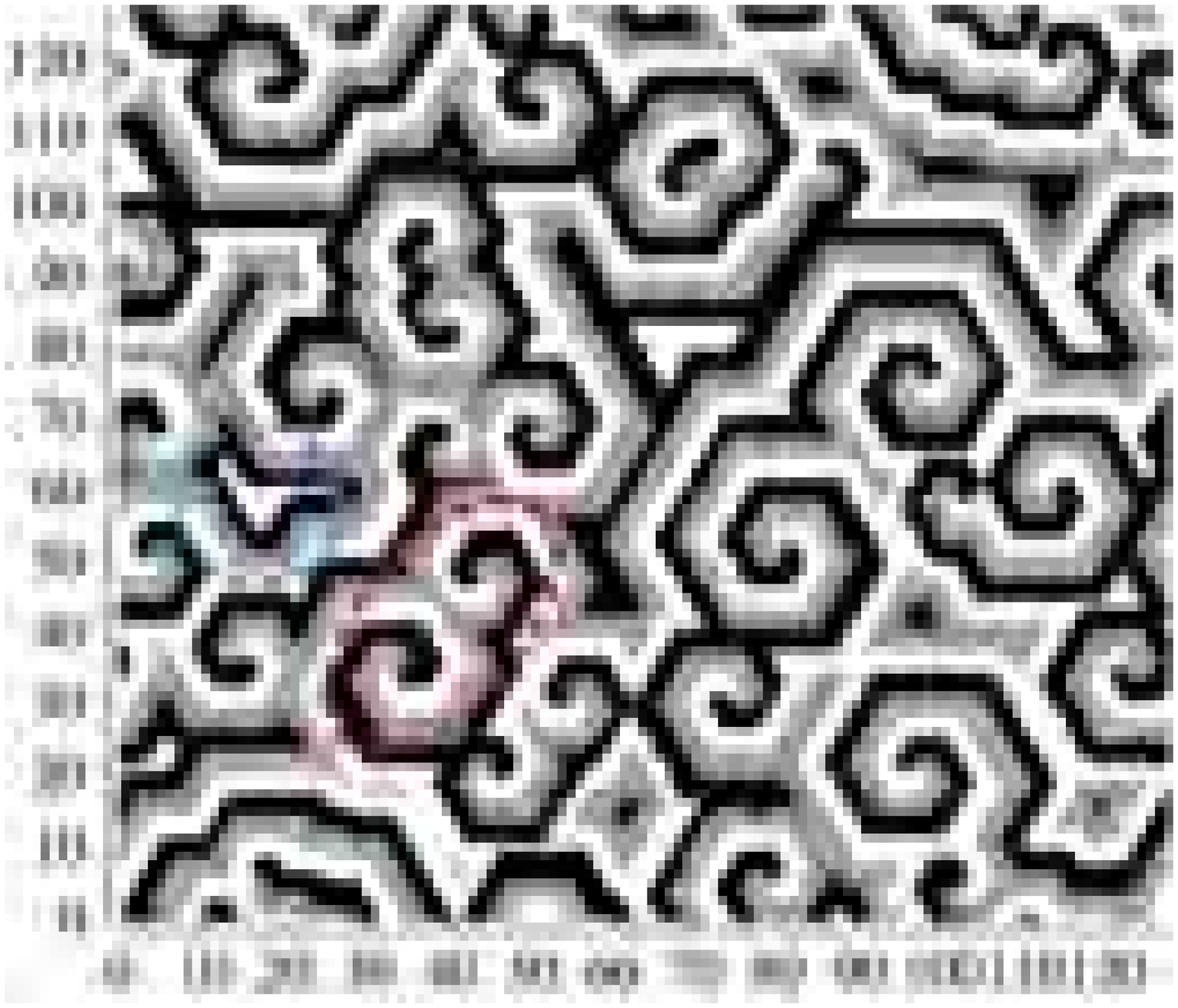}
\centering (a)
\end{minipage}
\hfill
\begin{minipage}[b]{0.45\textwidth}
\vspace{0pt}
\includegraphics[width=\textwidth]{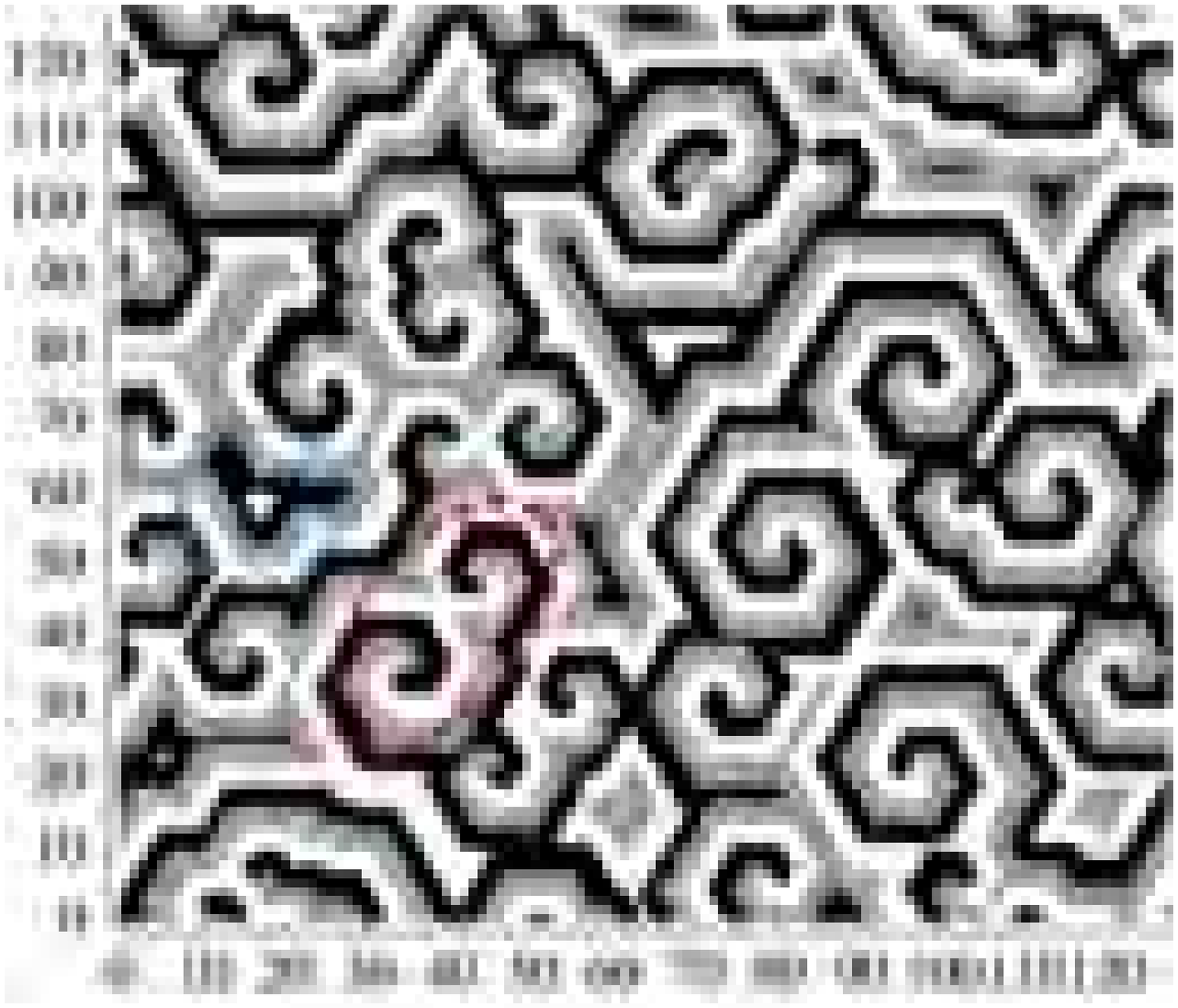}
\centering (b)
\end{minipage}
\\
\begin{minipage}[b]{0.45\textwidth}
\vspace{\baselineskip}
\includegraphics[width=\textwidth]{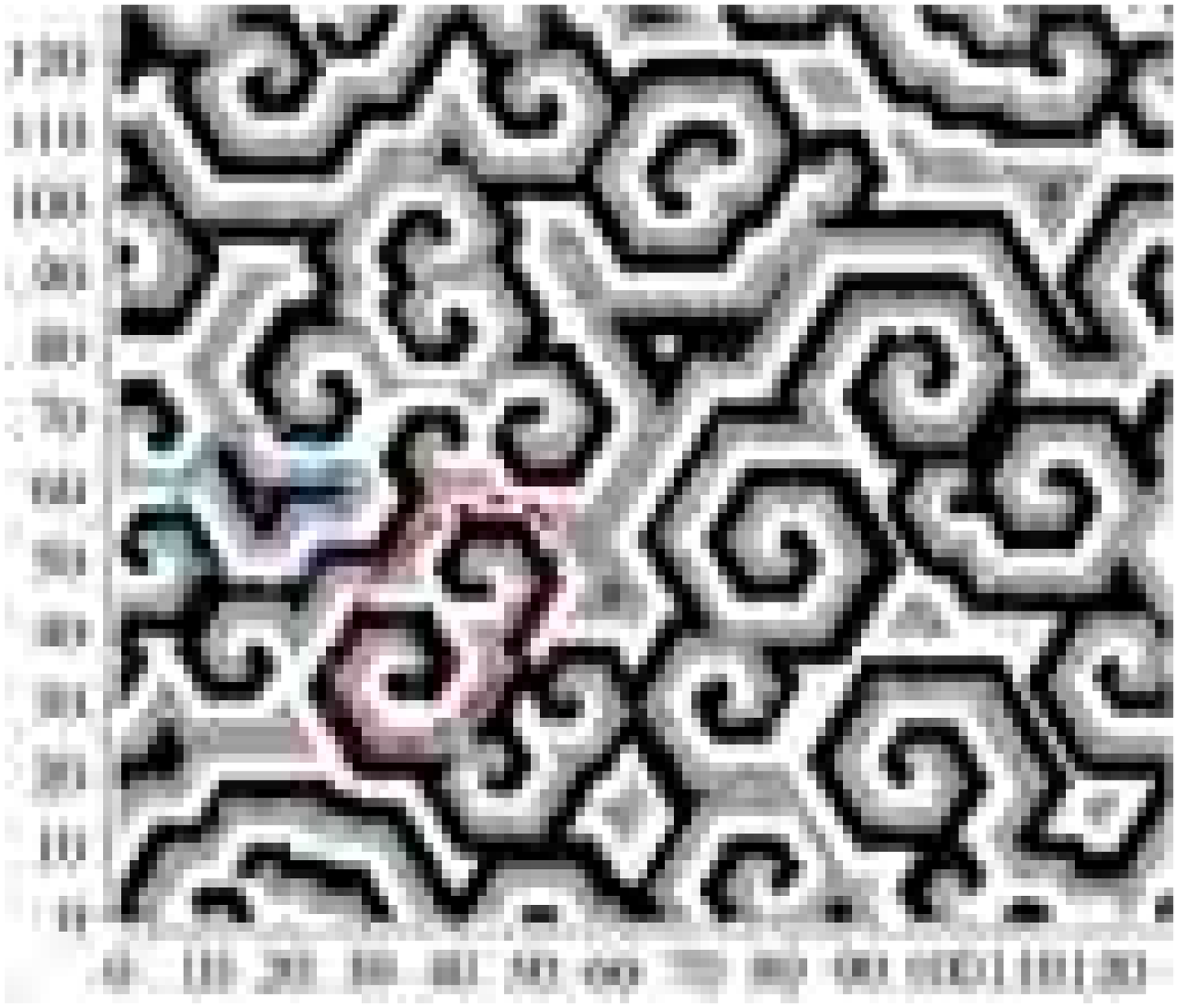}
\centering (c)
\end{minipage}
\hfill
\begin{minipage}[b]{0.45\textwidth}
\vspace{\baselineskip}
\includegraphics[width=\textwidth]{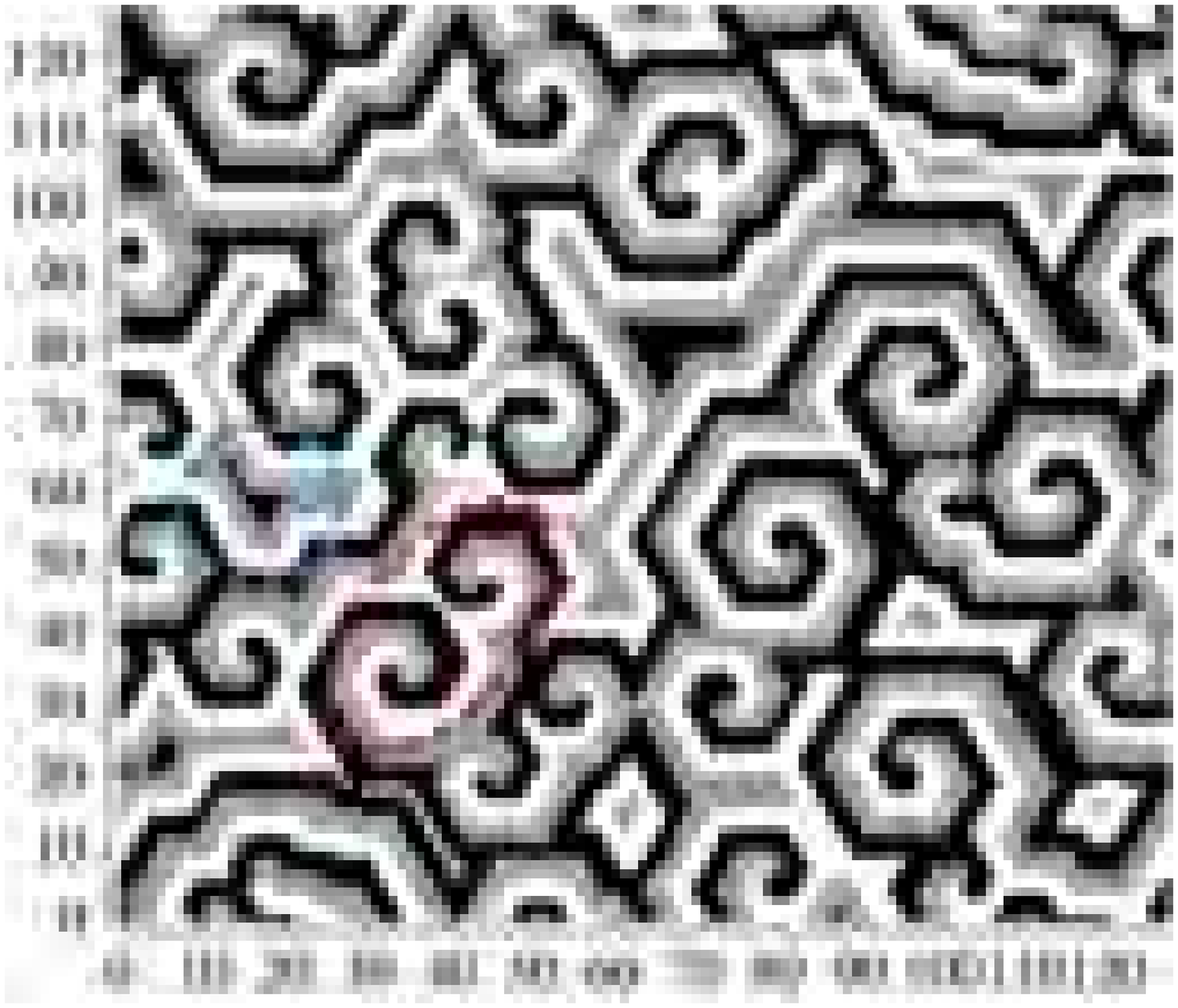}
\centering (d)
\end{minipage}
\caption{The configurations at the 997th step (a), the 998th step (b), the 999 step (c) and the 1~000th step (d) of a simulation on the triangular lattice. with $2^{14}$ players
The black hexagons denote the player with the hand~0, the gray hexagons the hand~1, and the white hexagons the hand~2.}
\label{fig6}
\end{figure}

We first show the convergence to a steady pattern, presenting snapshots of the simulations.
The initial configuration Fig.~\ref{fig4}~(a) was chosen randomly.
The domains of the three hands are quickly formed in the first few iterations as shown in Fig.~\ref{fig4}~(b)--(d).
A typical pattern consisting of vortices and sinks emerge by the 20th step as shown in Fig.~\ref{fig5}.
In the 17th step (Fig.~\ref{fig5}~(a)), we have, for example, a pair of a positive vortex around $(40,54)$ and a negative vortex around $(42,52)$, which is indicated by a red circle.
These vortices have cancelled each other by the 22nd step (Fig.~\ref{fig5}~(f)).
After such cancellations, the pattern settles into a fairly steady state by the 1~000th step as shown in Fig.~\ref{fig6}.
There is a vortex pair, for example, around $(34, 34)$ and $(44, 47)$, which is indicated by a red circle.
There is also a sink, for example, around $(15, 58)$, which is indicated by a blue circle.

\begin{figure}
\begin{center}
\includegraphics[width=0.55\textwidth]{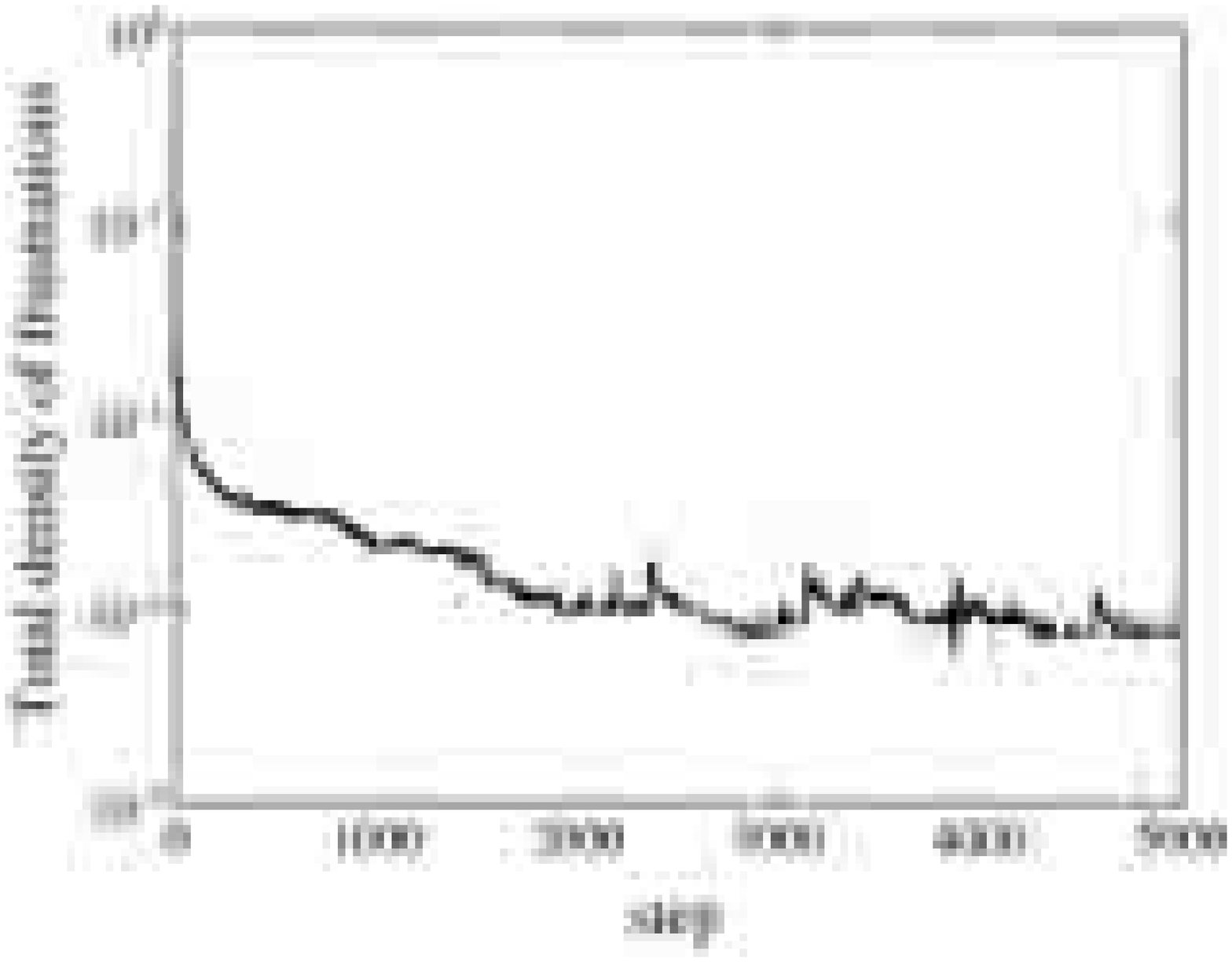}
\end{center}
\caption{The time dependence of the number of (positive and negative) frustrations per player.}
\label{fig7}
\end{figure}

In order to look into details of the convergence, we plot the time dependence of the total number of frustrations per player in Fig.~\ref{fig7}.
We can see that the number of frustrations becomes almost constant after the 3~000th step.

\subsection{Structure of the steady pattern}
\label{sec3-2}

\begin{figure}
\begin{minipage}[b]{0.45\textwidth}
\vspace{0pt}
\includegraphics[width=\textwidth]{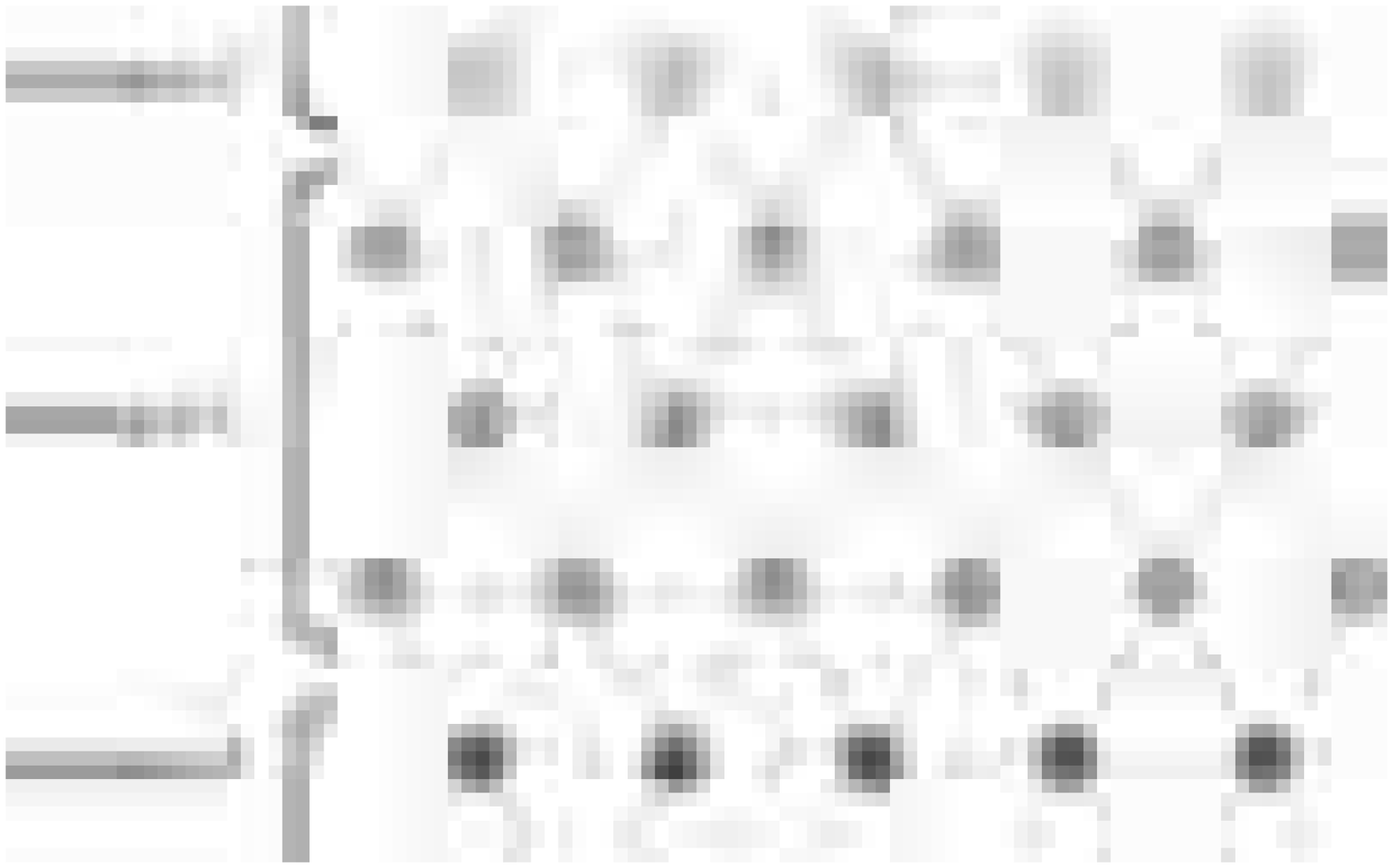}
\centering (a)
\end{minipage}
\hfill
\begin{minipage}[b]{0.45\textwidth}
\vspace{0pt}
\includegraphics[width=\textwidth]{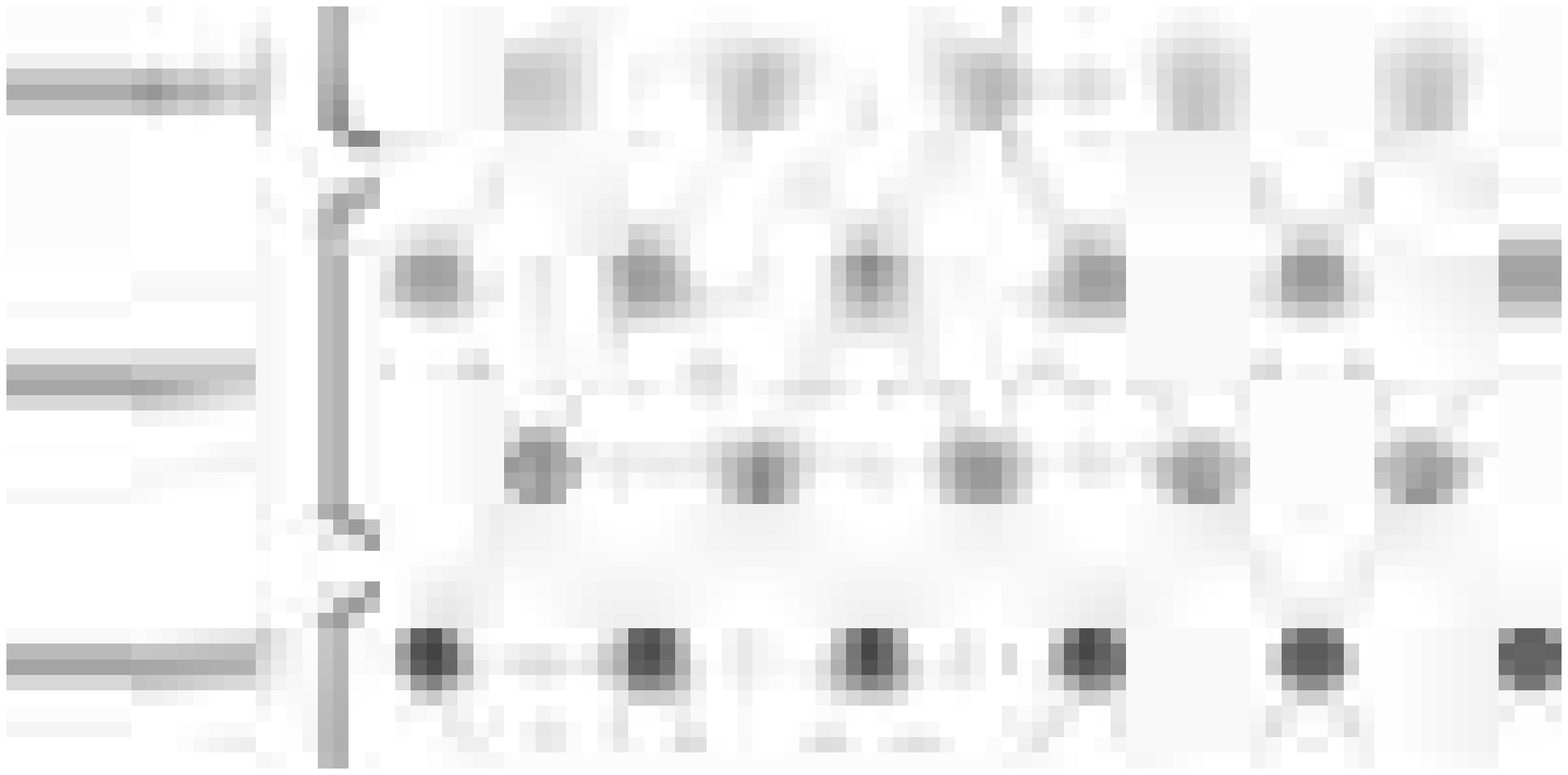}
\centering (b)
\end{minipage}
\caption{Schematic views of a part of the domain structure.
(a) A domain with three layers. (b) A domain with two layers.}
\label{fig8}
\end{figure}

We now discuss the structure of the steady pattern.
The snapshots in Fig.~\ref{fig6} indicate that most of the domains in the steady pattern consist of three layers of the players, where a layer means a straight line on the triangular lattice; see Fig.~\ref{fig8}~(a).
The situation is the same for lattices of different sizes.

\begin{figure}
\begin{minipage}[b]{0.3\textwidth}
\vspace{0pt}
\includegraphics[width=\textwidth]{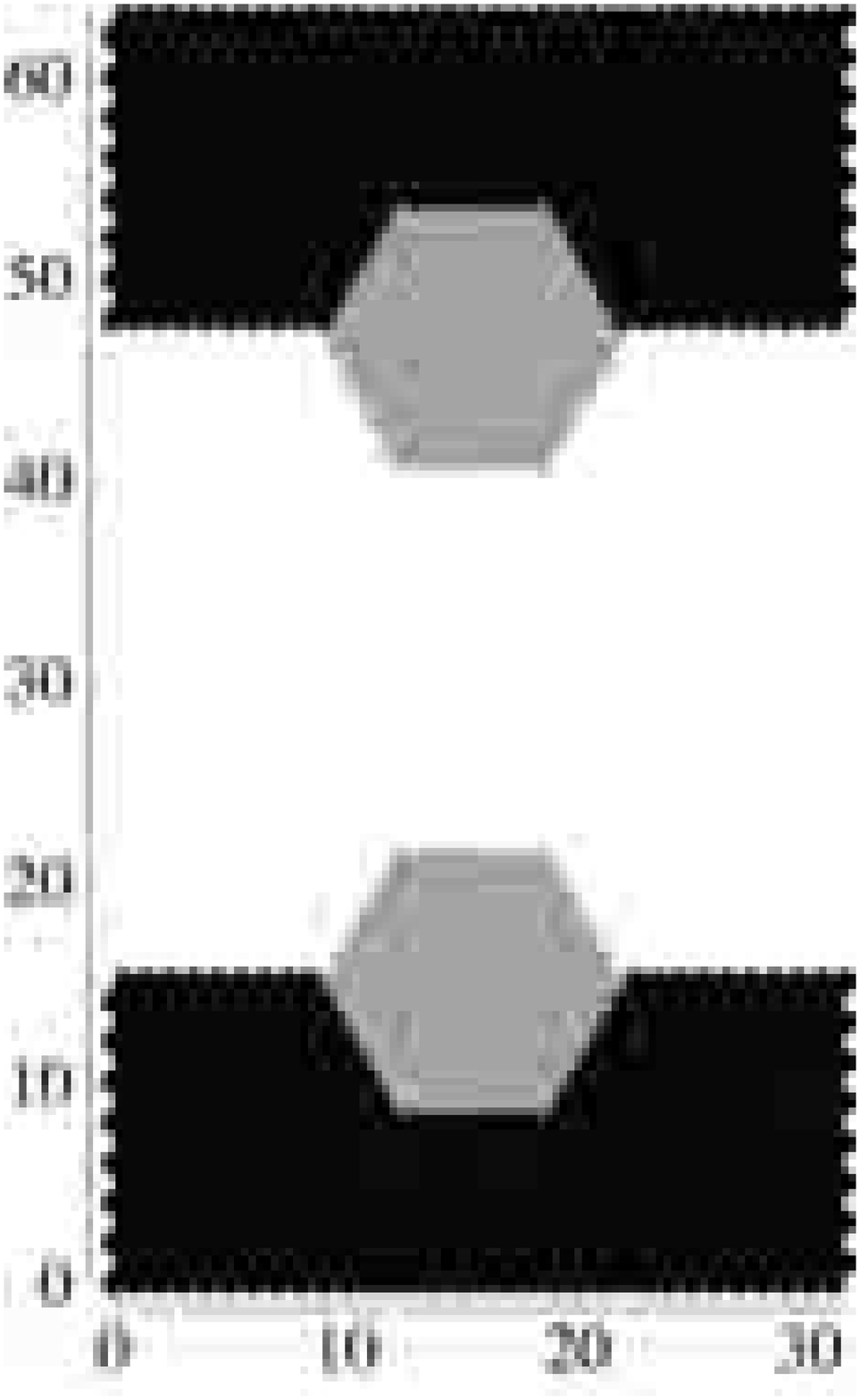}
\centering (a)
\end{minipage}
\hfill
\begin{minipage}[b]{0.3\textwidth}
\vspace{0pt}
\includegraphics[width=\textwidth]{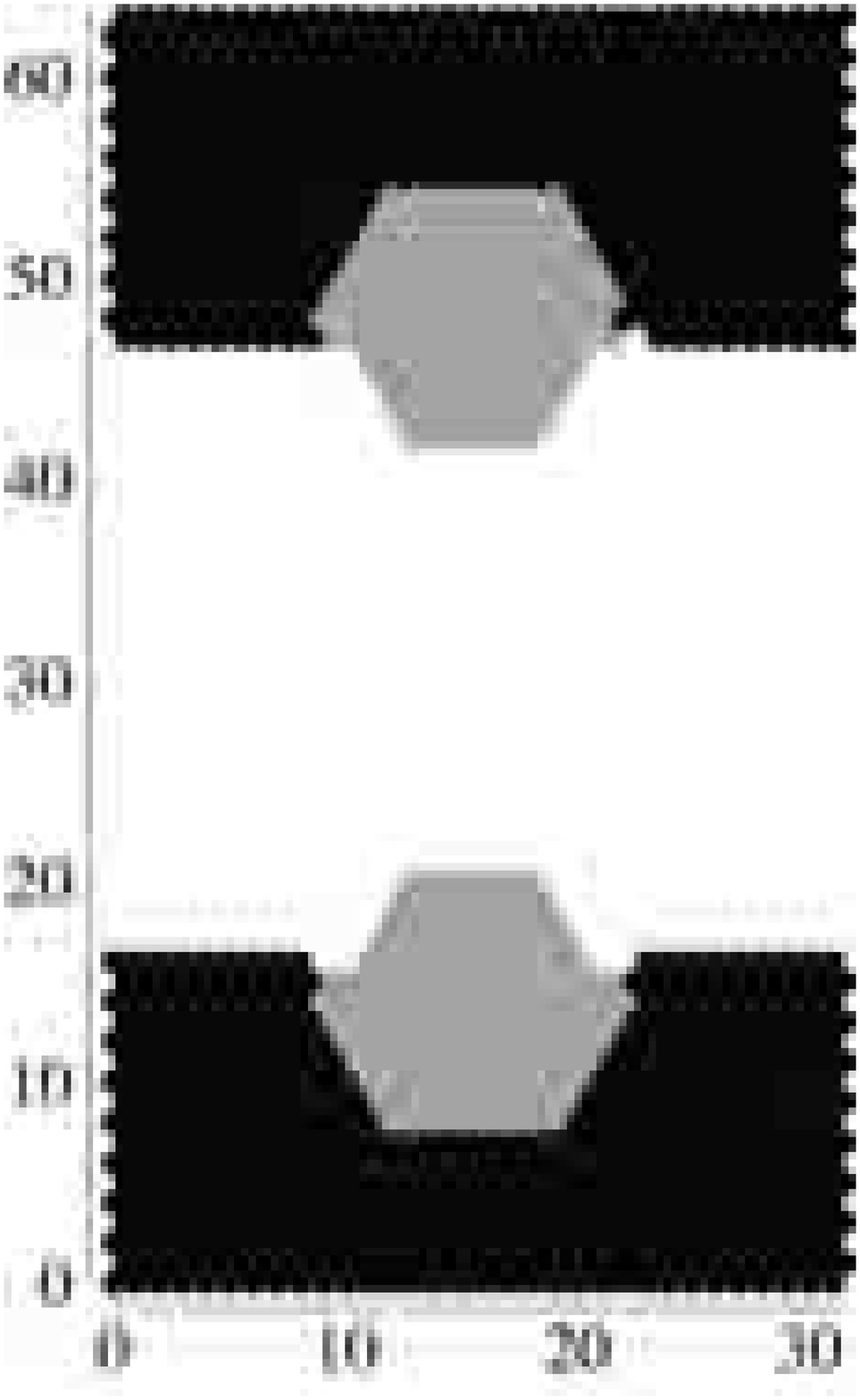}
\centering (b)
\end{minipage}
\hfill
\begin{minipage}[b]{0.3\textwidth}
\vspace{0pt}
\includegraphics[width=\textwidth]{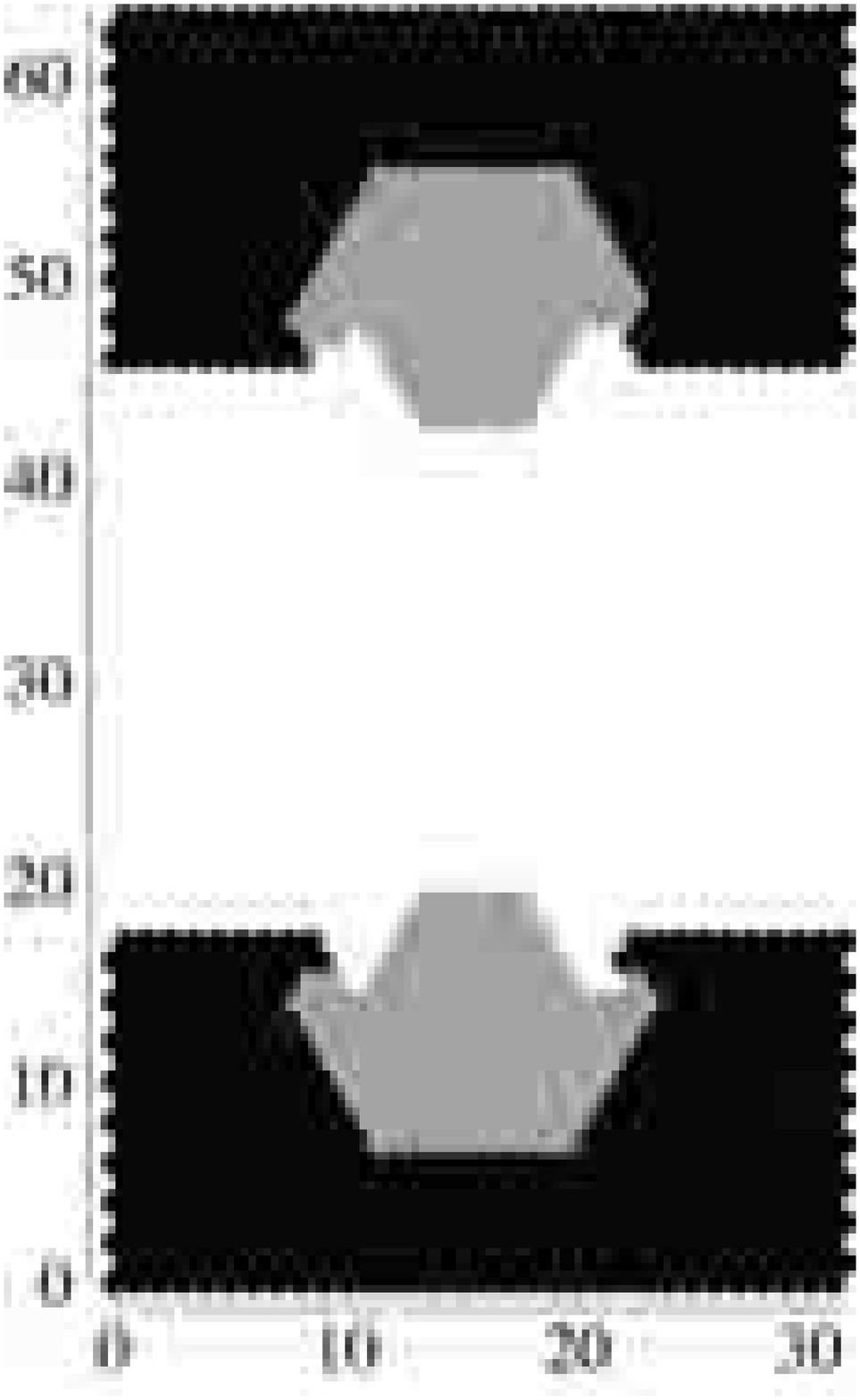}
\centering (c)
\end{minipage}
\vspace{\baselineskip}
\\
\begin{minipage}[b]{0.3\textwidth}
\vspace{0pt}
\includegraphics[width=\textwidth]{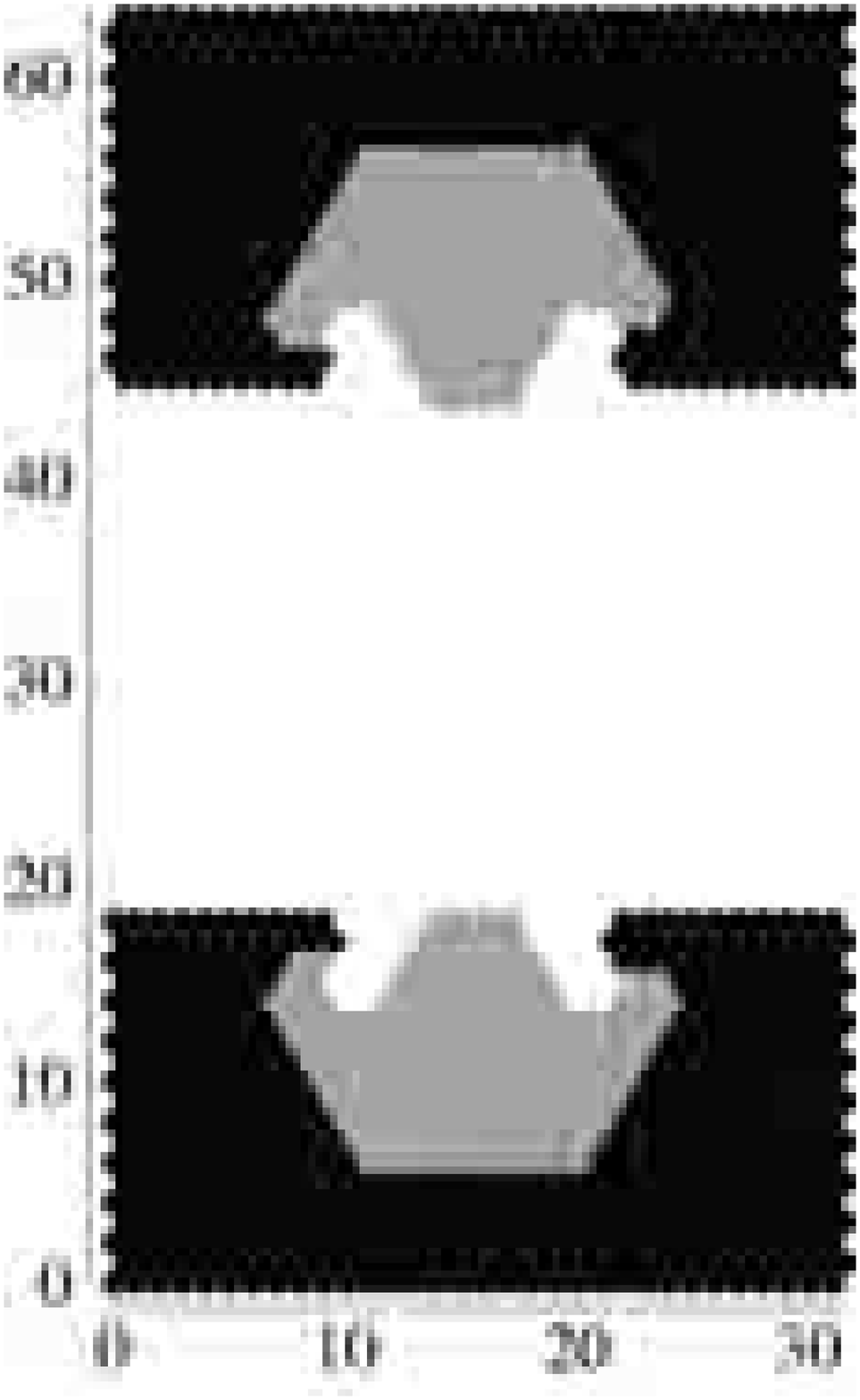}
\centering (d)
\end{minipage}
\hfill
\begin{minipage}[b]{0.3\textwidth}
\vspace{0pt}
\includegraphics[width=\textwidth]{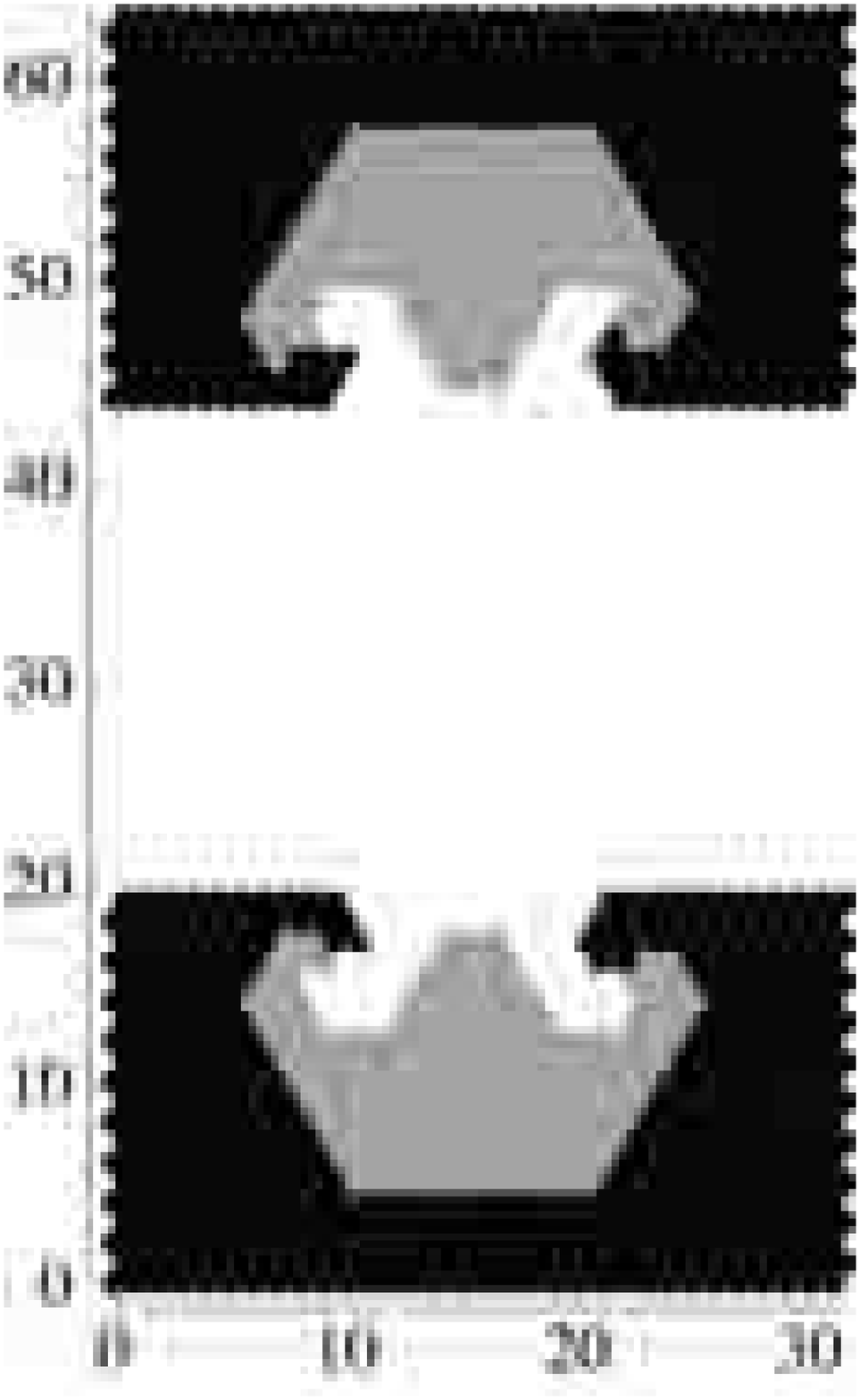}
\centering (e)
\end{minipage}
\hfill
\begin{minipage}[b]{0.3\textwidth}
\vspace{0pt}
\includegraphics[width=\textwidth]{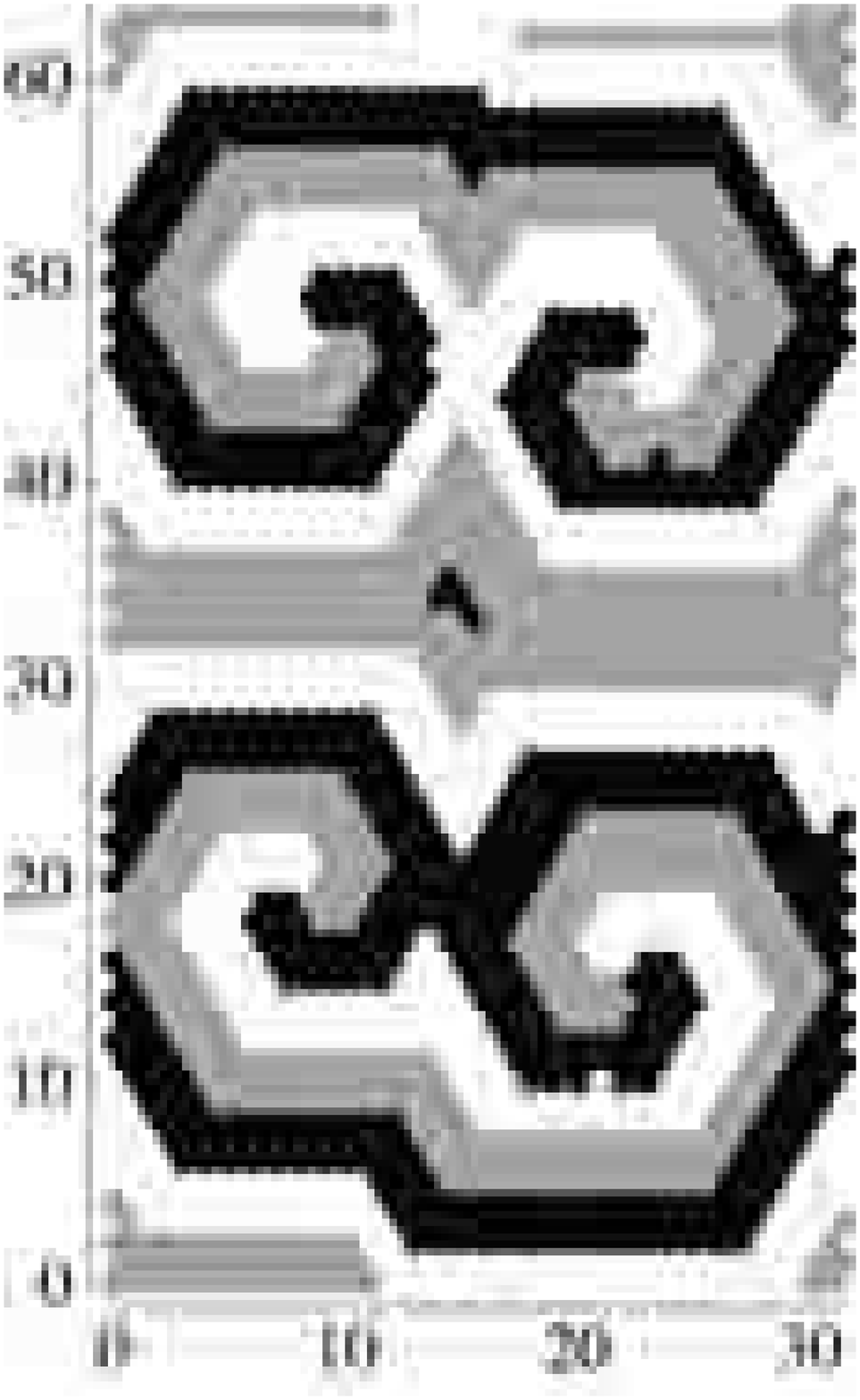}
\centering (f)
\end{minipage}
\caption{A simulation on the triangular lattice with $2^{11}$ players with the initial configuration shown in (a).
The con figurations at the first step (b), the second step (c), the third step (d), the fourth step (e), and  the 45th step (f).
The black hexagons denote the player with the hand~0, the gray hexagons the hand~1, and the white hexagons the hand~2.}
\label{fig9}
\end{figure}

\begin{figure}
\begin{center}
\includegraphics[width=0.8\textwidth]{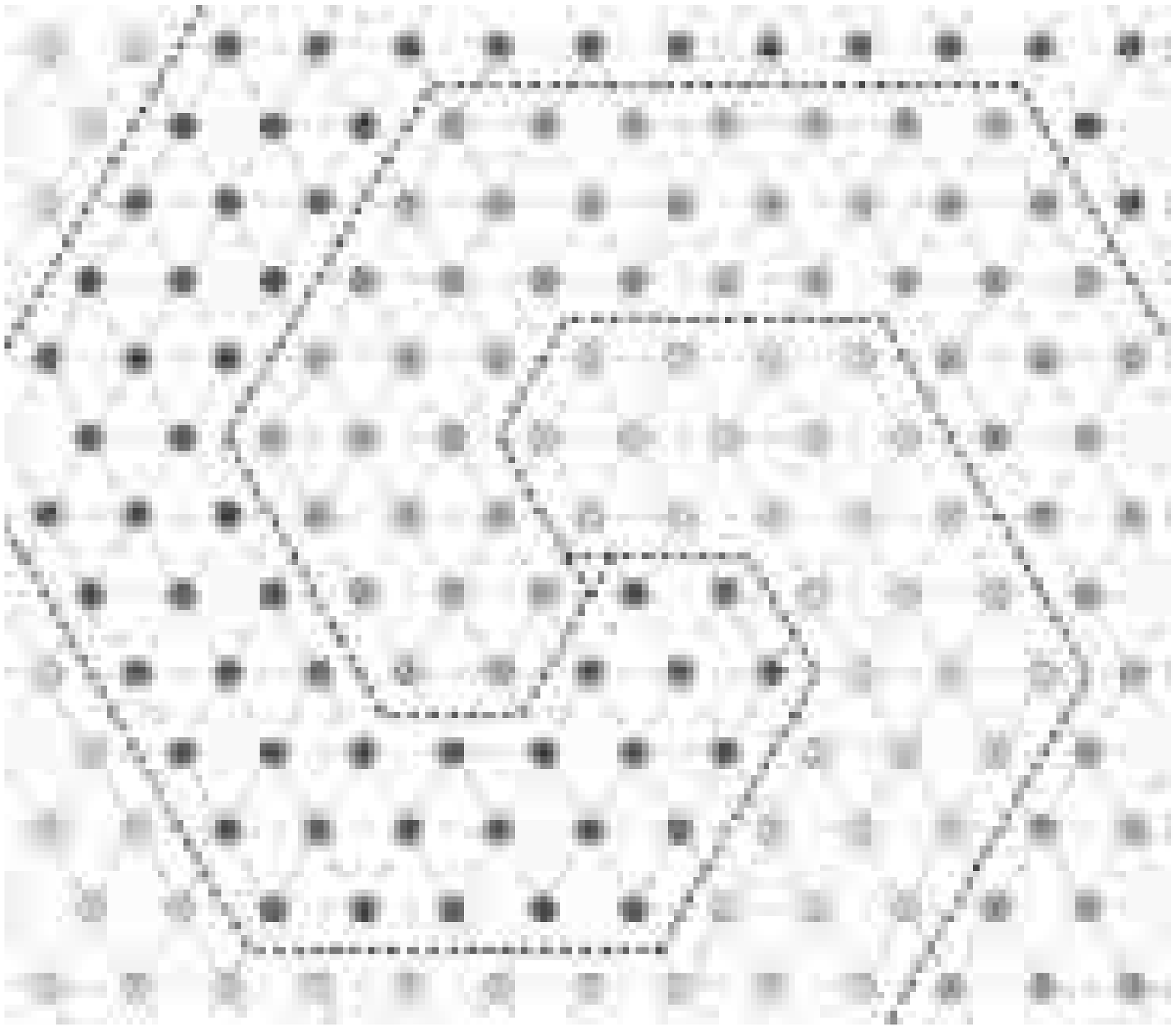}
\end{center}
\caption{A typical structure of a vortex core, spontaneously formed.}
\label{fig10}
\end{figure}

The three-layer domains are generated at a vortex; see Fig.~\ref{fig9}.
Here we started a simulation from a configuration of the form in Fig.~\ref{fig2}~(a), specifically the configuration in Fig.~\ref{fig9}~(a).
We can concretely see the vortex structure schematically shown in Fig.~\ref{fig2}~(c).
Furthermore, we can see that a vortex spontaneously takes the core structure of Fig.~\ref{fig10} with tails of three-layer domains.
Vortices thus generate domains with three layers.

A domain with three layers, once generated, is stabilized because it has a layer to be beaten by a stronger hand, a layer to remain unchanged and a layer to win over a weaker hand.
In Fig.~\ref{fig8}~(a), 
\begin{itemize}
\item every player in the layer of the domain~2 scores the point~$2$,
\item every player in the uppermost layer of the domain~1 scores the point~$-2$,
\item every player in the mid layer of the domain~1 scores the point~$0$,
\item every player in the lowermost layer of the domain~1 scores the point~$2$, and
\item every player in the layer of the domain~0 scores the point~$-2$,
\end{itemize}
under the assumption that the layer above the shown area in Fig.~\ref{fig8}~(a) belongs to the domain~2 and the layer below it belongs to the domain~0.
For every player in the uppermost layer of the domain~1, a neighbor with the highest score is one in the layer of the domain~2, and hence will change the hand to~2 in the next step as a copy player.
For every player in the mid layer of the domain~1, a neighbor with the highest score is one in the lowermost layer of the domain~1, and hence hence will remain unchanged in the next step.
For every player in the lowermost layer of the domain~1, a neighbor with the highest score is in the same layer, and hence will remain unchanged in the next step.
For every player in the layer of the domain~0, a neighbor with the highest score is one in the layer of the domain~1, and hence will change the hand to~1 in the next step as a copy player.
Therefore, the domain~1 will shift one layer below in the next step, remaining to be three layers.

A domain with only two layers can grow to a domain with three layers.
In Fig.~\ref{fig8}~(b),
\begin{itemize}
\item every player in the layer of the domain~2 scores the point~$2$,
\item every player in the upper layer of the domain~1 scores the point~$-2$,
\item every player in the lower layer of the domain~1 scores the point~$2$, and
\item every player in the layer of the domain~0 scores the point~$-2$,
\end{itemize}
under the assumption that the layer above the shown area in Fig.~\ref{fig8}~(b) belongs to the domain~2 and the layer below it belongs to the domain~0.
For each player in the upper layer of the domain~1, a neighbor with the highest score is \textit{either} one in the layer of the domain~2 or one in the lower layer of the domain~1.
Each player will choose either the hand~2 or the hand~1 randomly in the next step; \textit{i.e.}\ only half of the players will turn into the hand~2.
For every player in the lower layer of the domain~1, a neighbor with the highest score is in the same layer, and hence will remain unchanged in the next step.
For every player in the layer of the domain~0, a neighbor with the highest score is one in the layer of the domain~1, and hence will change the hand to~1 in the next step as a copy player.
Therefore, the upper boundary of the domain~1 shifts downward only halfway, whereas the lower boundary certainly shifts one layer below.
Thus the domain~1 grows gradually to a domain with three layers.

We do not have any arguments for the fact that domains with more than three layers are rare.
We speculate that a vortex, as in Fig.~\ref{fig10}, tends to generate the minimum stable domain, which is a domain with three layers.
Once its generated with just three layers, there is no mechanism that makes the domain grow to more than three layers.

\section{Simulation of the society of copy players: square and honeycomb lattice}
\label{sec4}

In this section, we show results of our simulations on the square lattice and the honeycomb lattice.
We simulated the society of copy players on a square lattice with $2^{14}$ players and on a honeycomb lattice with $2^{15}$ players.
We imposed periodic boundary conditions on both lattices.
We do not repeat discussion on the convergence to the steady pattern here for the square and honeycomb lattices; it is basically the same as the triangular lattice.
We demonstrate that the stationary vortex structure does \textit{not} appear in these lattices, thereby emphasizing that the frustration is essential to the stationary vortex structure.

\subsection{Square lattice}

\begin{figure}
\begin{minipage}[b]{0.45\textwidth}
\vspace{0pt}
\includegraphics[width=\textwidth]{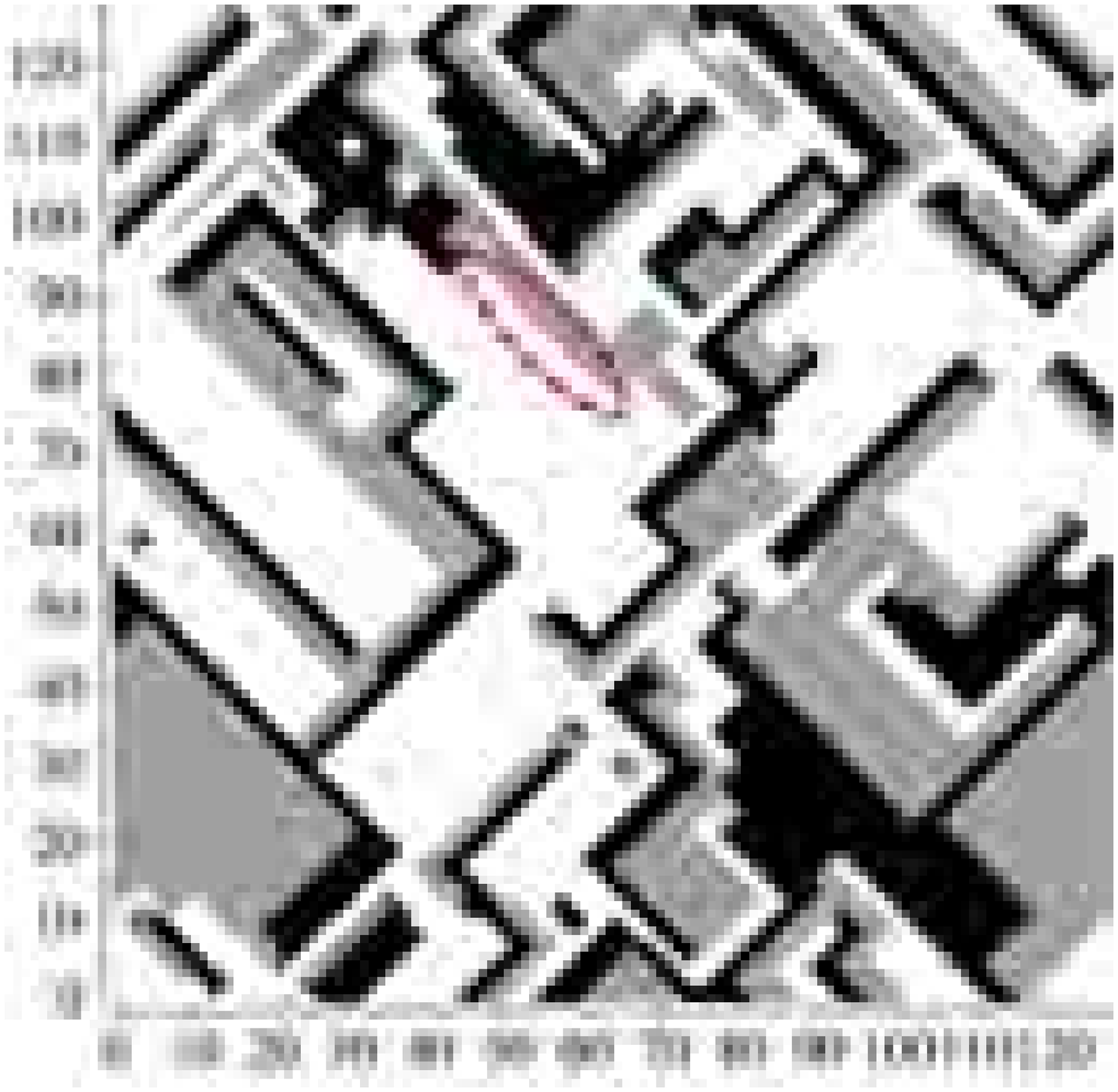}
\centering (a)
\end{minipage}
\hfill
\begin{minipage}[b]{0.45\textwidth}
\vspace{0pt}
\includegraphics[width=\textwidth]{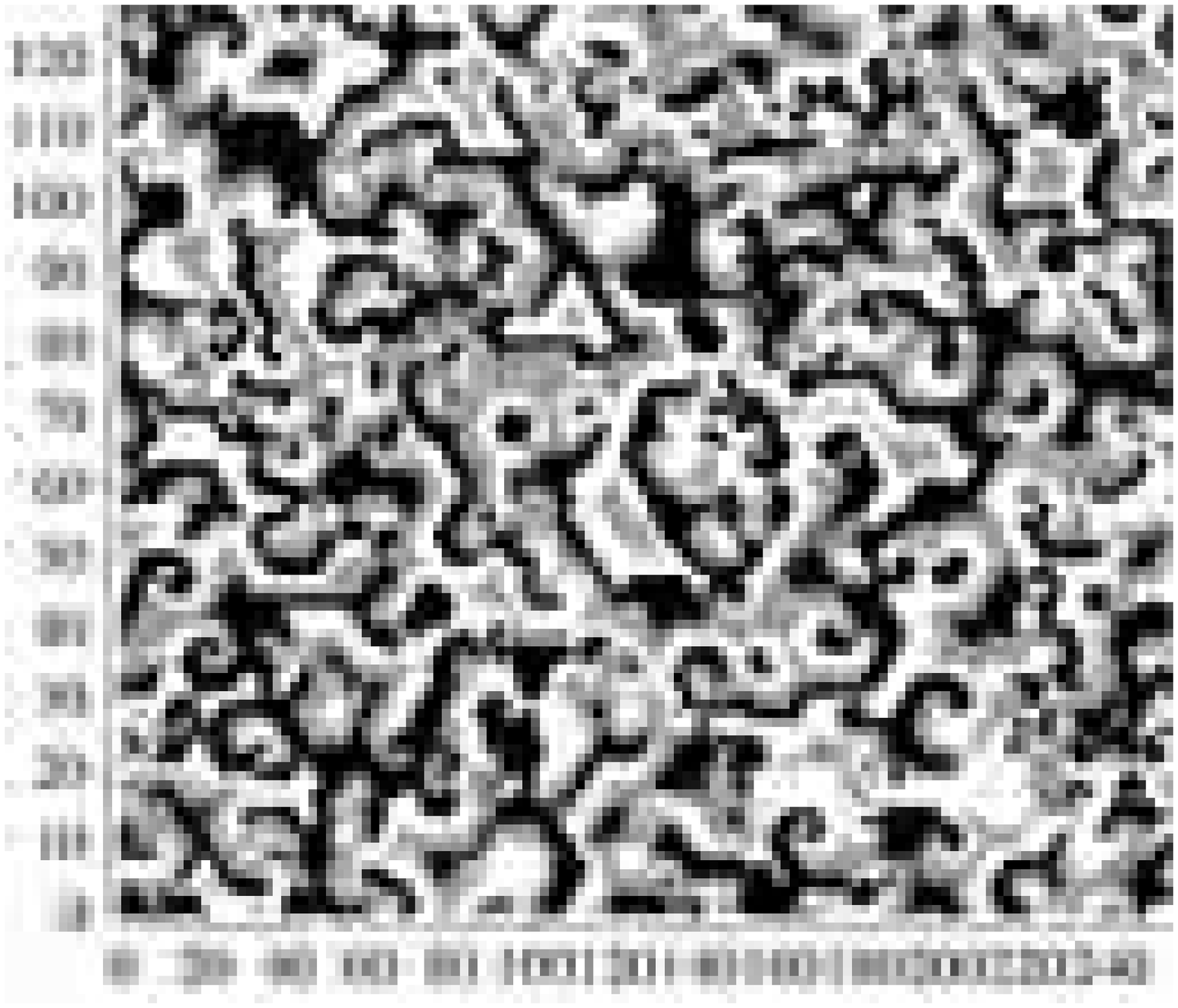}
\centering (b)
\end{minipage}
\\
\begin{minipage}[b]{0.45\textwidth}
\vspace{\baselineskip}
\includegraphics[width=\textwidth]{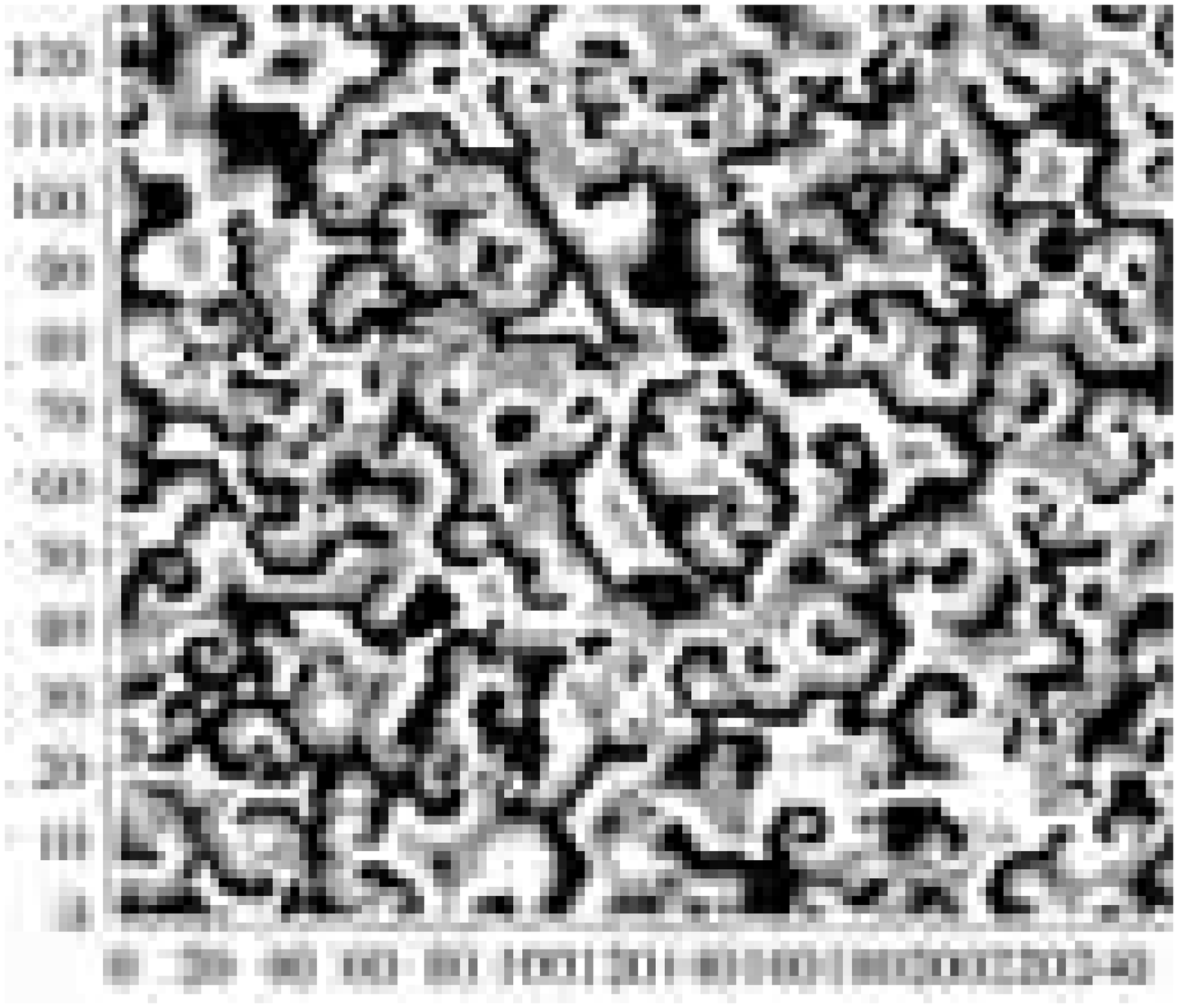}
\centering (c)
\end{minipage}
\hfill
\begin{minipage}[b]{0.45\textwidth}
\vspace{\baselineskip}
\includegraphics[width=\textwidth]{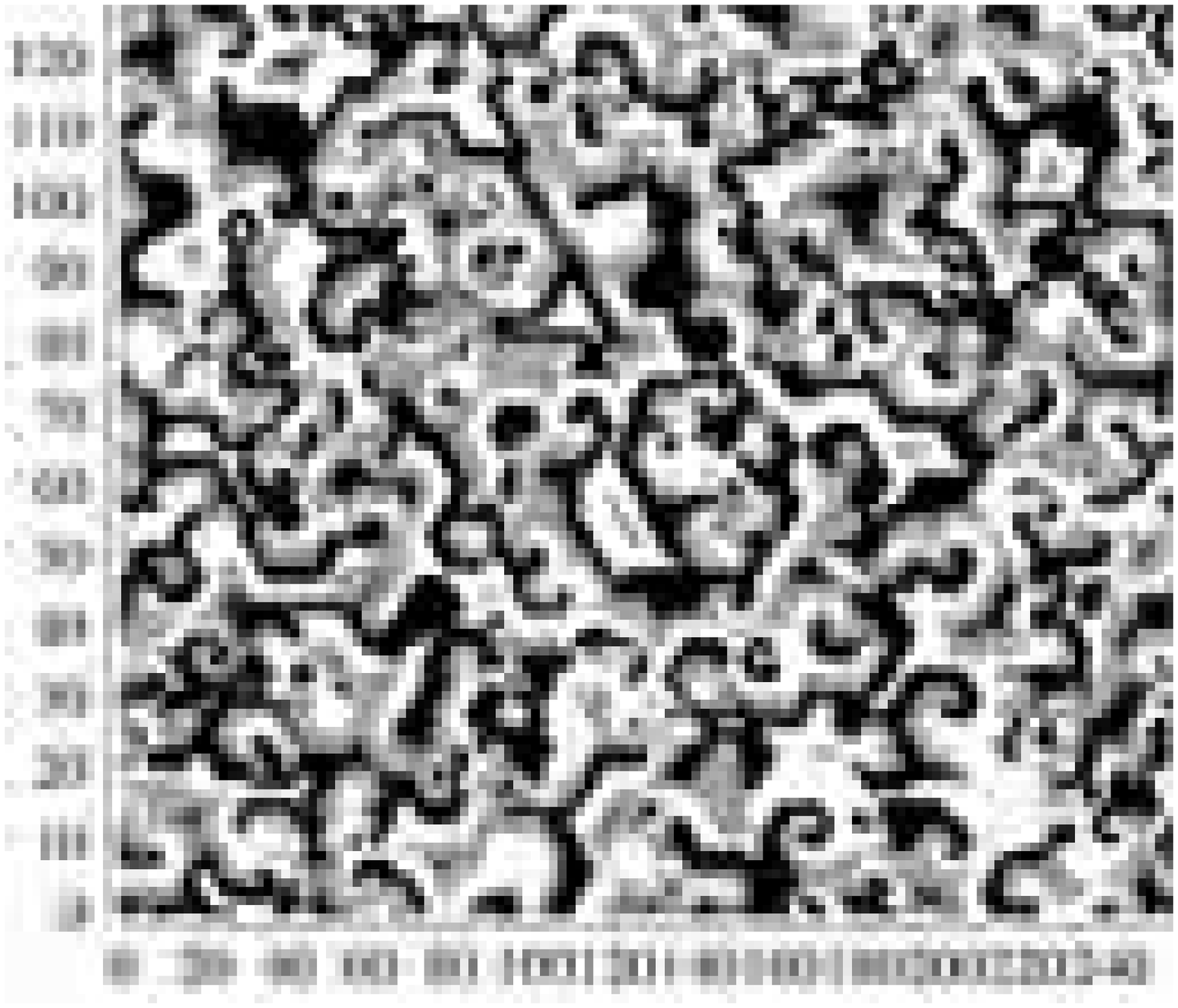}
\centering (d)
\end{minipage}
\caption{The configurations at the 997th step (a), the 998th step (b), the 999 step (c) and the 1~000th step (d) of a simulation on the square lattice. with $2^{14}$ players
The black squares denote the player with the hand~0, the gray squares the hand~1, and the white squares the hand~2.}
\label{fig11}
\end{figure}

Figure~\ref{fig11} shows snapshots of a simulation on the triangular lattice with $2^{14}$ players.
There is obviously no vortex structure on the square lattice.

The boundaries run diagonally in each snapshot.
This means that each boundary runs between the two sublattices of the square lattice;
on the boundary indicated by the red circle in Fig.~\ref{fig11}~(a), for example, the players with the hand~0 on the immediately lower left side of boundary is on the different sublattice from the players with the hand~1 on the immediately upper right side of the boundary.
The boundaries move either upward, downward or sideways in the next step
The upward and downward movements do not interfere with the sideway movements.

\begin{figure}
\begin{minipage}[b]{0.45\textwidth}
\vspace{0pt}
\includegraphics[width=\textwidth]{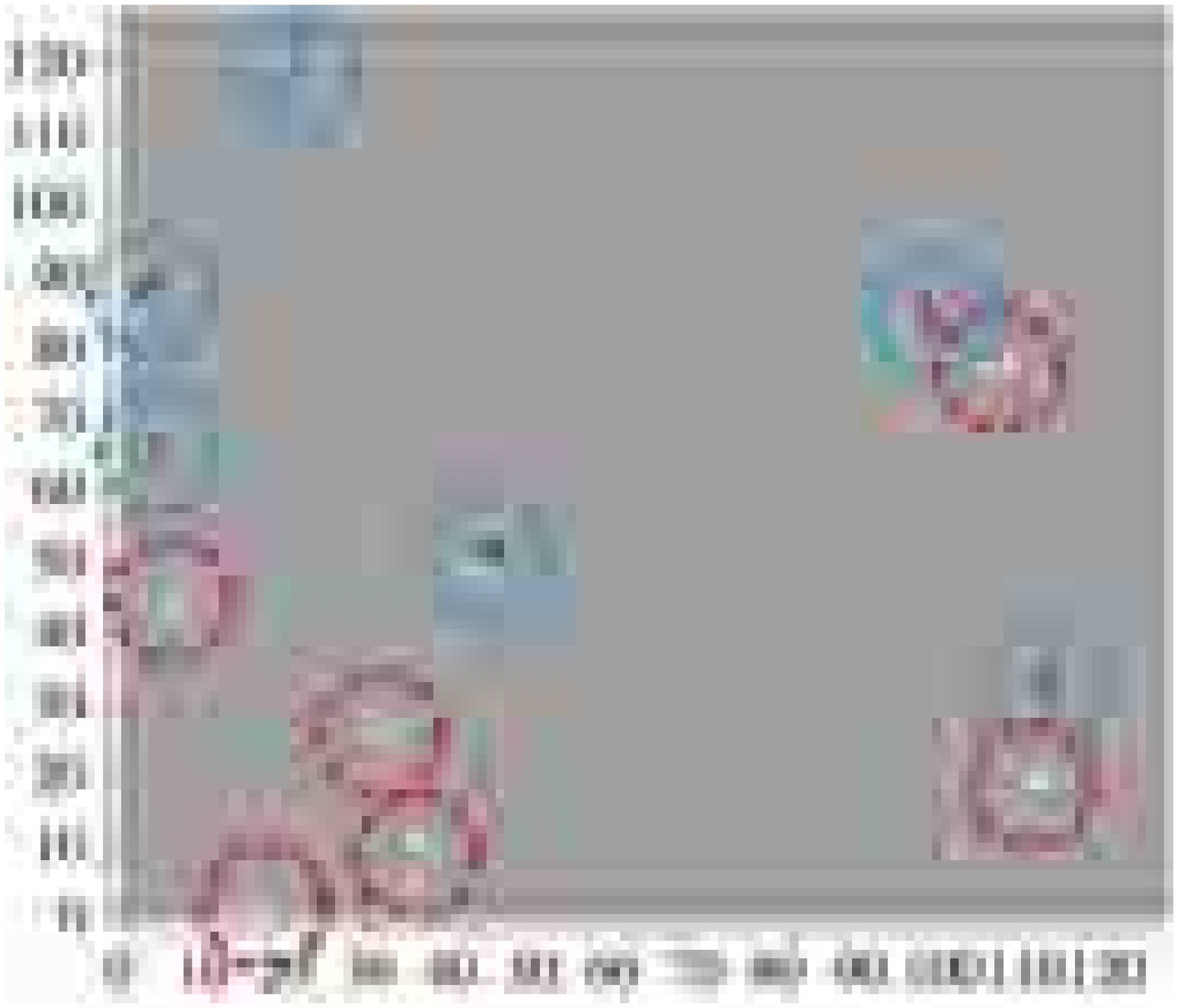}
\centering (a)
\end{minipage}
\hfill
\begin{minipage}[b]{0.45\textwidth}
\vspace{0pt}
\includegraphics[width=\textwidth]{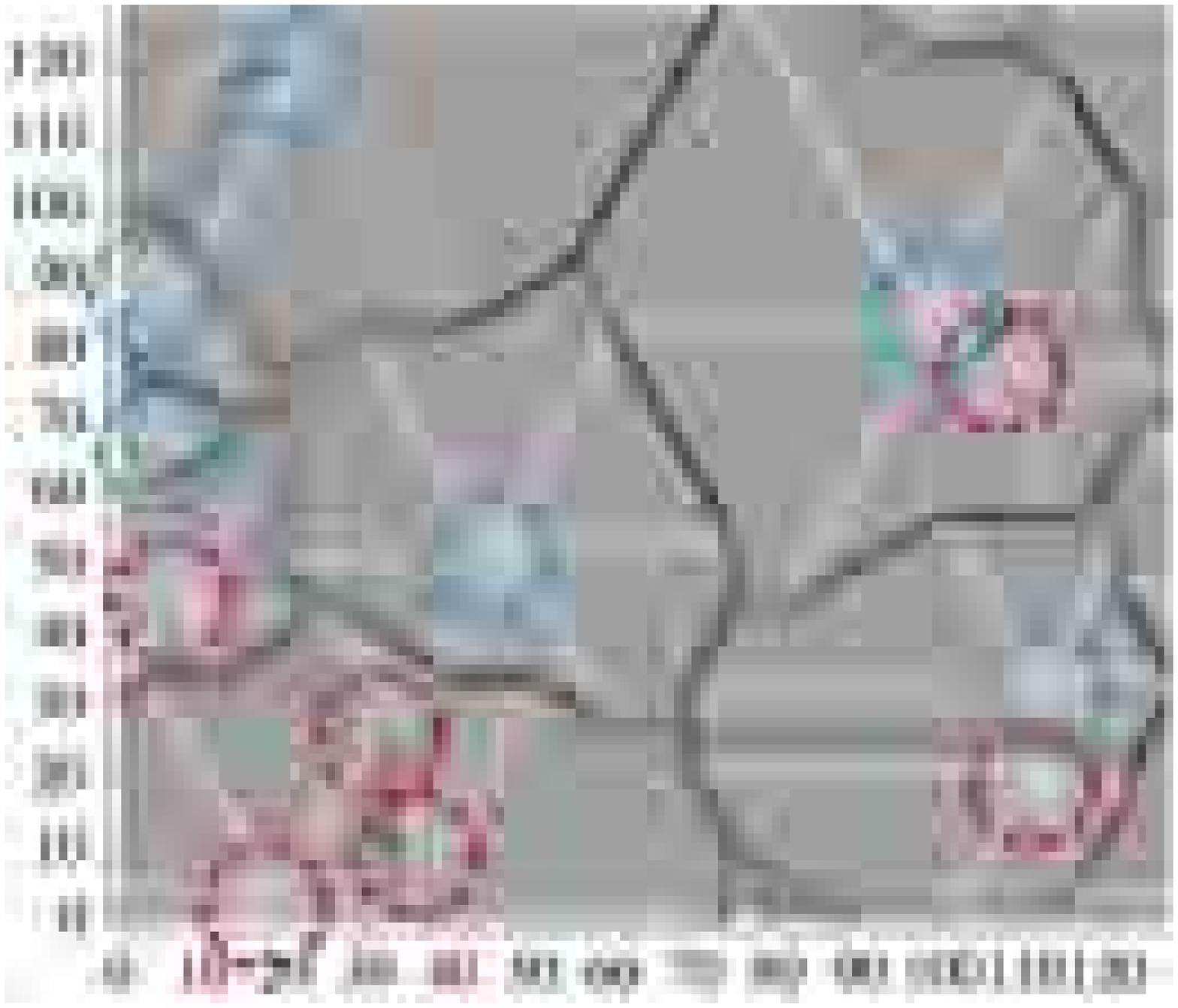}
\centering (b)
\end{minipage}
\\
\begin{minipage}[b]{0.45\textwidth}
\vspace{0pt}
\includegraphics[width=\textwidth]{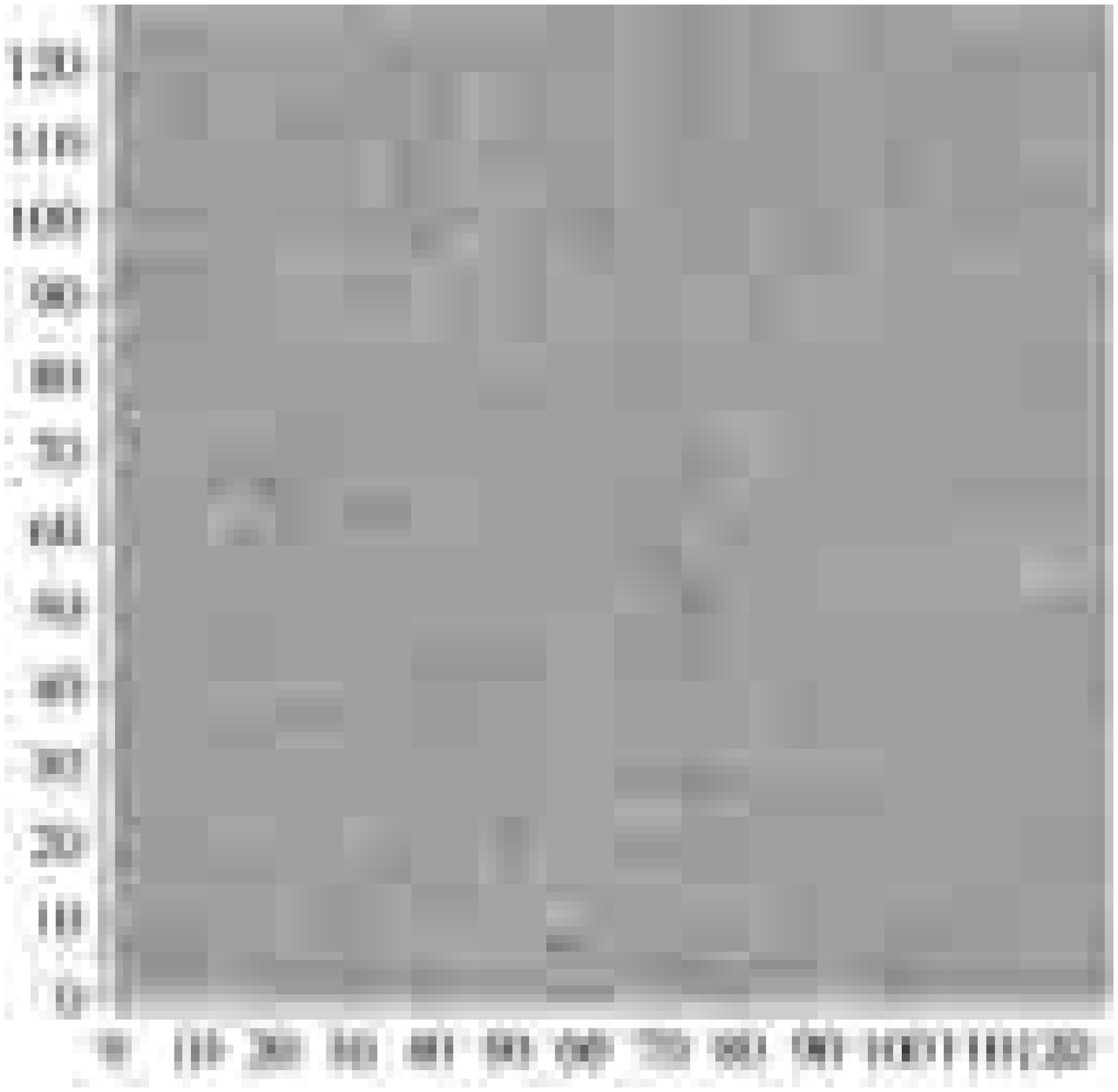}
\centering (c)
\end{minipage}
\vspace{\baselineskip}
\hfill
\begin{minipage}[b]{0.45\textwidth}
\vspace{0pt}
\includegraphics[width=\textwidth]{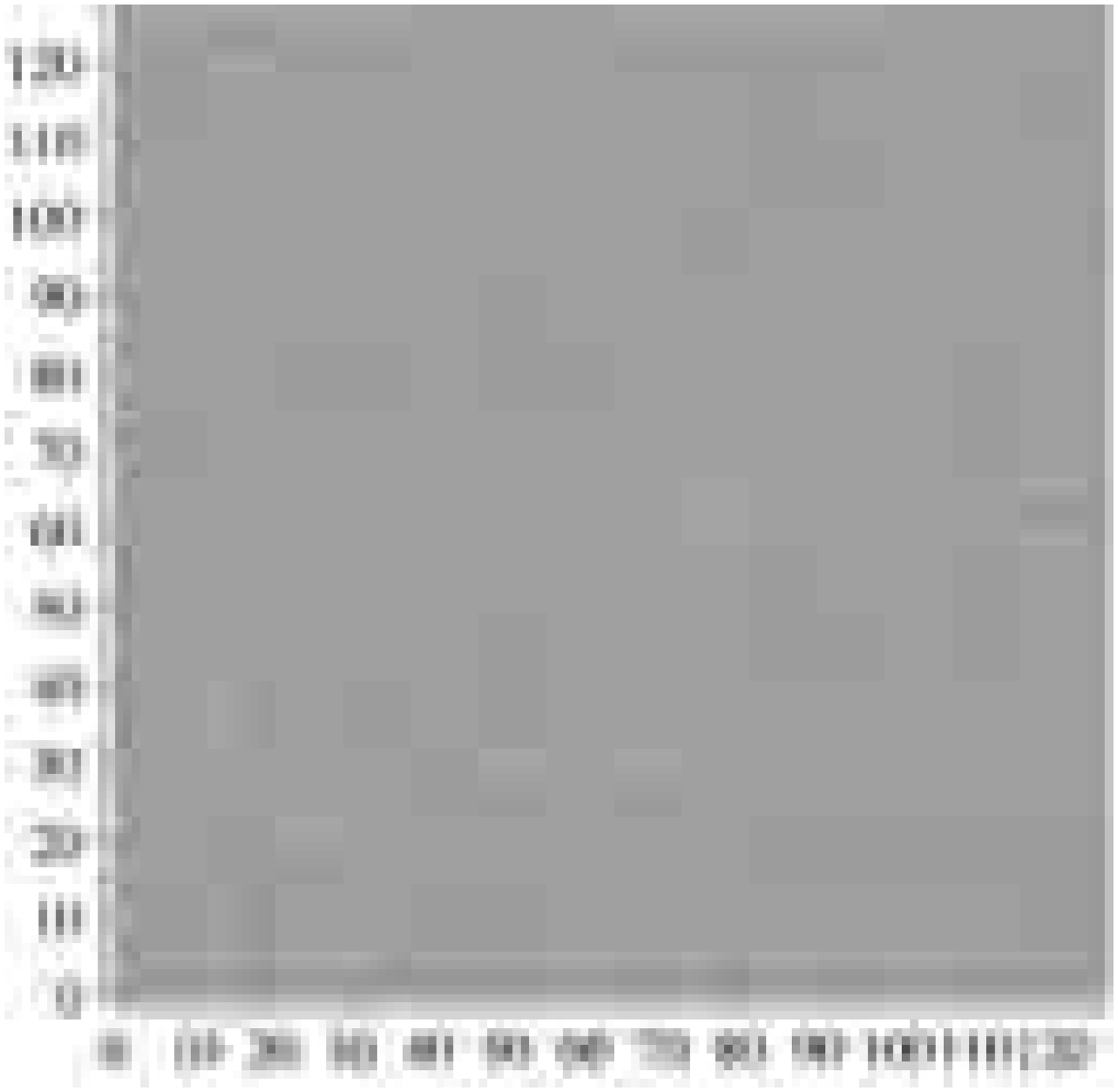}
\centering (d)
\end{minipage}
\\
\begin{minipage}[b]{0.45\textwidth}
\vspace{0pt}
\includegraphics[width=\textwidth]{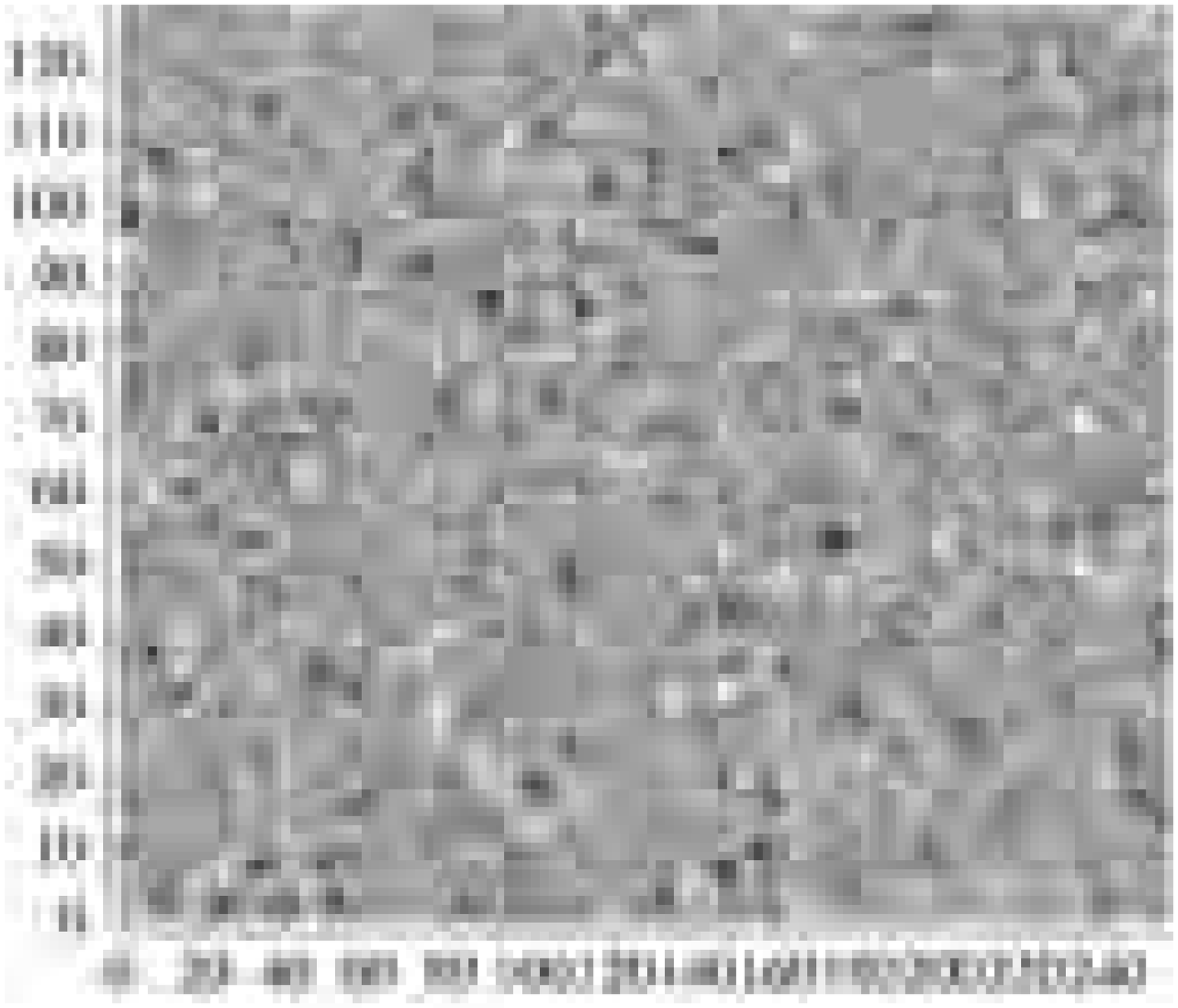}
\centering (e)
\end{minipage}
\hfill
\begin{minipage}[b]{0.45\textwidth}
\vspace{0pt}
\includegraphics[width=\textwidth]{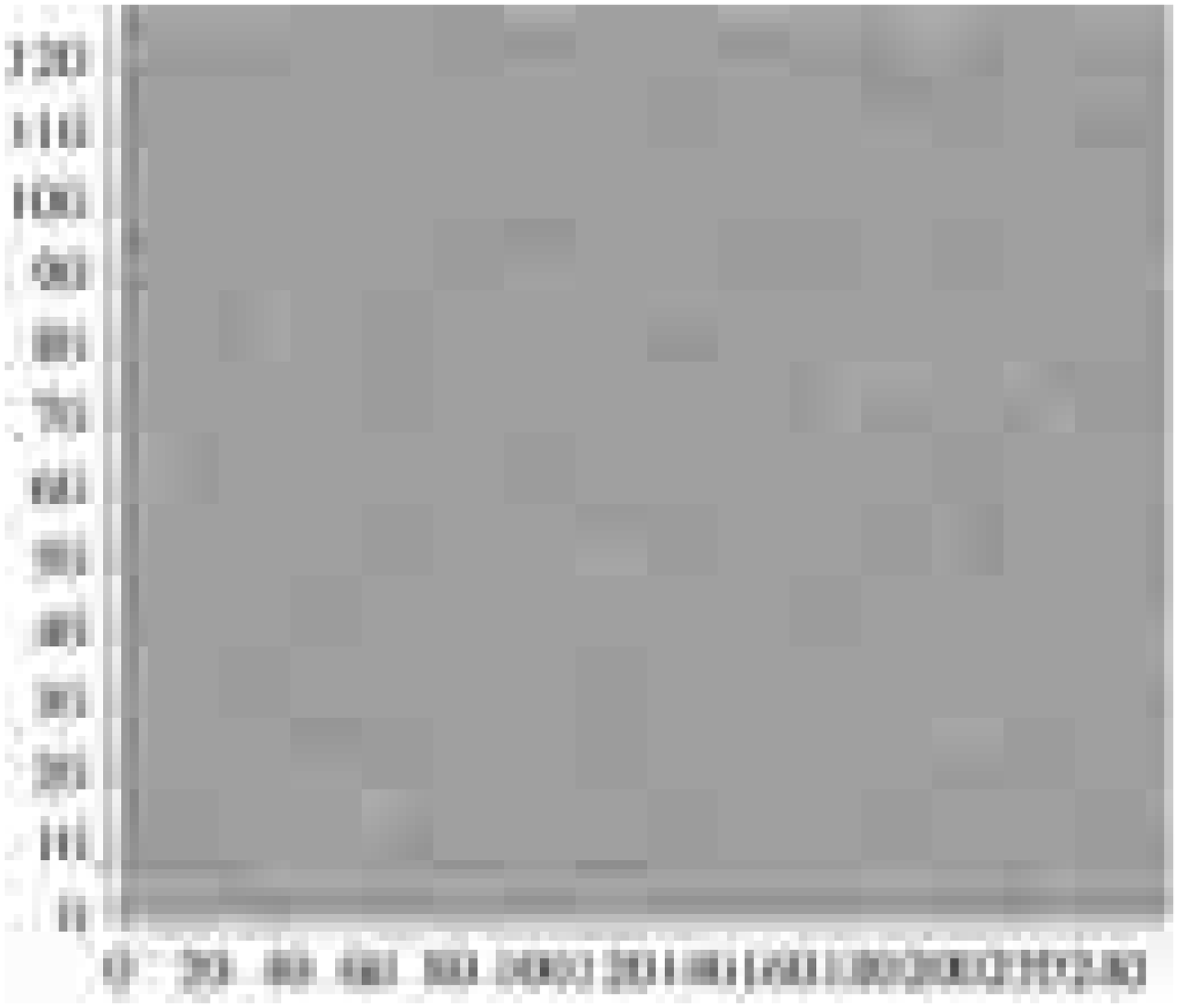}
\centering (f)
\end{minipage}
\caption{A simulation on the square lattice with $2^{11}$ players with the initial configuration shown in (a).
The configurations at the first step (b), the second step (c), the third step (d), the fourth step (e), and  the fifth step (f).
The black squares denote the player with the hand~0, the gray squares the hand~1, and the white squares the hand~2.}
\label{fig12}
\end{figure}

We demonstrate in Fig.~\ref{fig12} that an initial configuration of the type in Fig.~\ref{fig2}~(a) never generate vortices.
Here we started a simulation from the configuration in Fig.~\ref{fig12}~(a), which mimics Fig.~\ref{fig2}~(a).
We do not see any structure of the form schematically shown in Fig.~\ref{fig2}~(c).

\subsection{Honeycomb lattice}

\begin{figure}
\begin{minipage}[b]{0.45\textwidth}
\vspace{0pt}
\includegraphics[width=\textwidth]{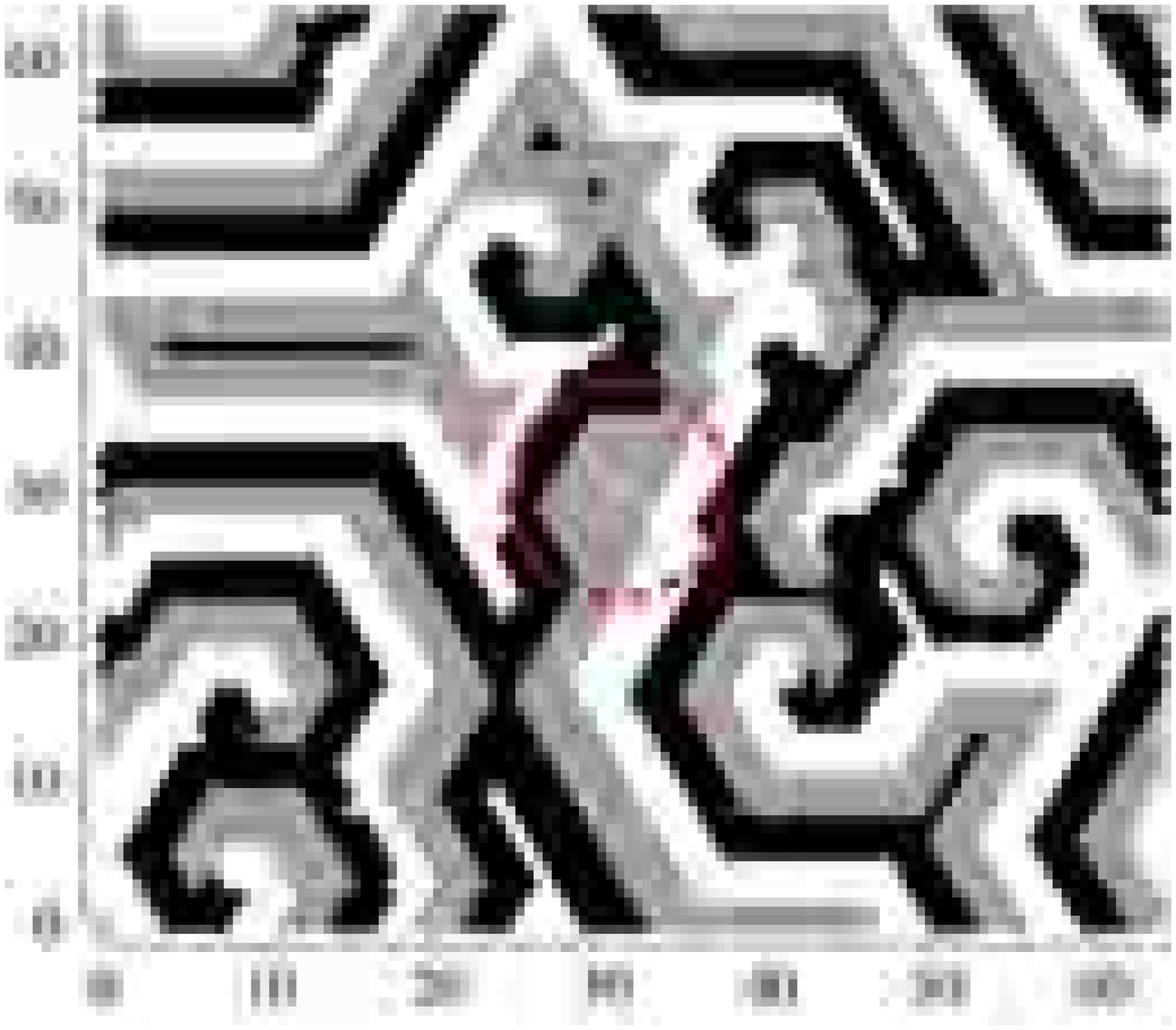}
\centering (a)
\end{minipage}
\hfill
\begin{minipage}[b]{0.45\textwidth}
\vspace{0pt}
\includegraphics[width=\textwidth]{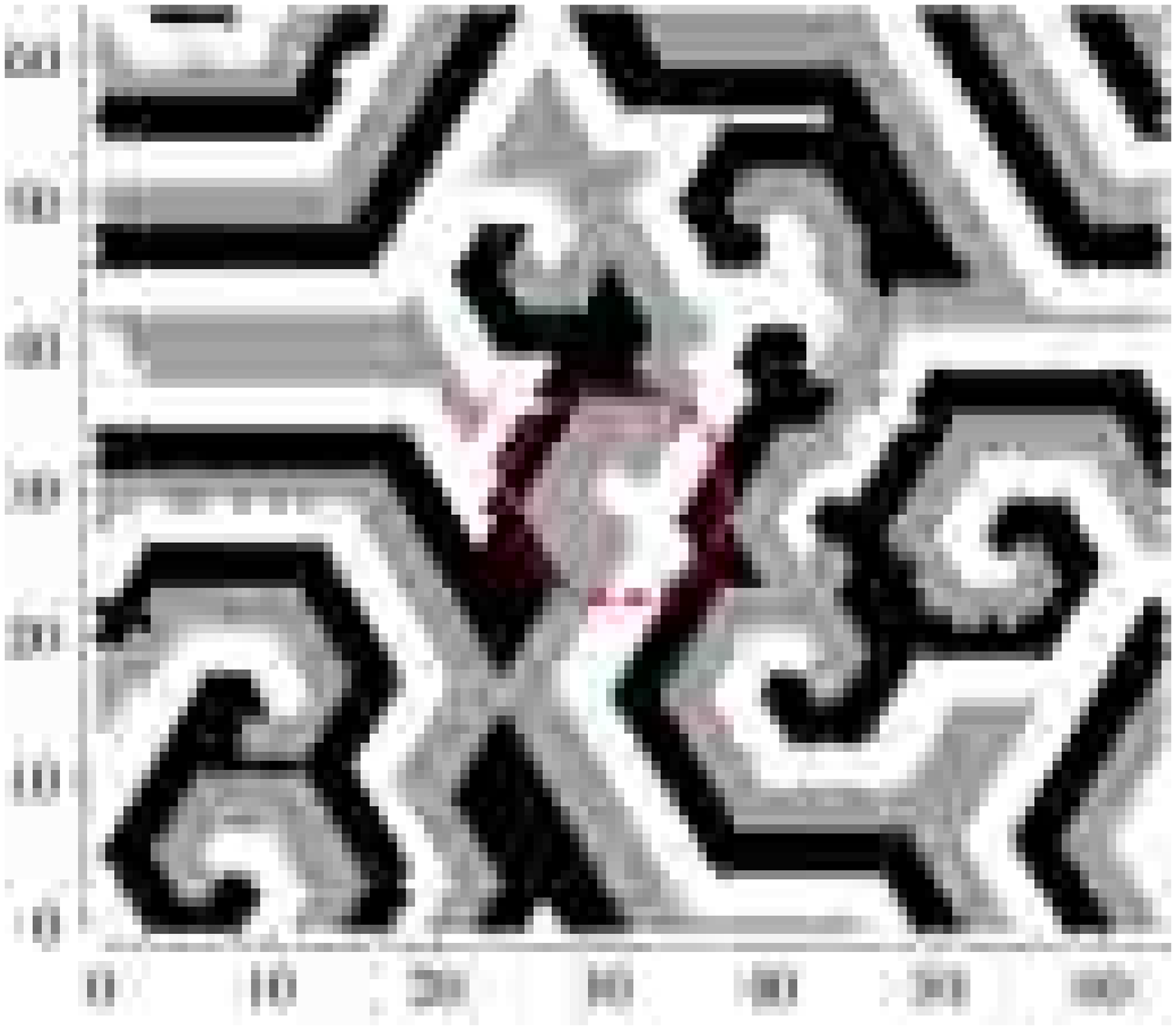}
\centering (b)
\end{minipage}
\\
\begin{minipage}[b]{0.45\textwidth}
\vspace{\baselineskip}
\includegraphics[width=\textwidth]{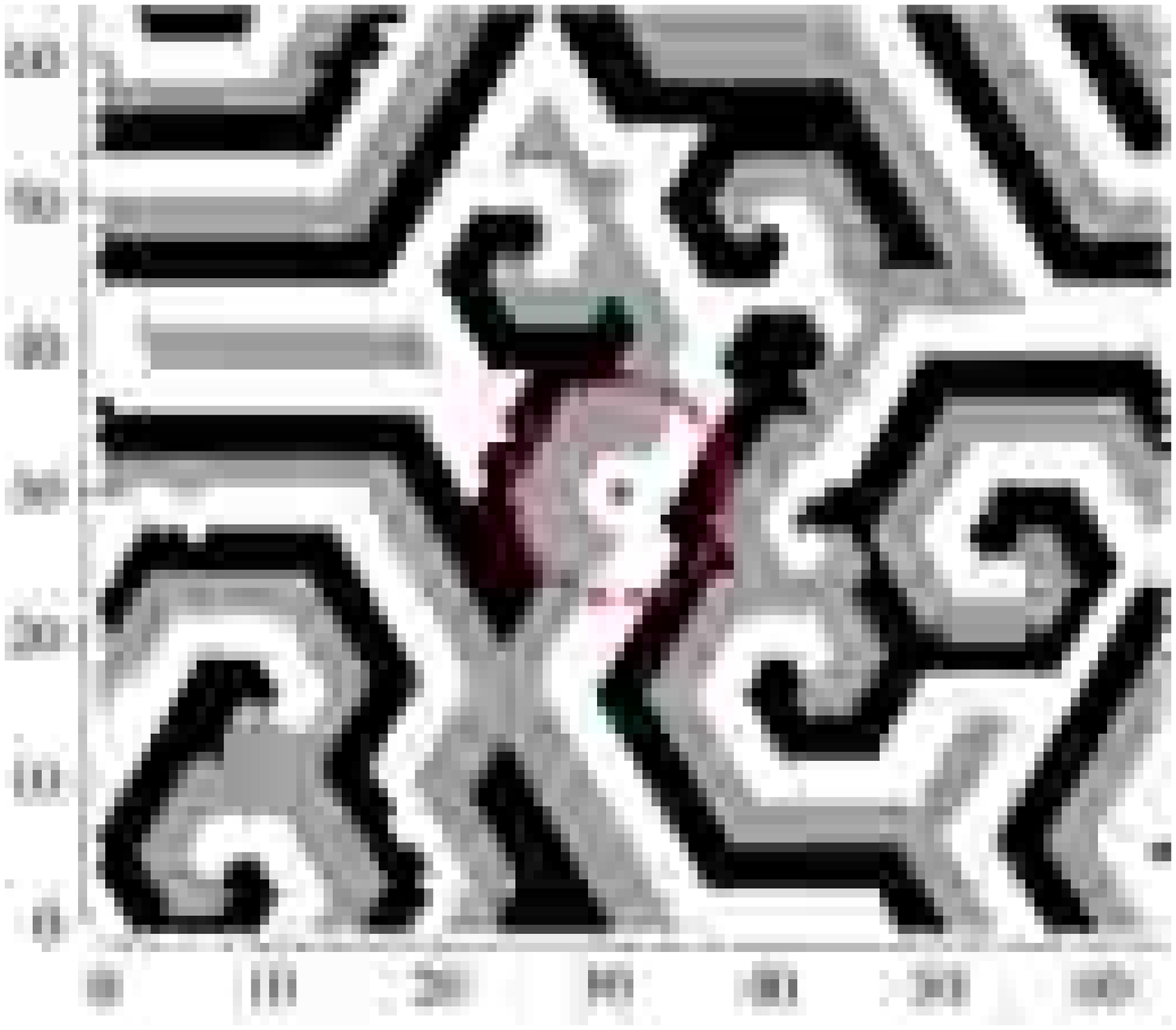}
\centering (c)
\end{minipage}
\hfill
\begin{minipage}[b]{0.45\textwidth}
\vspace{\baselineskip}
\includegraphics[width=\textwidth]{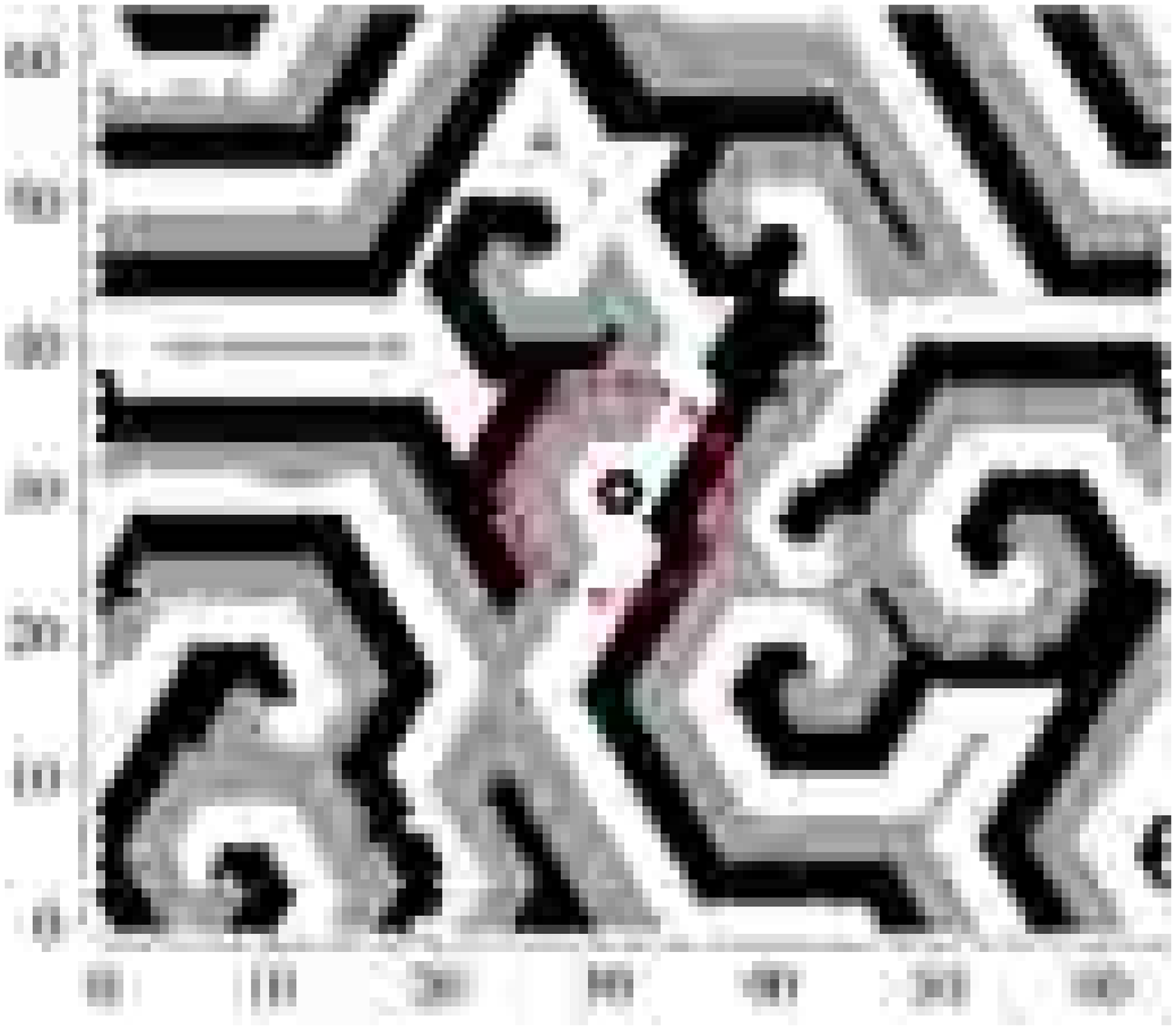}
\centering (d)
\end{minipage}
\caption{The configurations at the 997th step (a), the 998th step (b), the 999 step (c) and the 1~000th step (d) of a simulation on the honeycomb lattice. with $2^{15}$ players
The black triangles denote the player with the hand~0, the gray triangles the hand~1, and the white triangles the hand~2.}
\label{fig13}
\end{figure}

Figure~\ref{fig13} shows snapshots of a simulation on the honeycomb lattice with $2^{15}$ players.
The pattern may appear to have a vortex structure.
We hereafter argue that the seemingly vortex structure on the honeycomb lattice is not stationary and hence is essentially different from the stationary vortex structure on the triangular lattice.

\begin{figure}
\begin{minipage}[b]{0.45\textwidth}
\vspace{0pt}
\includegraphics[width=\textwidth]{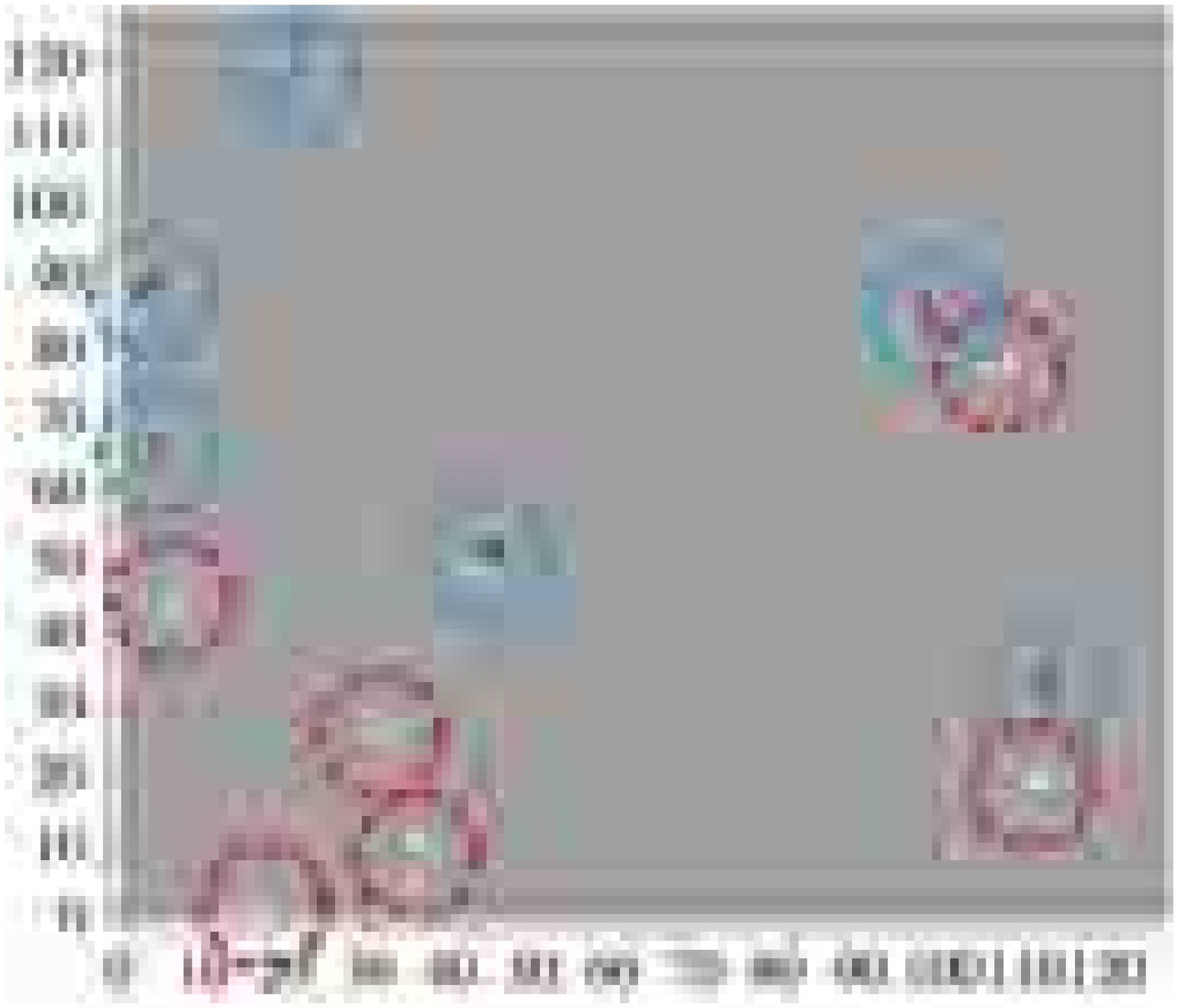}
\centering (a)
\end{minipage}
\hfill
\begin{minipage}[b]{0.45\textwidth}
\vspace{0pt}
\includegraphics[width=\textwidth]{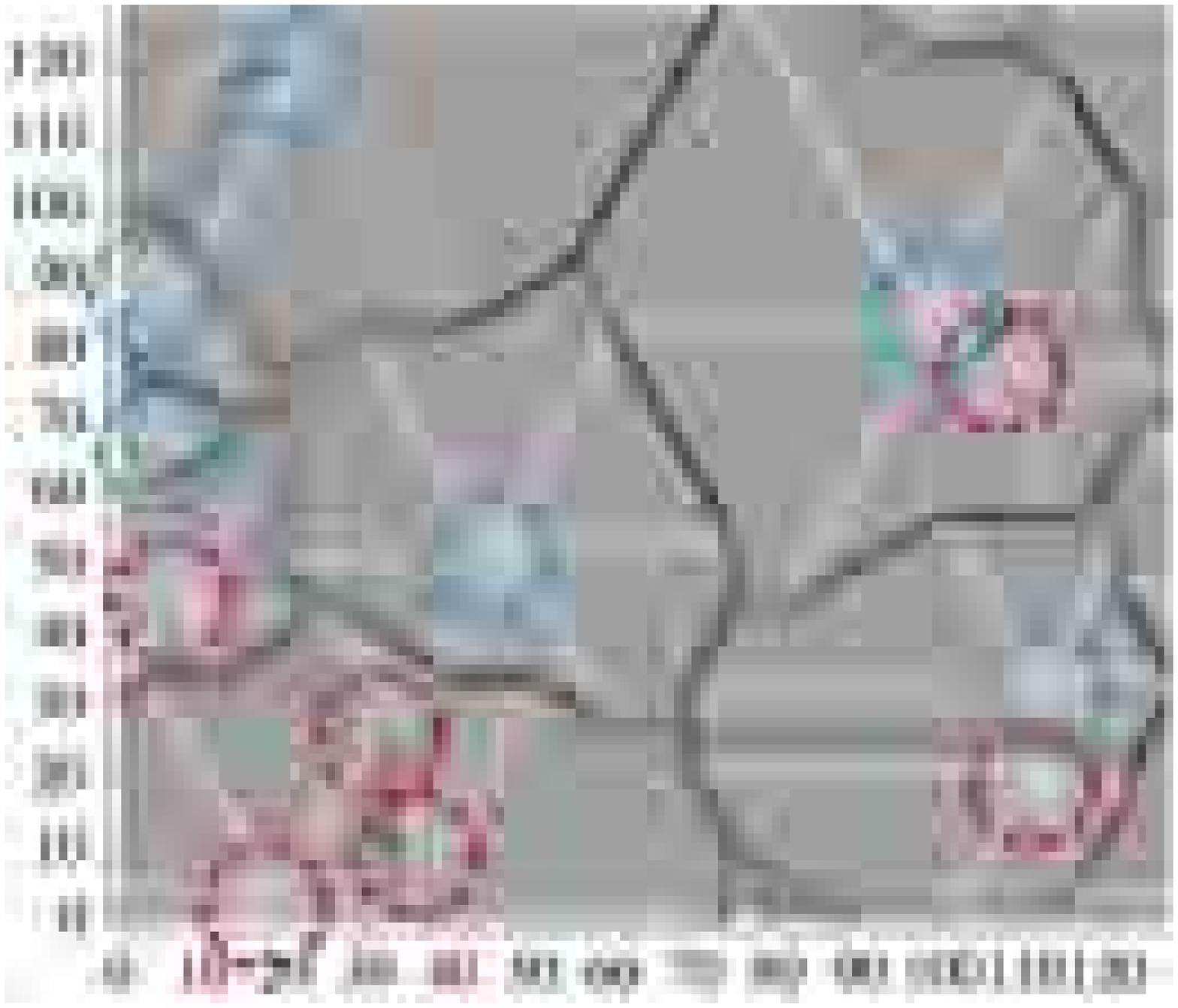}
\centering (b)
\end{minipage}
\\
\begin{minipage}[b]{0.45\textwidth}
\vspace{\baselineskip}
\includegraphics[width=\textwidth]{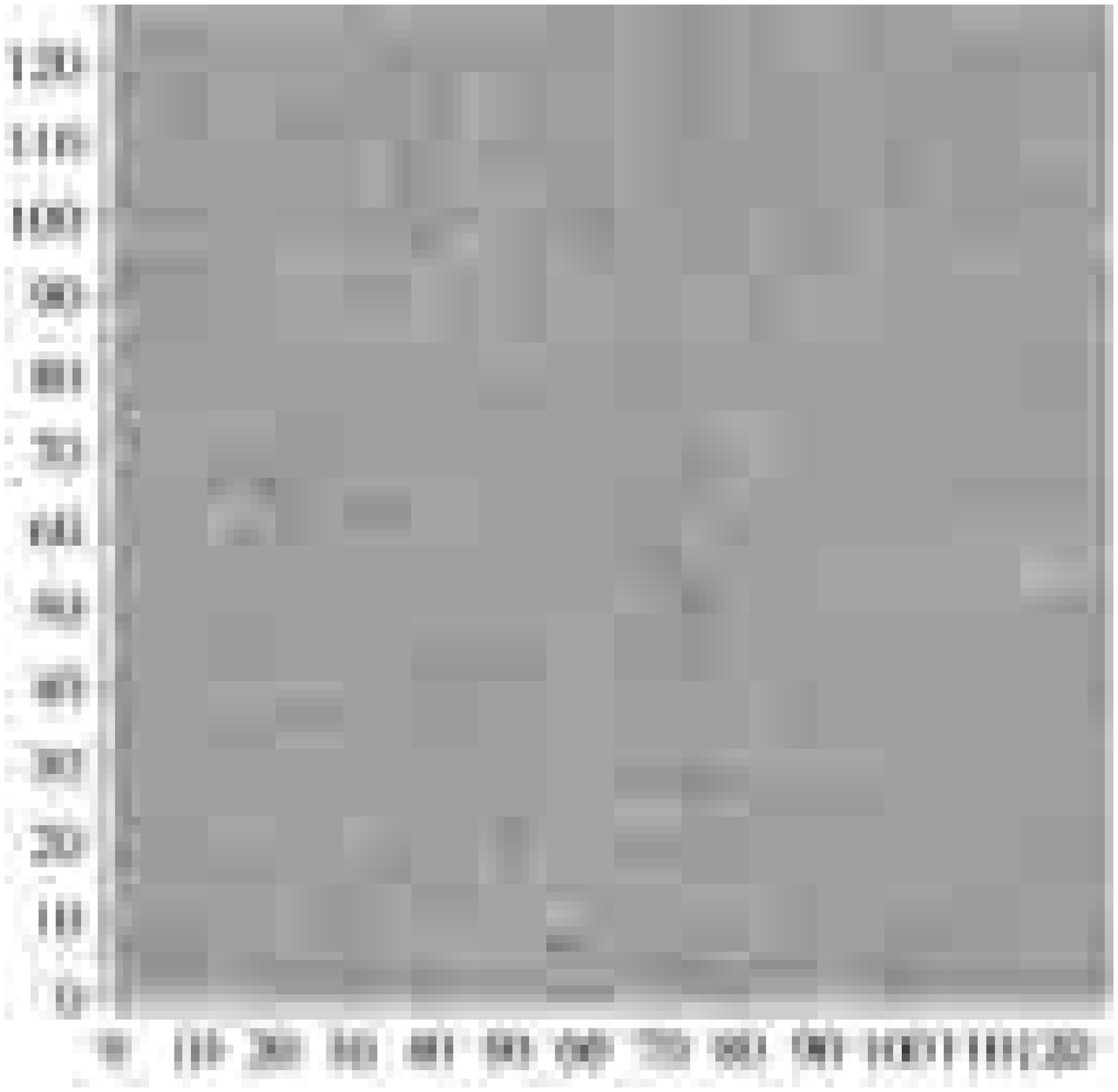}
\centering (c)
\end{minipage}
\hfill
\begin{minipage}[b]{0.45\textwidth}
\vspace{\baselineskip}
\includegraphics[width=\textwidth]{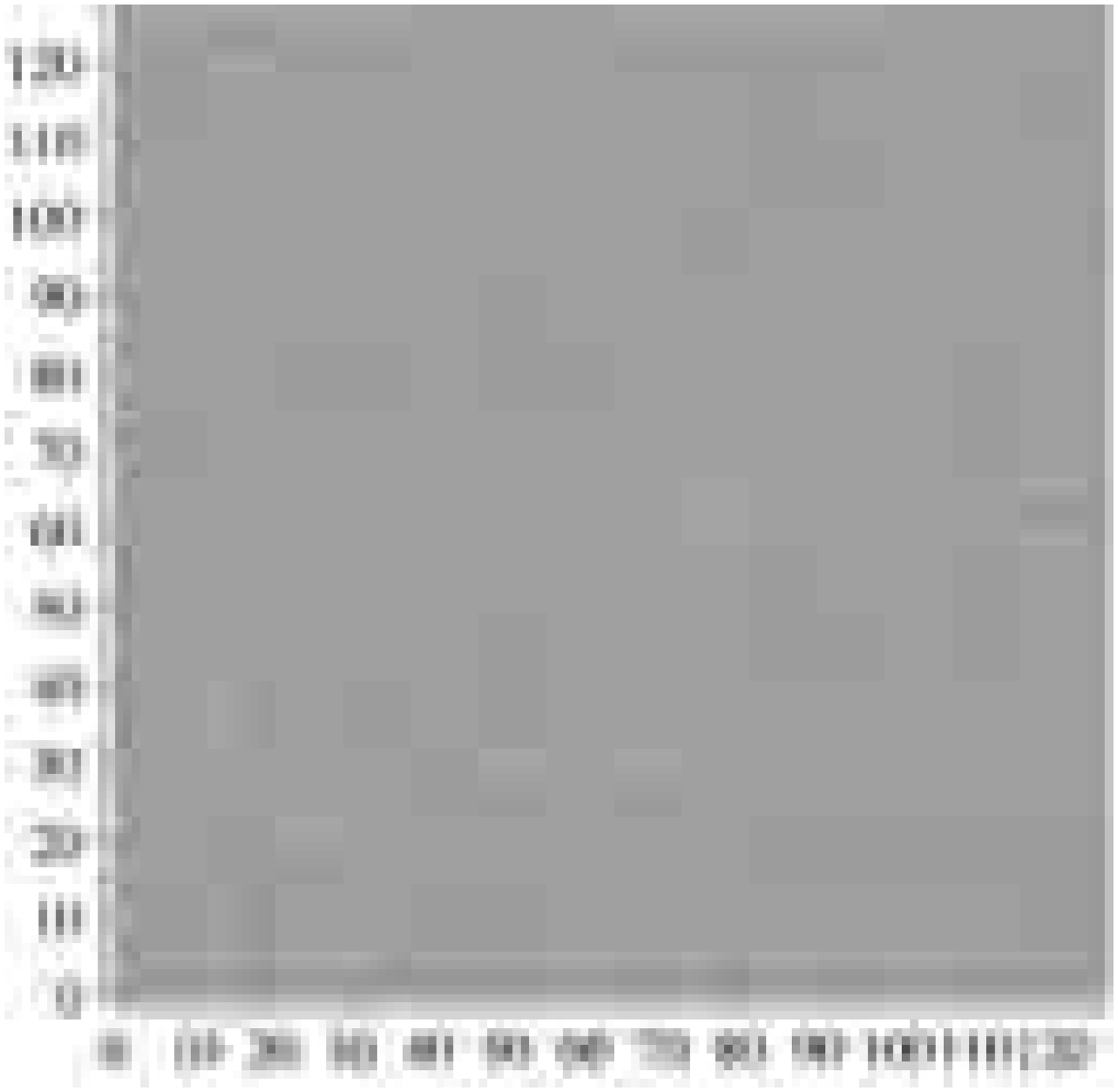}
\centering (d)
\end{minipage}
\\
\begin{minipage}[b]{0.45\textwidth}
\vspace{\baselineskip}
\includegraphics[width=\textwidth]{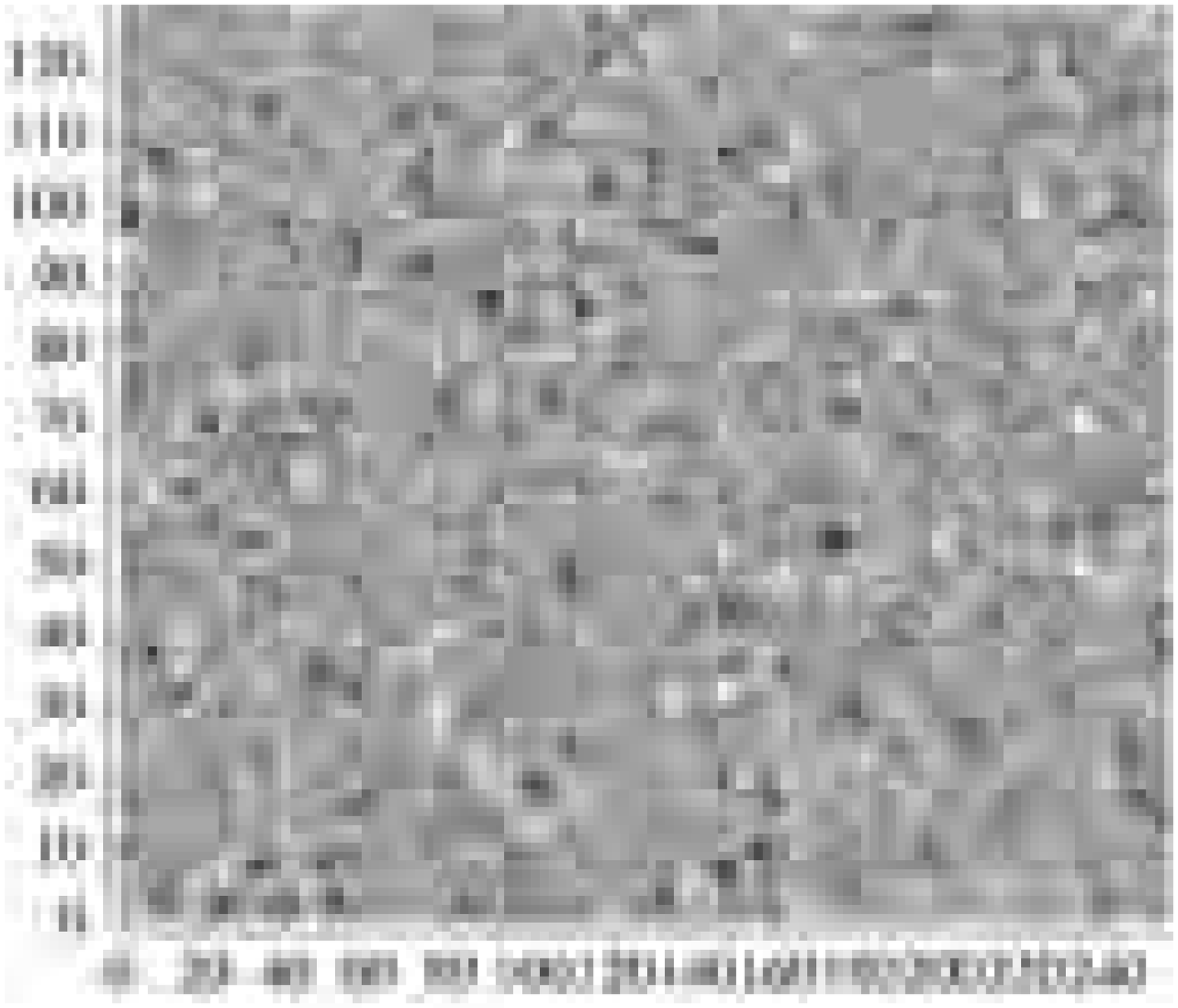}
\centering (e)
\end{minipage}
\hfill
\begin{minipage}[b]{0.45\textwidth}
\vspace{\baselineskip}
\includegraphics[width=\textwidth]{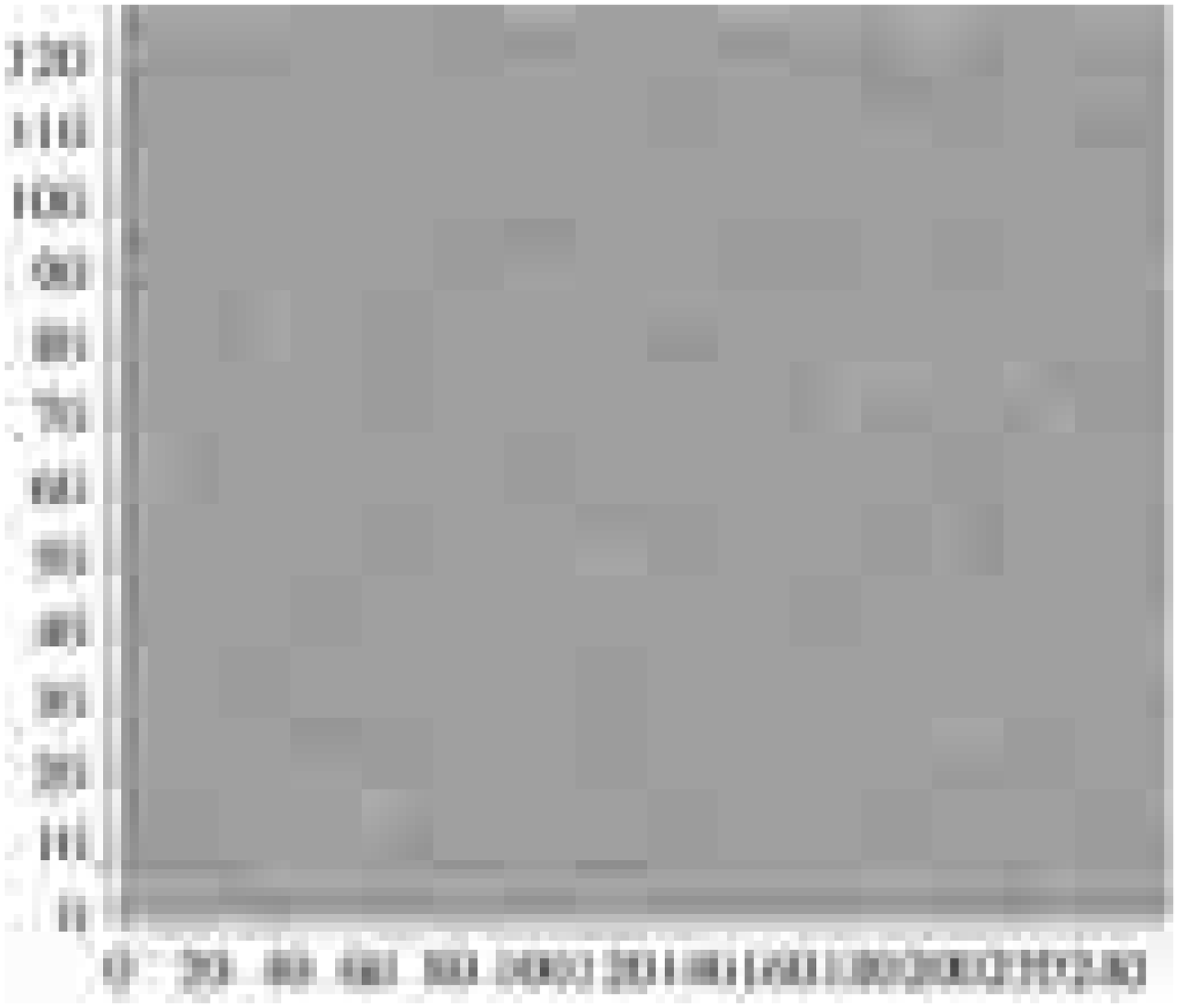}
\centering (f)
\end{minipage}
\caption{The spatial distribution of the time-averaged frustration and the time-averaged score. 
(a) and~(b) The triangular lattice with  $2^{14}$ players.
The red circles indicate positive frustrations and the blue circles indicate negative frustrations.
(c) and~(d) The square lattice with  $2^{14}$ players.
(e) and~(f) The honeycomb lattice with  $2^{15}$ players.
In the panels~(a), (c) and~(e), white symbols indicate the time-averaged frustration more than $0.1$, black symbols indicate the time-averaged frustration less than $-0.1$ and gray symbols with gradation indicate the time-averaged frustration in between.
In the panels~(b), (d) and~(f), white symbols indicate the time-averaged score more than $1.0$, black symbols indicate the time-averaged score less than $-1.0$ and gray symbols with gradation indicate the time-averaged score in between.}
\label{fig14}
\end{figure}

As a piece of evidence for the essential difference, we first show the spatial distribution of the time-averaged frustration.
Figure~\ref{fig14}~(a) shows the time-averaged frustration on the \textit{triangular} lattice with $2^{14}$ players.
The time average was taken over 200 steps after the 5~000th step.
We can see that the frustrations on the triangular lattice (the red and blue circles) remain mostly at the same positions over the 200 steps.
The spatial distribution of the time-averaged score (Fig.~\ref{fig14}~(b)) shows a corresponding structure, where players closer to the vortex cores get higher scores.

We do not see such structures on the square and honeycomb lattices (Fig.~\ref{fig14}~(c)--(f)).
(We define the frustration on the square and honeycomb lattices similarly to the definition for the triangular lattice as in Sec.~\ref{sec2-1}.
On the honeycomb lattice, we can have from a $-2$ frustration to a $+2$ frustration on a honeycomb plaquette.)
In Fig.~\ref{fig14}~(c) and~(e), we can vaguely see vortex cores crawl around over the 200 steps.
They do not stay at the same positions.
The corresponding spatial distributions of the time-averaged score do not show steady patterns.

\begin{figure}
\begin{center}
\includegraphics[width=0.55\textwidth]{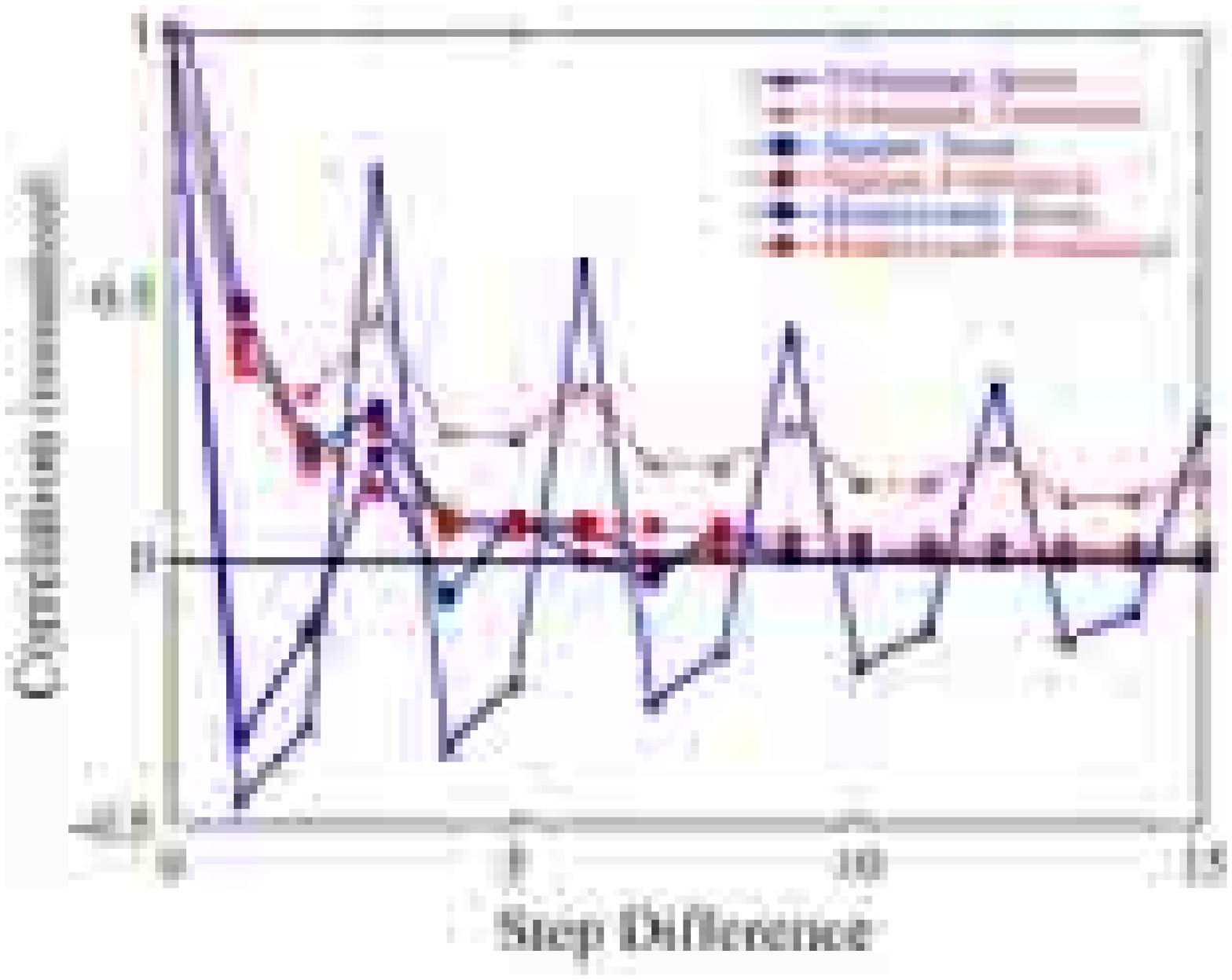}
\end{center}
\caption{The auto-correlations of the score and the frustration, normalized so that they may be unity on the left end of the graph.
Triangles on a solid (dotted) line denotes the auto-correlation of the score (the frustration) of a simulation on the triangular lattice with $2^{14}$ players.
Squares on a solid (dotted) line denotes the auto-correlation of the score (the frustration) of a simulation on the square lattice with $2^{14}$ players.
Hexagons on a solid (dotted) line denotes the auto-correlation of the score (the frustration) of a simulation on the honeycomb lattice with $2^{15}$ players.
Each data point represents the spatial average as well as the time average over the 10~000 steps after the 3~000th step.}
\label{fig15}
\end{figure}

In order to show further the difference between the non-bipartite triangular lattice and the bipartite lattices, we plot in Figure~\ref{fig15} the auto-correlations of the score and the frustration on the three lattices.
We can clearly see that the auto-correlations of the triangular lattice are one-order magnitude greater than the auto-correlations of the square lattice and the honeycomb lattice.
We thereby conclude that vortices seen in Fig.~\ref{fig13} on the honeycomb lattice are not stationary in time and does not grow spatially.

\section{Effect of random players on the triangular lattice}
\label{sec5}

We now introduce random players among the copy players on the triangular lattice.
We show that a random player can be a source, which was not present in the system with copy players only.

\subsection{One random player}

\begin{figure}
\begin{minipage}[b]{0.45\textwidth}
\vspace{0pt}
\includegraphics[width=\textwidth]{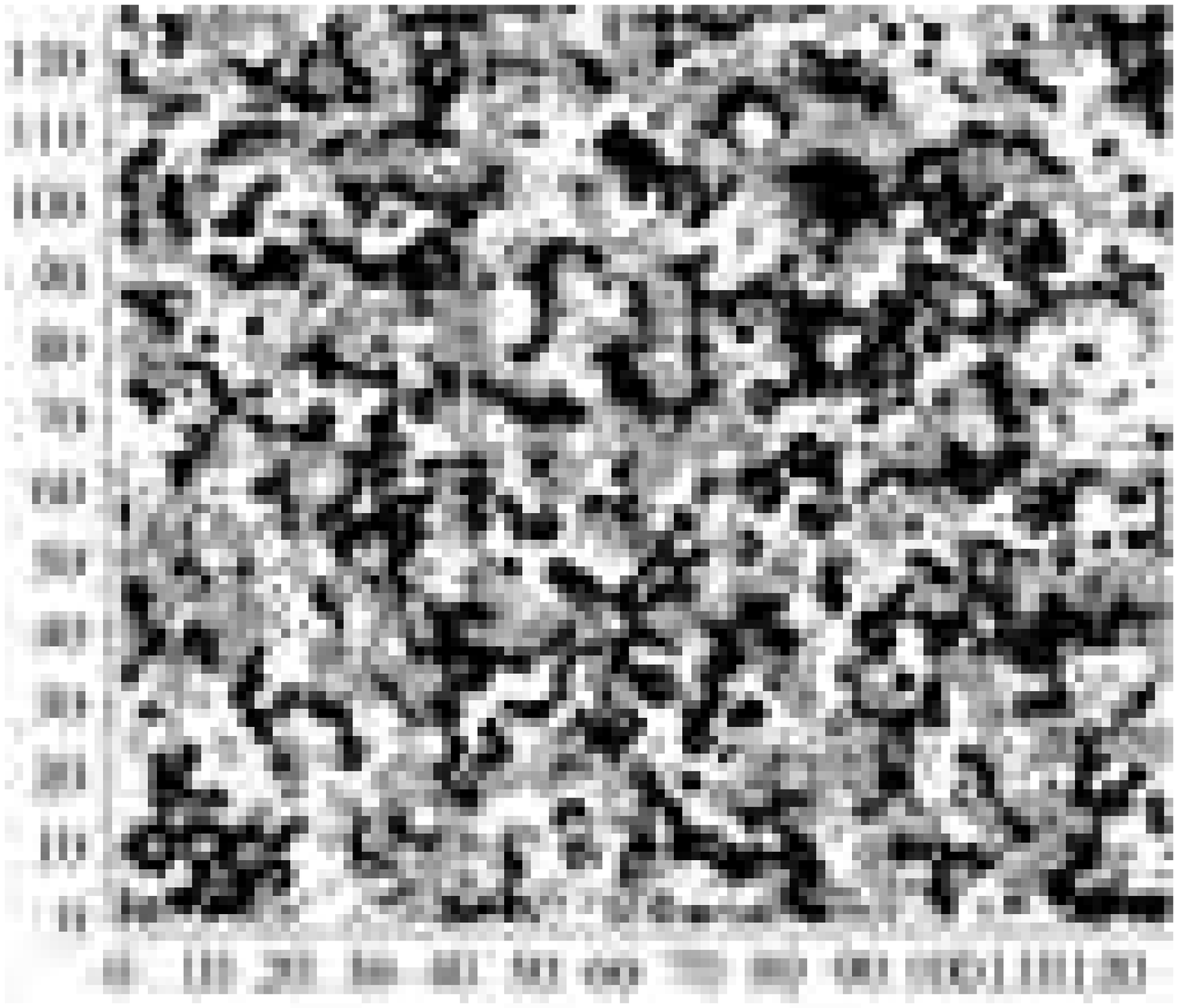}
\centering (a)
\end{minipage}
\hfill
\begin{minipage}[b]{0.45\textwidth}
\vspace{0pt}
\includegraphics[width=\textwidth]{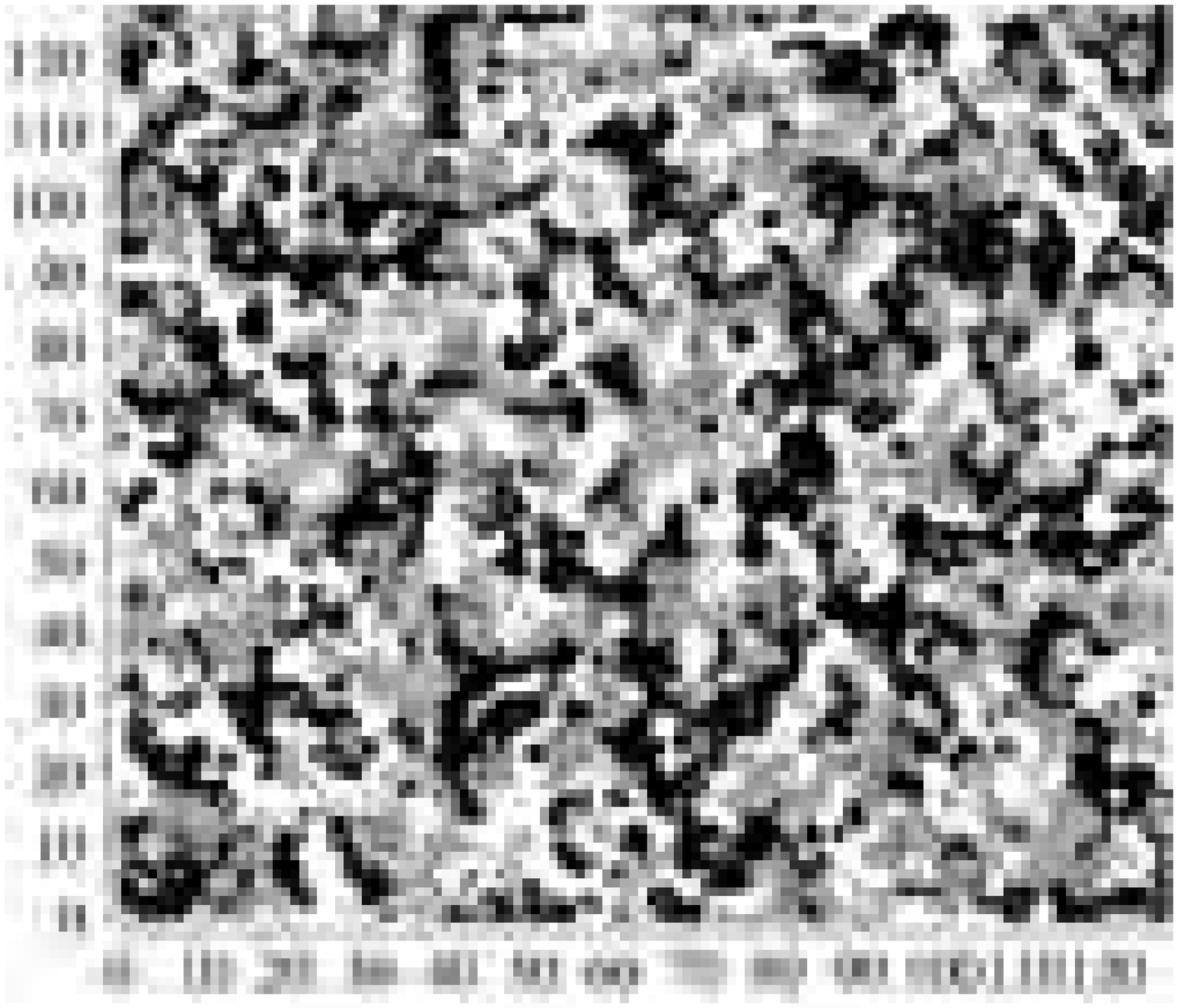}
\centering (b)
\end{minipage}
\\
\begin{minipage}[b]{0.45\textwidth}
\vspace{\baselineskip}
\includegraphics[width=\textwidth]{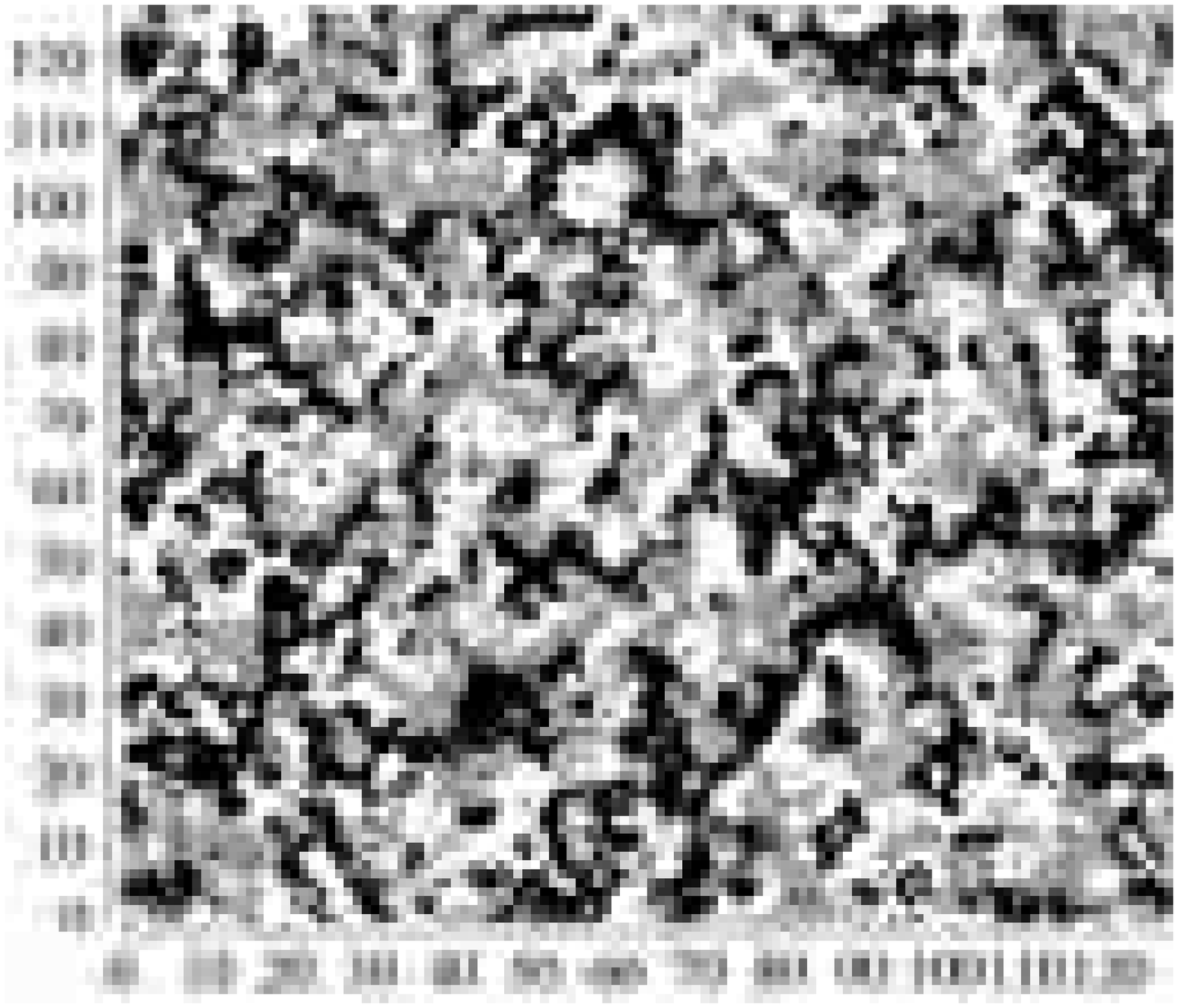}
\centering (c)
\end{minipage}
\hfill
\begin{minipage}[b]{0.45\textwidth}
\vspace{\baselineskip}
\includegraphics[width=\textwidth]{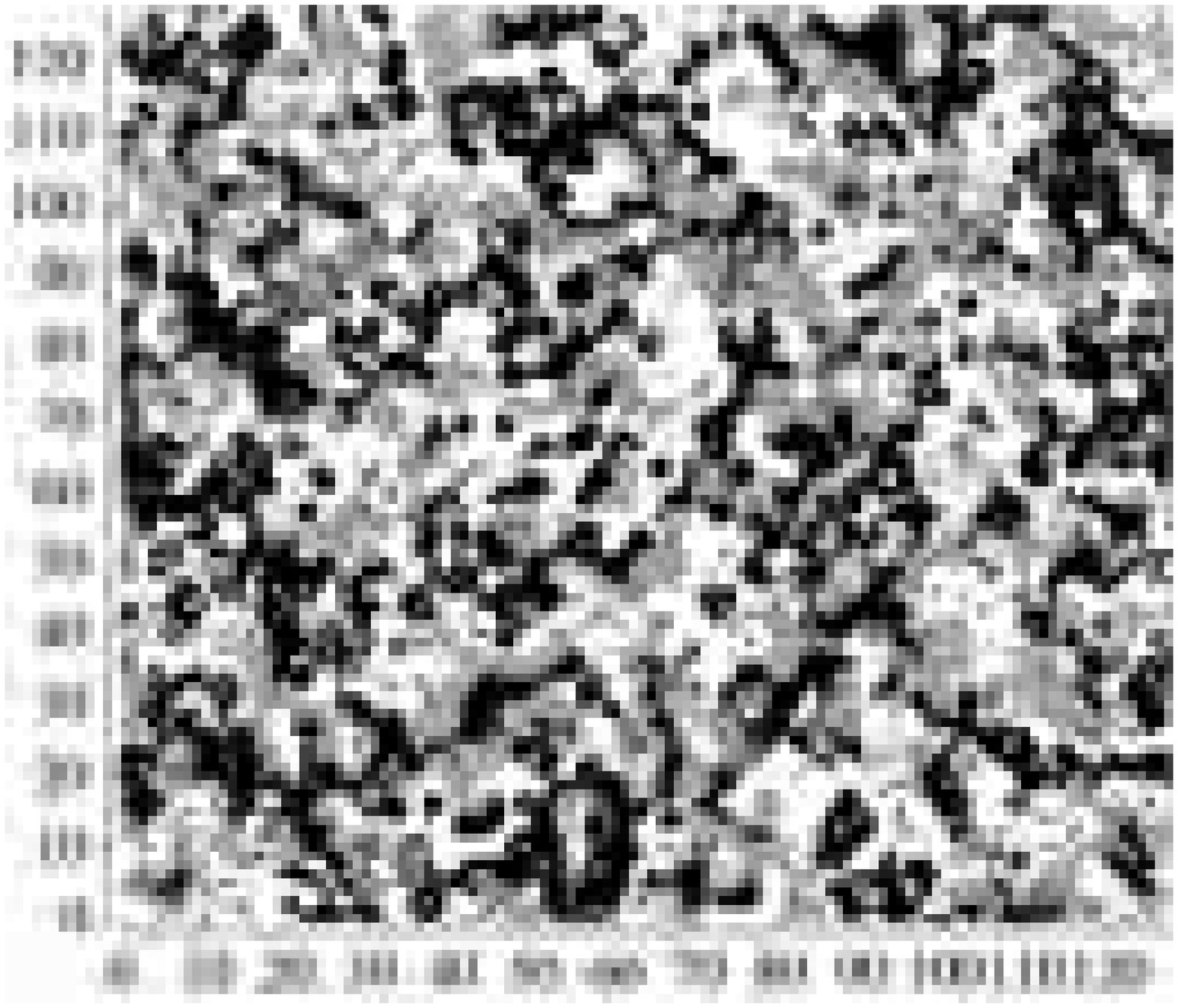}
\centering (d)
\end{minipage}
\\
\begin{minipage}[b]{0.45\textwidth}
\vspace{\baselineskip}
\includegraphics[width=\textwidth]{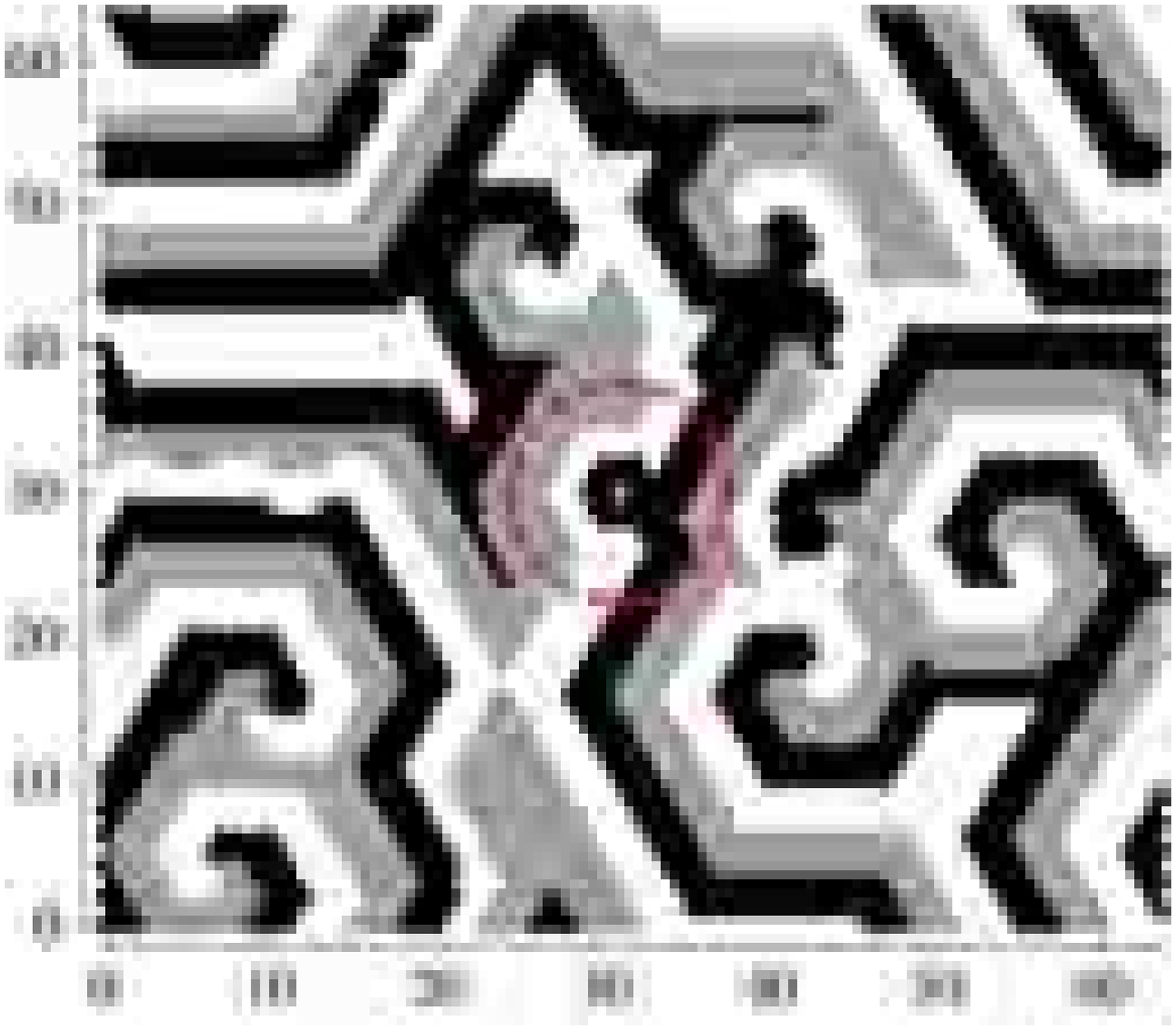}
\centering (e)
\end{minipage}
\hfill
\begin{minipage}[b]{0.45\textwidth}
\vspace{\baselineskip}
\includegraphics[width=\textwidth]{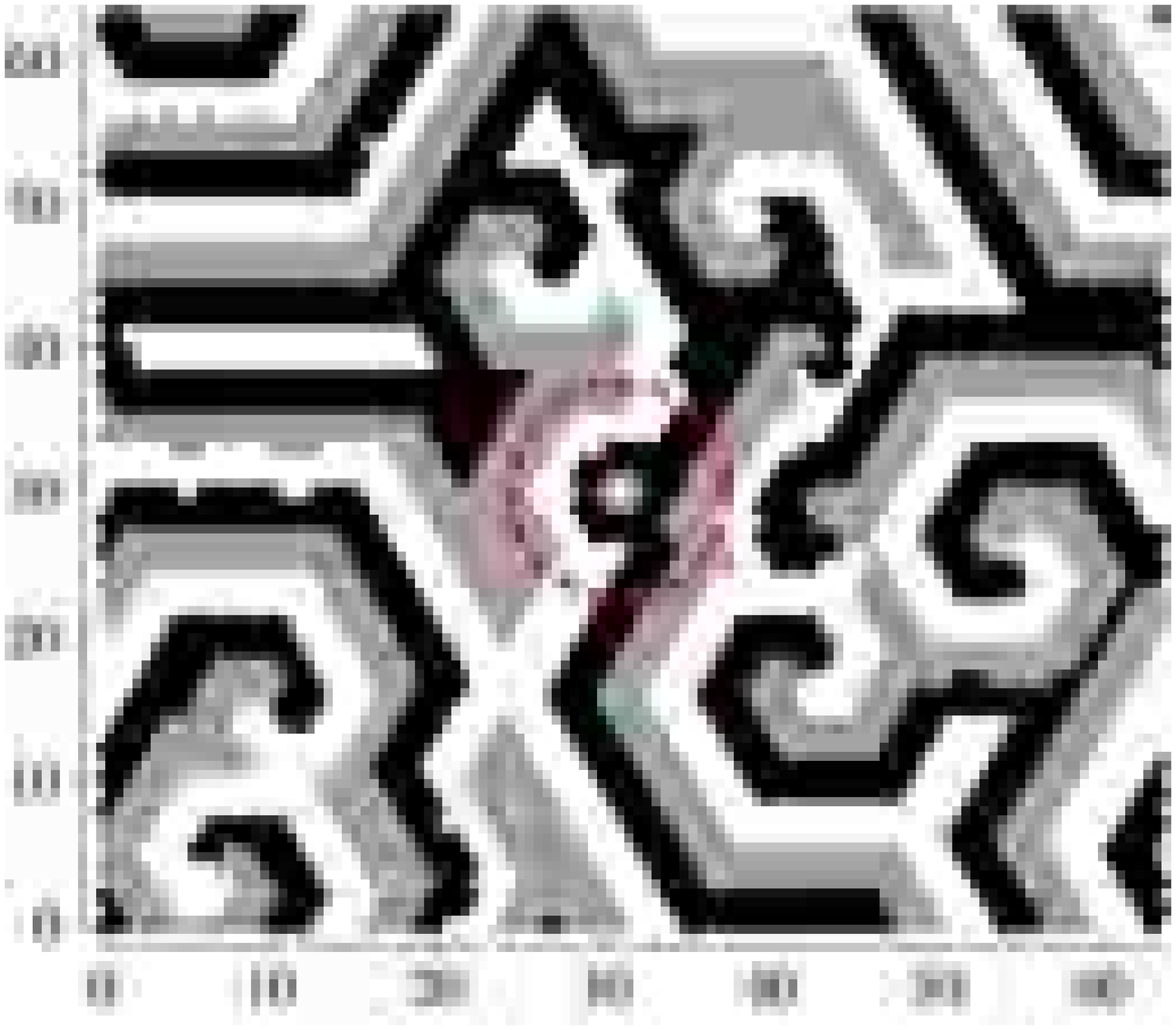}
\centering (f)
\end{minipage}
\caption{The configurations at (a) the 4~870th step, (b) the 4~871st step, (c) the 4~872nd step, (d) the 4~873rd step, (e) the 4~874th step and (f) the 4~875th step of a simulation on the triangular lattice with $2^{12}$ players.
There is only one random player at the position indicated by the red circle; the rest are copy players.
The black hexagons denote the player with the hand~0, the gray hexagons the hand~1, and the white hexagons the hand~2.}
\label{fig16}
\end{figure}

We demonstrate in Fig.~\ref{fig16} that a random player can be a source.
Figure~\ref{fig16} shows snapshots of a simulation on the triangular lattice with one random player and $2^{12}-1$ copy players.

The random player chooses its hand randomly at every step.
When its hand happens to be stronger than the hand of the copy players around the random player (such as in Fig.~\ref{fig16}~(a) when the random player chooses the hand~2 among the copy players of the hand~1), the copy players neighboring the random player will mimic the random player's hand in the next step.
This may propagate as demonstrated in Fig.~\ref{fig16}.
Thus the random player can be a source with the probability of about $1/3$.

\begin{figure}
\begin{center}
\includegraphics[width=0.45\textwidth]{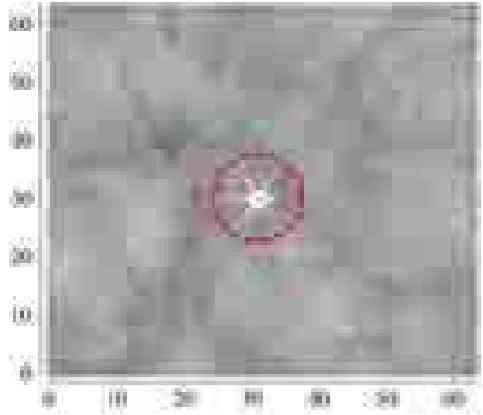}
%
\end{center}
\caption{The spatial distribution of the time-averaged score of a simulation on the triangular lattice with one random player at the center (indicated by the red circle) and $2^{12}-1$ copy players.
The average was take over 10~000 steps after the 3~000th step.
White hexagons indicate the time-averaged score more than $0.5$ and black hexagons indicate the time-averaged score less than $-0.5$.
Gray hexagons with gradation indicate the time-averaged score in between $-0.5$ and $0.5$.
}
\label{fig17}
\end{figure}

We argued in Sec.~\ref{sec2-3} that the players near a sink get lower scores.
Because of the same reason working in the opposite direction, the players around a random player, or a possible source, get higher scores than the average.
(The random player itself obviously gets the average.)
Figure~\ref{fig17} shows that the players around the random player have higher scores than the average.

\subsection{Many random players}

\begin{figure}
\begin{minipage}[b]{0.45\textwidth}
\vspace{0pt}
\includegraphics[width=\textwidth]{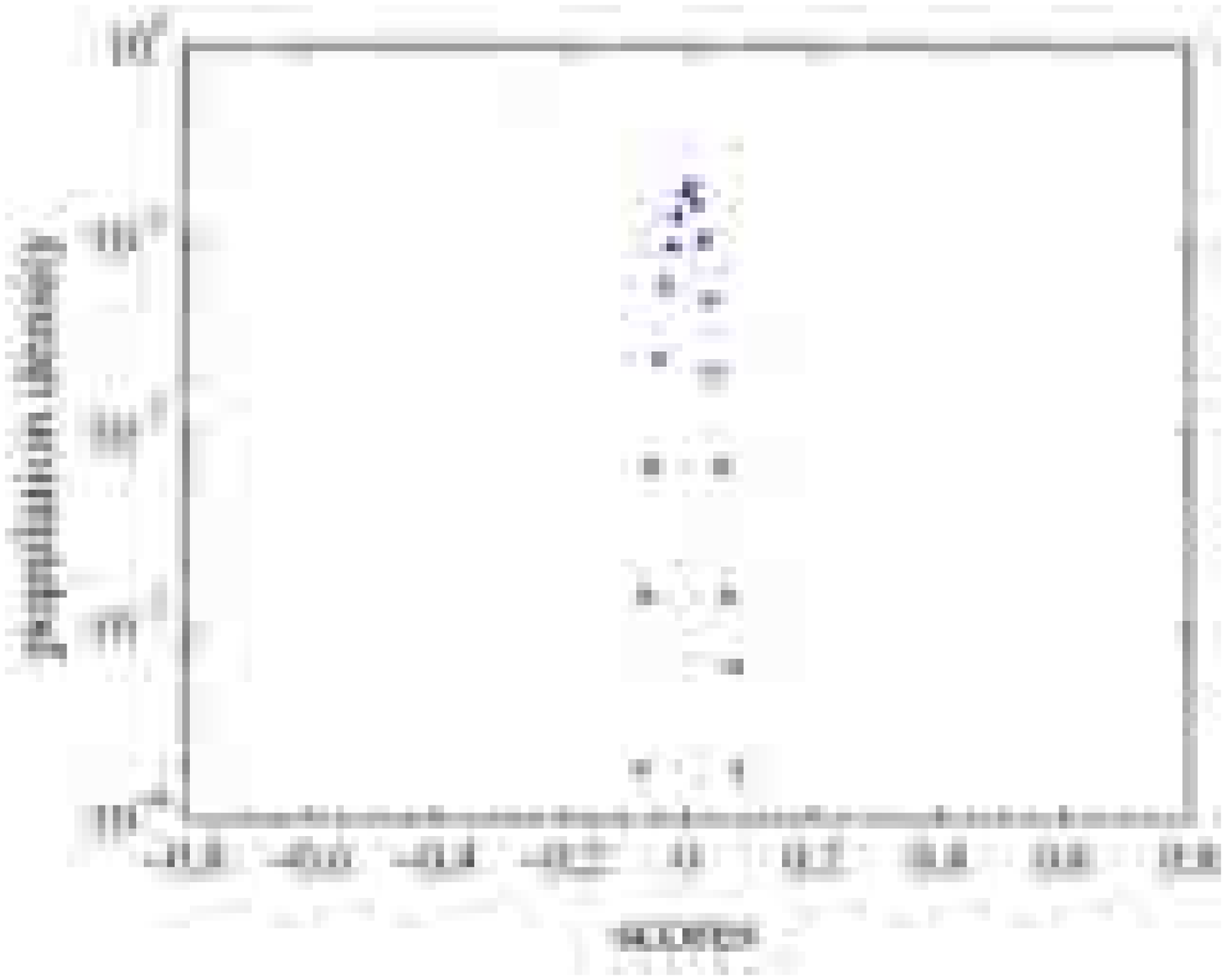}
\centering (a)
\end{minipage}
\hfill
\begin{minipage}[b]{0.45\textwidth}
\vspace{0pt}
\includegraphics[width=\textwidth]{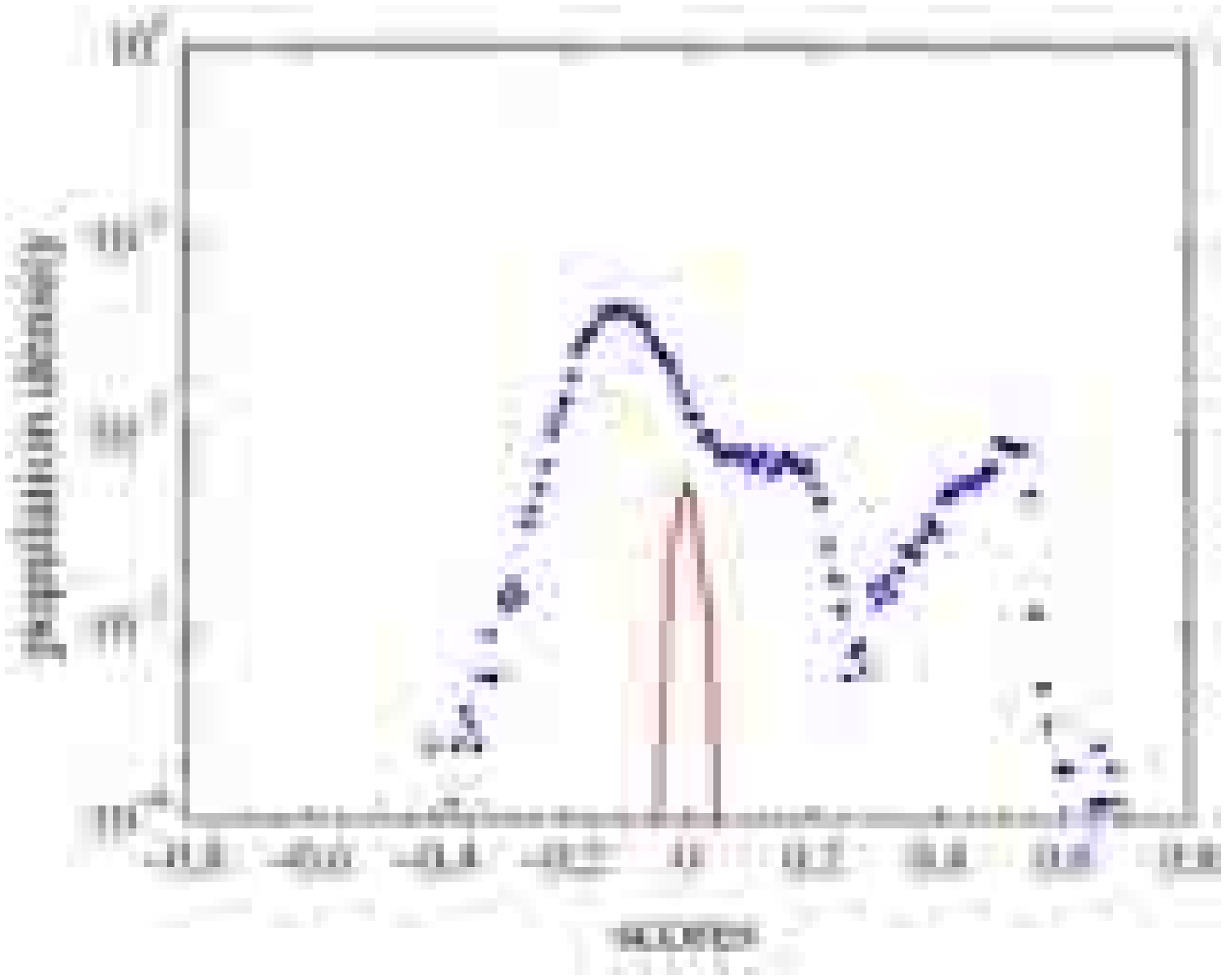}
\centering (b)
\end{minipage}
\\
\begin{minipage}[b]{0.45\textwidth}
\vspace{\baselineskip}
\includegraphics[width=\textwidth]{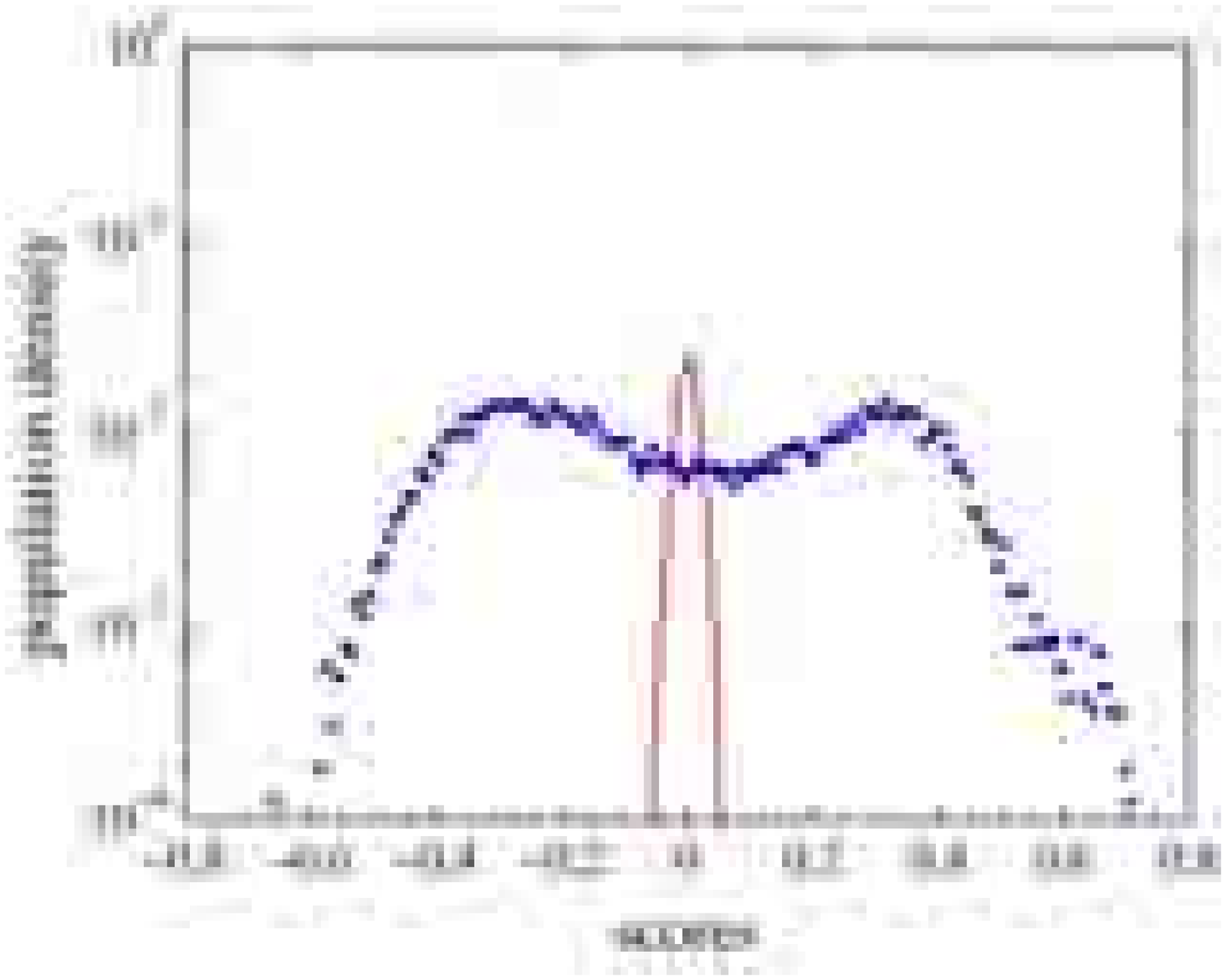}
\centering (c)
\end{minipage}
\hfill
\begin{minipage}[b]{0.45\textwidth}
\vspace{\baselineskip}
\includegraphics[width=\textwidth]{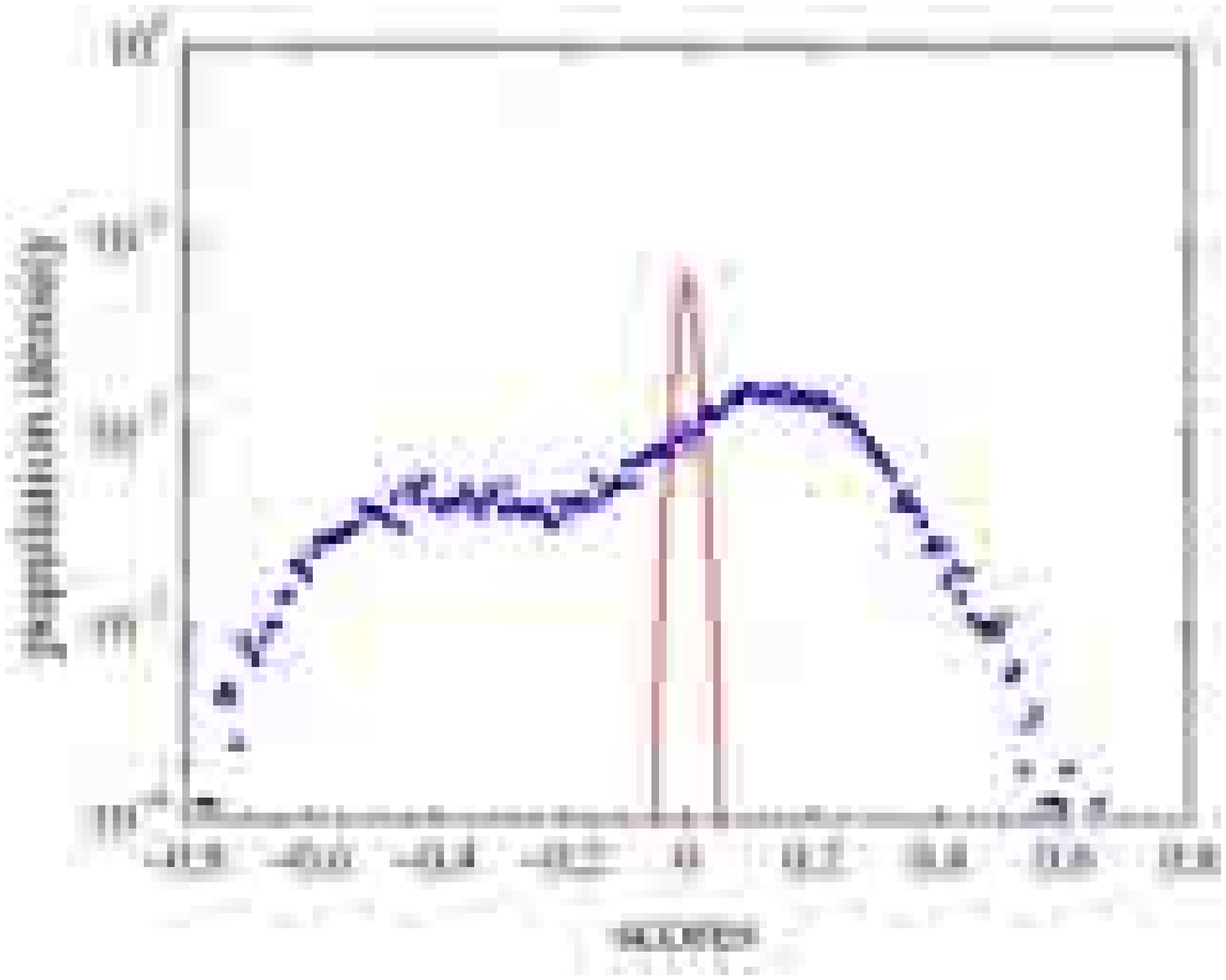}
\centering (d)
\end{minipage}
\\
\begin{center}
\includegraphics[width=0.45\textwidth]{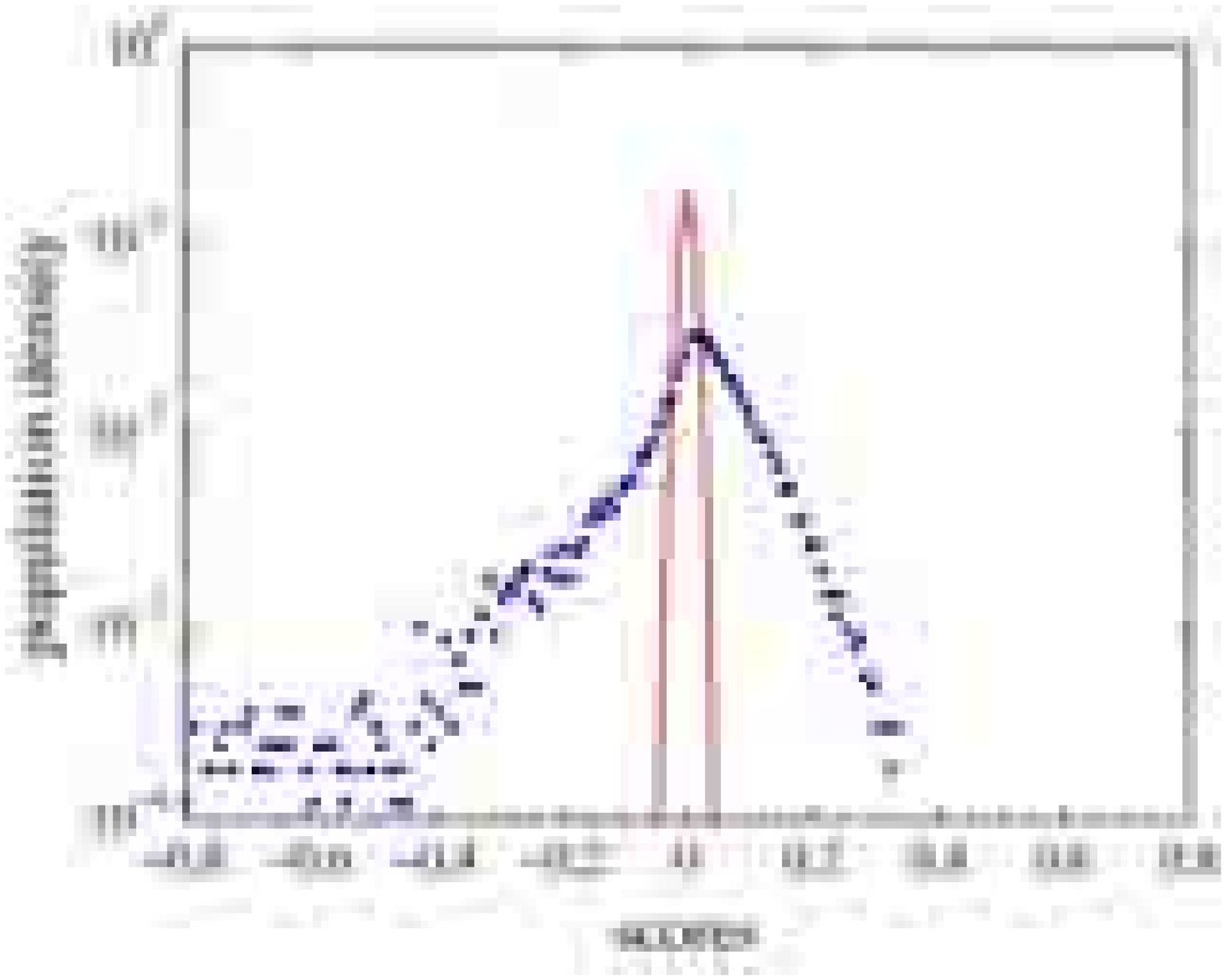}

(e)
\end{center}
\caption{Semi-logarithmic plots of the population density distribution of the time-averaged score of the copy players and the random players over 50~000 steps after the 3~000th step of a simulation on the triangular lattice with $2^{14}$ players in total.
In every panel, the solid circles indicate the distribution of the copy players while the solid lines indicate the distribution of the random players.
(a) No random players and $2^{14}$ copy players. 
(b) 400 random players ($2.4\%$). 
(c) 1~600 random players ($9.8\%$).
(d) 3~600 random players ($22.0\%$).
(e) 8~100 random players ($49.4\%$).
Note that a parabola on a semi-logarithmic plot is a Gaussian distribution.}
\label{fig18}
\end{figure}

In this subsection, we randomly scatter many random players over the triangular lattice.
Figure~\ref{fig18} shows the population density distribution of the time-averaged scores of the copy players and the random players of a simulation on the triangular lattice with $2^{14}$ players.
When there is no random players, the distribution of the score is almost Gaussian (Fig.~\ref{fig18}~(a)).
The fluctuation of the score is due to the fact that each player is occasionally close to a vortex, getting high scores, and occasionally close to a sink, getting low scores.
As we introduce a few random players, the copy players just around the random players get scores higher than the average owing to the same reason described in the previous subsection.
This generates the additional peak on the right of the highest peak in Fig.~\ref{fig18}~(b).

\begin{figure}
\begin{minipage}[b]{0.45\textwidth}
\vspace{0pt}
\includegraphics[width=\textwidth]{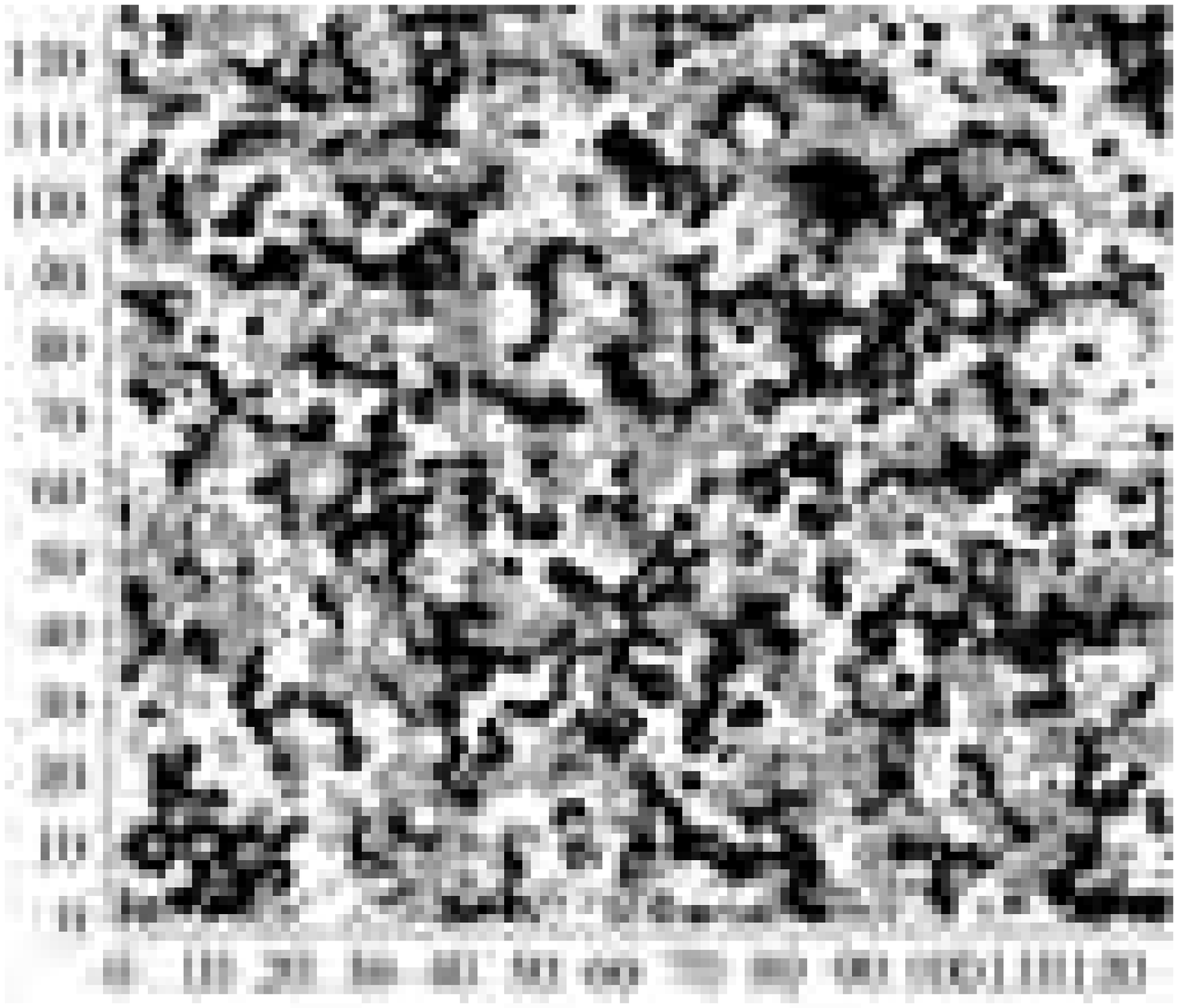}
\centering (a)
\end{minipage}
\hfill
\begin{minipage}[b]{0.45\textwidth}
\vspace{0pt}
\includegraphics[width=\textwidth]{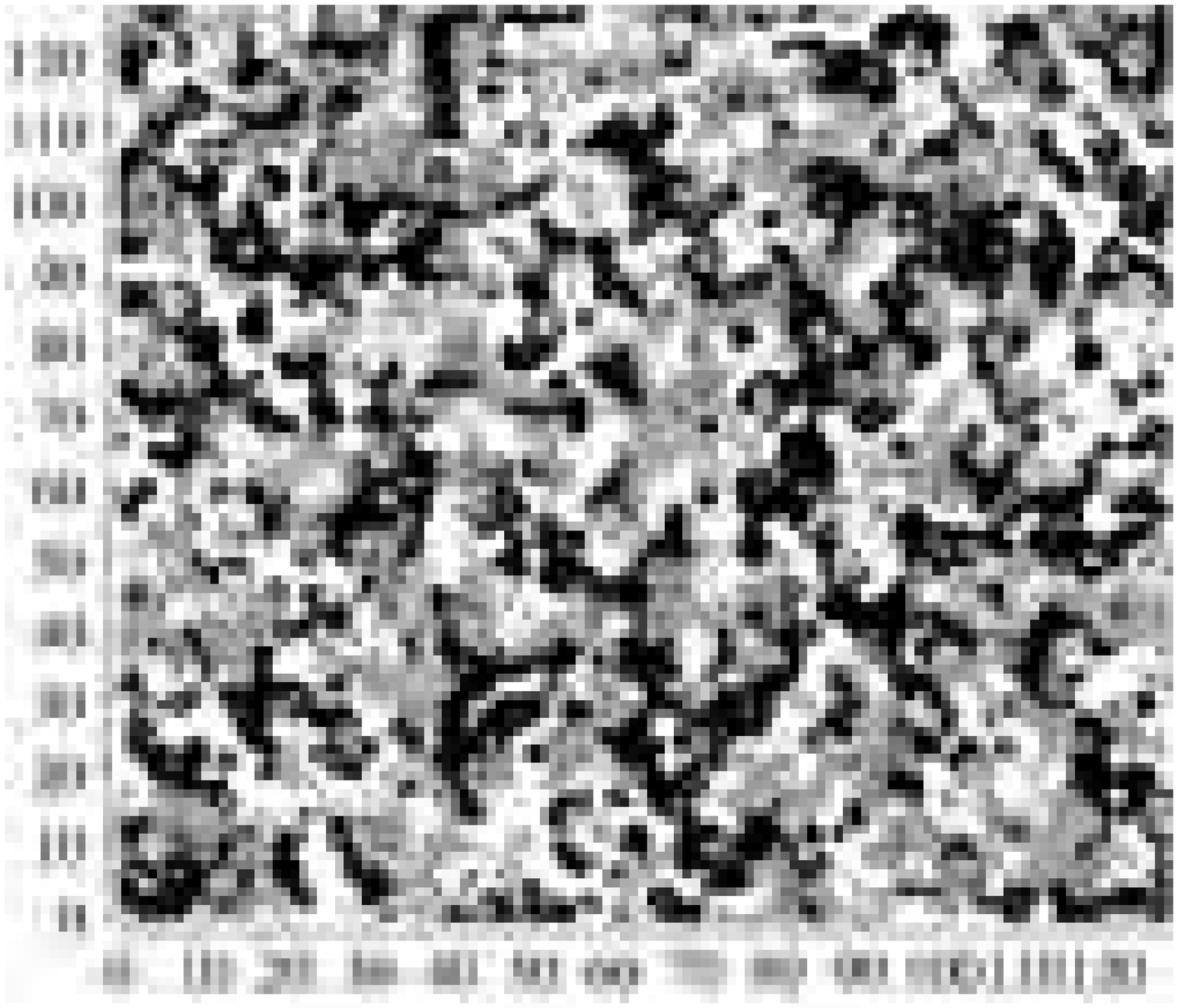}
\centering (b)
\end{minipage}
\\
\begin{minipage}[b]{0.45\textwidth}
\vspace{\baselineskip}
\includegraphics[width=\textwidth]{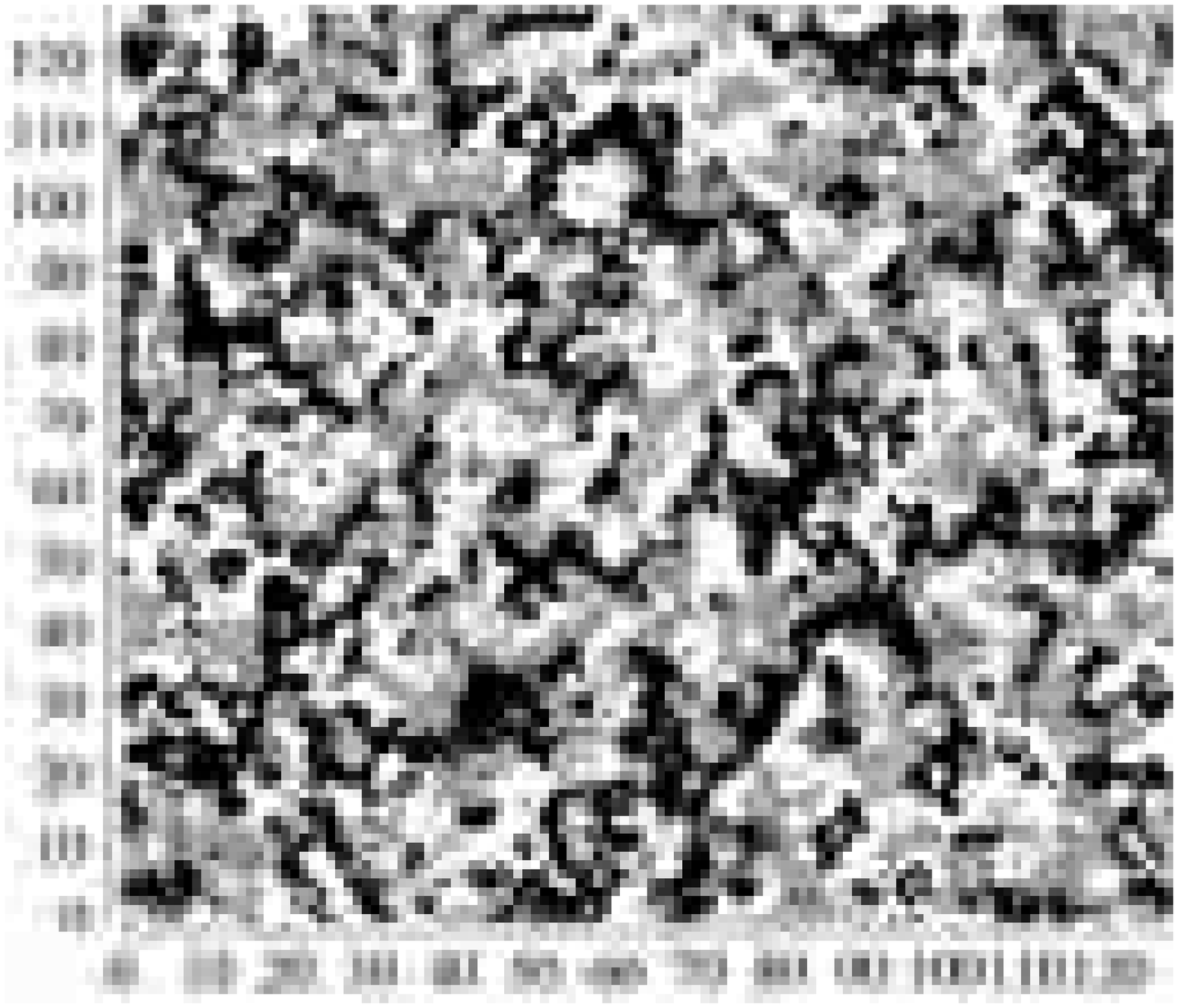}
\centering (c)
\end{minipage}
\hfill
\begin{minipage}[b]{0.45\textwidth}
\vspace{\baselineskip}
\includegraphics[width=\textwidth]{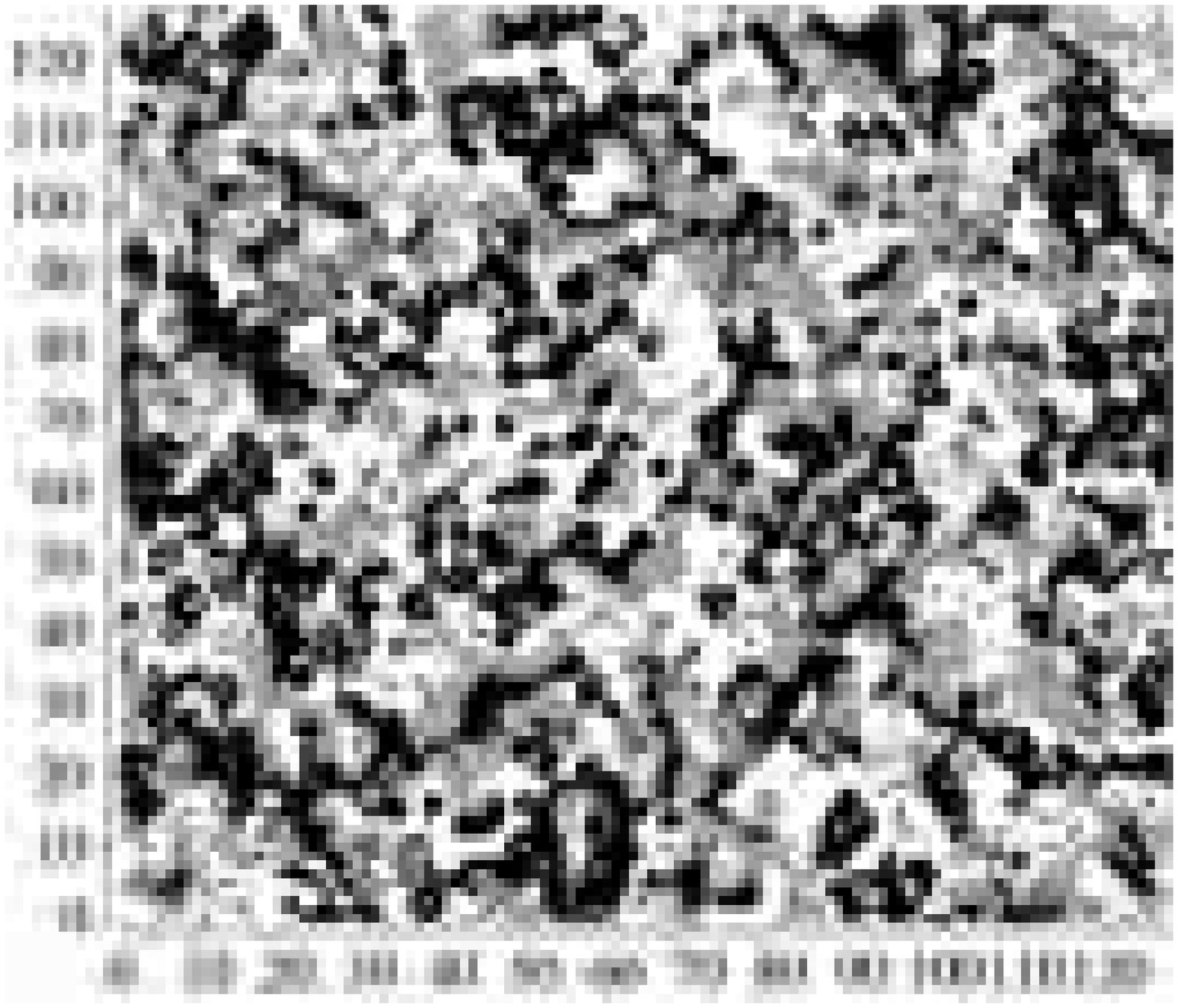}
\centering (d)
\end{minipage}
\caption{The configurations at the 997th step (a), the 998th step (b), the 999 step (c) and the 1~000th step (d) of a simulation on the triangular lattice. with 1~600 random players ($9.8\%$) and $2^{14}-1600$ copy players
The black squares denote the player with the hand~0, the gray squares the hand~1, and the white squares the hand~2.}
\label{fig19}
\end{figure}

As we increase the number of random players, the vortex structure disappears from the steady pattern (Fig.~\ref{fig19}).
When the number of random players is 1~600 (Fig,~\ref{fig18}~(c)), random players are scattered in the system every three lattice points on average.
This is enough to destroy a vortex which generates the three-layer structure explained in Sec.~\ref{sec3-2}.
The copy players not neighboring the random players cannot get high scores generated by vortices and keep losing scores because of sinks. 
Hence the highest peak in Fig.~\ref{fig18}~(b) shifts in the direction of the lower score in Fig.~\ref{fig18}~(c).
The copy players just around the random players, on the other hand, keep getting scores higher than the average.
The number of such players is increased and hence the side peak in Fig.~\ref{fig18}~(b) has grown in Fig.~\ref{fig18}~(c).
As we further increase the number of random players, the peak on the side of the lower score keeps shrinking and the other peak keeps growing until the latter dominates as in Fig.~\ref{fig18}~(e).

\begin{figure}
\begin{center}
\includegraphics[width=0.55\textwidth]{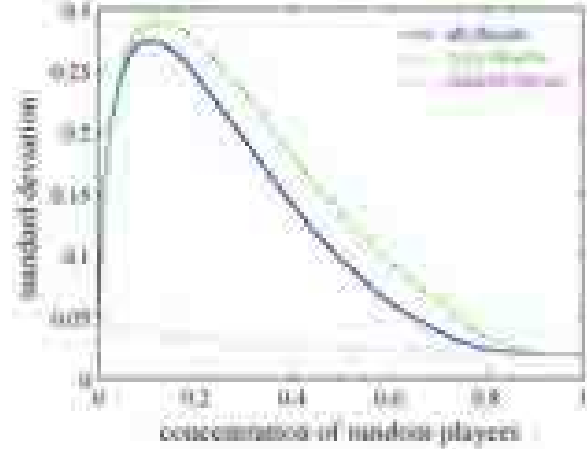}
\end{center}
\caption{The dependence of the standard deviation of the distribution of the time-averaged score on the concentration of random players.
The time-averaged scores of all players (solid line), copy players only (dashed line) and random players only (dotted line).
The average was take over 10~000 steps after the 3~000th step of a simulation on the triangular lattice with $2^{14}$ players in total.}
\label{fig20}
\end{figure}

The width of the distribution is the greatest when random players are about $10\%$ of all players, or in the case Fig.~\ref{fig18}~(c).
Figure ~\ref{fig20} shows how the standard deviation of the distribution of the time-averaged score depends on the concentration of random players.

\section{Summary}
\label{sec-sum}

We introduced a new lattice model of the RSP game with copy players, who mimic the hand of the player with the maximum score.
The key feature is the existence of the frustration, which is the three-sided situation where the hands of the three players on a triangle are all different;
then the hand~1 wins over the hand~0, the hand~2 wins over the hand~1 and the hand~0 wins over the hand~2.
We showed that the frustration generates a stationary vortex on the triangular lattice.

We argued that the structure which consists of vortex pairs, sinks and domains of three layers is stable on the triangular lattice.
The structure does not appear on the square lattice nor on the honeycomb lattice.

Finally, we introduced random players, each of which chooses the hand randomly at every step.
A random player can be a source, which was not existent in the copy society.
Random players of about $10\%$ destroy the structure of vortex pairs.

\section*{Acknowledgement}
The authors express their sincere gratitude to Dr.~Naoki Masuda for many valuable comments to the earlier version of the paper.

\end{document}